\documentclass[preprint, pra]{revtex4}
\usepackage{graphicx}
\usepackage{dcolumn}
\usepackage{bm}
\usepackage{setspace}
\usepackage{booktabs}

\graphicspath{{../arXiv/}}

\def\ep{|e+ \rangle}
\def\em{|e-\rangle}
\def\ap{|A+\rangle}
\def\am{|A-\rangle}

\usepackage{xcolor}
\definecolor{darkGreen}{rgb}{0, 0.7, 0}

\newcommand{\be}{\begin{equation}}
\newcommand{\ee}{\end{equation}}

\begin{document}

\title{Realization of a complete Stern-Gerlach interferometer}

\author{Yair Margalit}
	\affiliation{Department of Physics, Ben-Gurion University of the Negev, Be'er Sheva 84105, Israel}	
\author{Zhifan Zhou}	
	\affiliation{Department of Physics, Ben-Gurion University of the Negev, Be'er Sheva 84105, Israel}
\author{Or Dobkowski}
	\affiliation{Department of Physics, Ben-Gurion University of the Negev, Be'er Sheva 84105, Israel}	
\author{Yonathan Japha}
	\affiliation{Department of Physics, Ben-Gurion University of the Negev, Be'er Sheva 84105, Israel}	
\author{Daniel Rohrlich}
	\affiliation{Department of Physics, Ben-Gurion University of the Negev, Be'er Sheva 84105, Israel}	
\author{Samuel Moukouri}
	\affiliation{Department of Physics, Ben-Gurion University of the Negev, Be'er Sheva 84105, Israel}	
\author{Ron Folman}
	\thanks{Corresponding author}
	\email{folman@bgu.ac.il}
	\affiliation{Department of Physics, Ben-Gurion University of the Negev, Be'er Sheva 84105, Israel}

\begin{abstract}
The Stern-Gerlach (SG) effect, discovered almost a century ago, has become a paradigm of quantum mechanics. Surprisingly there is little evidence that the original scheme with freely propagating atoms exposed to gradients from macroscopic magnets is a fully coherent quantum process. Specifically, no high-visibility spatial interference pattern has been observed with such a scheme, and furthermore no full-loop SG interferometer has been realized with the scheme as envisioned decades ago. On the contrary, numerous theoretical studies explained why it is a near impossible endeavor. Here we demonstrate for the first time both a high-visibility spatial SG interference pattern and a full-loop SG interferometer, based on an accurate magnetic field, originating from an atom chip, that ensures coherent operation within strict constraints described by previous theoretical analyses. This also allows us to observe the gradual emergence of time-irreversibility as the splitting is increased. Finally, achieving this high level of control over magnetic gradients may facilitate technological applications such as large-momentum-transfer beam splitting for metrology with atom interferometry, ultra-sensitive probing of electron transport down to shot-noise and squeezed currents, as well as nuclear magnetic resonance and compact accelerators.
\end{abstract}

\maketitle

The discovery of the Stern-Gerlach (SG) effect\,\cite{stern-gerlach,SLB} was followed by ideas concerning a SG interferometer (SGI) consisting of a freely propagating atom exposed to magnetic gradients from macroscopic magnets \,\cite{Wigner}. However, starting with Heisenberg, Bohm and Wigner \cite{briegel} a coherent SGI was considered
impractical because it was thought that the macroscopic device could not be precise enough to ensure a reversible splitting process.
Bohm, for example, noted that the magnet would need to have ``fantastic" accuracy\,\cite{Bohm}.
Englert, Schwinger and Scully analyzed the effect in more detail and coined it the Humpty-Dumpty (HD)
effect \cite{ESS_1,ESS_2,ESS_3}. They too concluded that for
significant coherence to be observed, exceptional precision would be required.
The HD effect illustrates how the difficulty in achieving coherence, due to imprecise experimental quantum operations, is related to the practical irreversibility of quantum processes, and indeed Englert has emphasized more recently the role played in the SGI by the emergence of time-irreversibility (TI) in quantum theory \cite{englert_1,englert_2}. (In \cite{SM} we define TI rigorously.)
Later work added the effect of dissipation and suggested that low-temperature magnetic field
sources would enable an operational SGI\,\cite{caldeira}. Claims have even been made that no coherent splitting is possible at all \cite{Devereux}.

As shown in Fig.\,1, we have measured a SGI coherence of 99\% and 95\% with spatial and spin interference signals, respectively. We achieve both with highly accurate macroscopic magnets at room temperature, whereby a freely propagating atom is exposed to magnetic gradients. Following in the footsteps of impressive endeavors\,\cite{OldSG0,OldSG2,OldSG3,OldSG4,OldSG5,OldSG6,OldSG7,OldSG8,OldSG9,OldSG10, Marechal2000} this is, to the best of our knowledge, the first complete realization of a SG interferometer analogous to that originally envisioned, showing that indeed this textbook example of a quantum device is sound. In addition, we note that in the regime of weak quantum decoherence as we have, the two experiments presented here demonstrate the emergence of TI from two origins: instability and imprecision of quantum operations. As shown in the following, we are able to suppress instability to a high degree and thus our signal is a measure of imprecision, the focus of the HD effect. To the best of our knowledge, this is a first demonstration of the HD effect.

Let us briefly note that spin-dependent forces have been realized many years ago (e.g. \cite{Monroe1996}) and are still frequently used (e.g. \cite{Mizrahi2013, Mandel2003, Steffen2012, Kienzler2016}). However, these utilize laser fields. Entanglement between spin and motional degrees of freedom with magnetic gradients is also used, for example, for quantum gates (e.g. \cite{Johanning2009}), or for precision measurement (e.g. \cite{Werth2002}), but to the best of our knowledge, no spatial interferometry has been realized.

\begin{figure}
\begin{center}
\includegraphics[width=\textwidth]{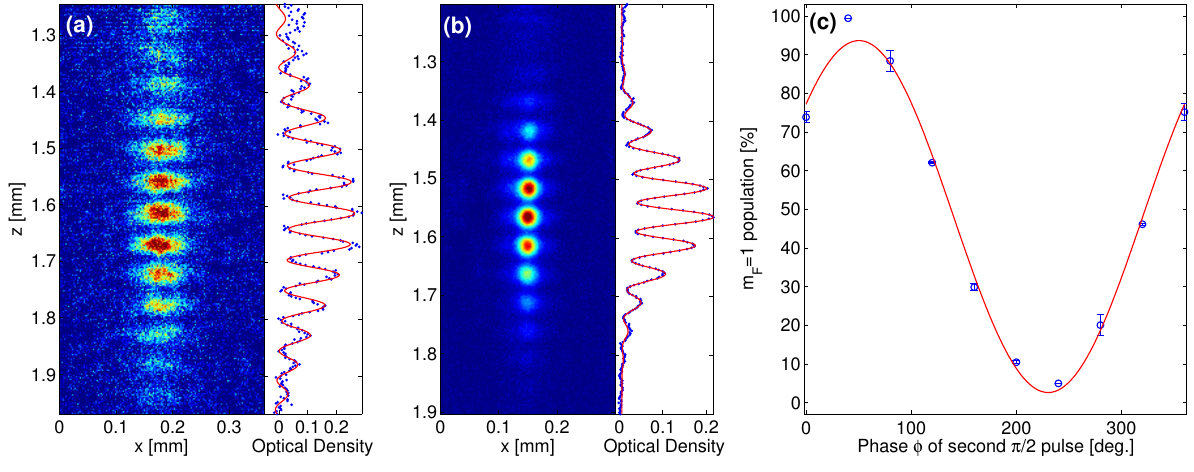}
\end{center}
\caption{ Interference patterns of a Stern-Gerlach interferometer (SGI). (a) Half-loop interferometer (see Fig.\,2 for definition): A single-shot interference pattern of a thermal cloud (BEC fraction of $\sim$0\%), with a visibility of 0.65 (only slightly lower than the single-shot visibility of a BEC). This shows that our SGI is robust to initial state uncertainties and does not rely on the inherent coherence of a BEC (${z}$ is the distance from the chip in the direction of gravity).
(b) A multi-shot image made of a sum (average) of 40 consecutive interference images of a half-loop SGI with a BEC (no correction or post-selection).  The normalized visibility is approximately $99\%$ (see\,\cite{SM} for a polar plot of the phases).
This high stability, together with the low decoherence rate (see text), allows testing the precision of the magnet.
(c) Full-loop interferometer: a high-visibility interference pattern of spin population showing precision and time reversibility (see text). The normalized visibility is $95\%$. See Figs.\,3 and 4 (and \,\cite{SM}) for the experimental parameters of the 3 plots.
}
\label{fringes_new}
\end{figure}

\begin{figure}
\centerline{
\includegraphics*[angle=0, width=12.cm]{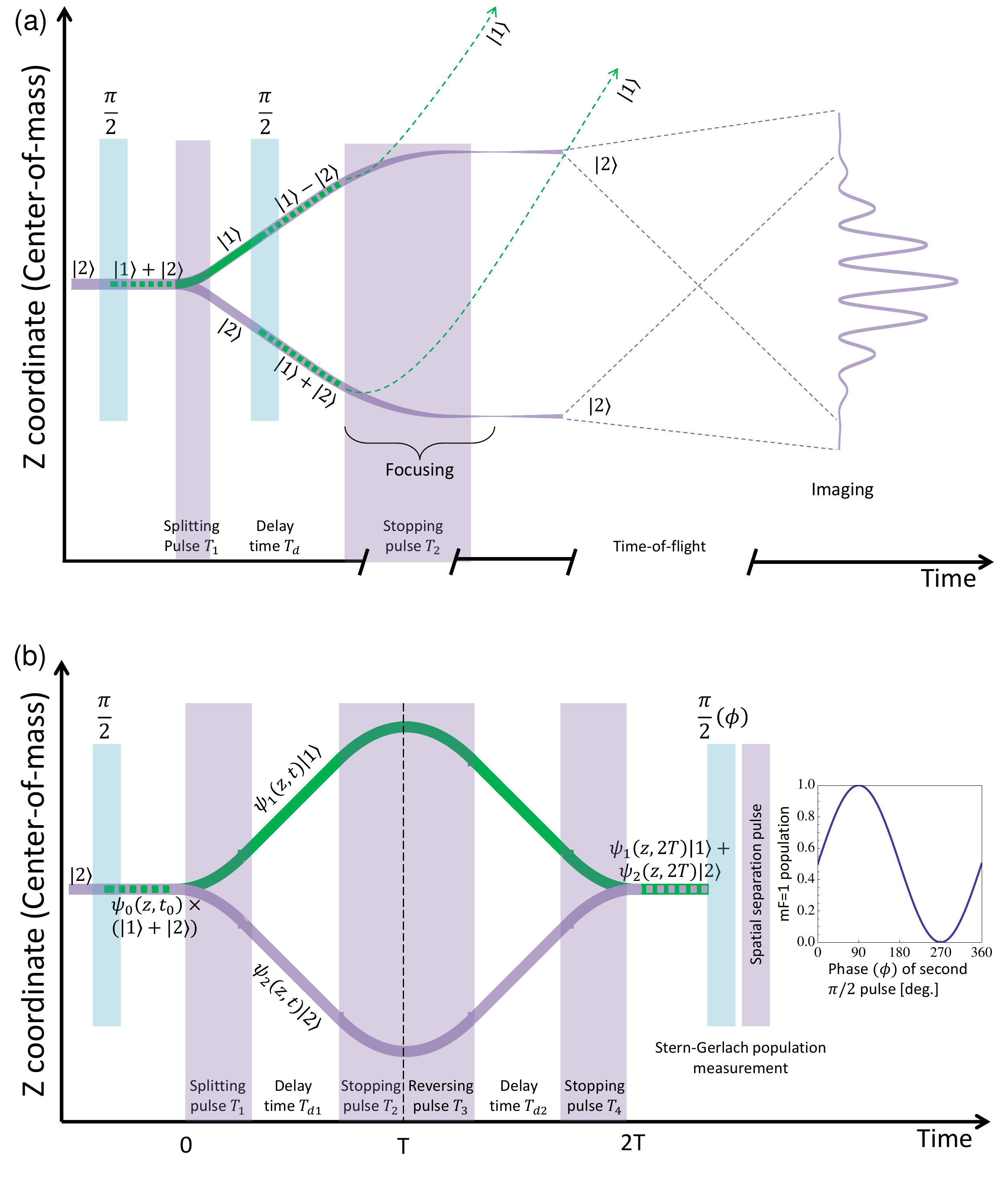}}
\caption{The longitudinal SGI (z position vs. time). (a) The half-loop interferometer. Here the signal is made of spatial interference fringes. For interference to occur the two wavepackets are made to have the same spin with a $\pi/2$ pulse and a selection of two of the four emerging wavepackets. This configuration does not require high precision and it is mainly sensitive to stability. Note that as the two wavepackets have the same spin, a long stopping pulse giving rise to an harmonic potential is required. This creates a tight focus for the wavepackets \cite{SM}. (b) The full-loop interferometer. Here the signal is made of spin population fringes. This configuration requires high precision in order to maintain coherence, as in contrast to (a) it uses active recombination. Here the stopping pulse [again in contrast to (a)] may be short as the magnetic gradient is very effective in stopping the relative motion of two different spins.
Both figures are plotted in the center-of-mass frame. As the half-loop sequence requires wavepacket expansion to achieve overlap, the half-loop sequence is much longer and  the difference between the two experiments in eventual wavepacket size at the time of recombination and overlap is 1-2 orders of magnitude. \label{configurations}}
\end{figure}

In Fig.\,2 we present our two longitudinal SGI configurations. The {\it half-loop} consists only of splitting and stopping. After the initial splitting, we manipulate the wavepackets to have the same spin so that they may spatially interfere. Despite having the same spin, their relative velocity may be stopped as they are at different locations in which the magnetic gradient differs. Recombination occurs as the separated wavepackets expand and overlap after time-of-flight (TOF). The {\it full-loop}, in which the entanglement of spin with spatial degrees of freedom persists throughout the SGI, actively recombines the two wavepackets in both position and momentum (i.e. four regions or pulses including splitting, stopping, accelerating back and stopping), and uses the spin state of the recombined wavepacket as the interference signal.

Three factors determine coherence (and similarly the level of TI) in a SGI: First, the initial state may be too spread in position or momentum. To suppress this effect we utilize a minimal uncertainty wavepacket in the form of a Bose-Einstein condensate (BEC).
In addition, we use an optimized sequence that is most robust against initial state uncertainties, as apparent even for a thermal cloud (Fig.\,1a).

The second factor is instability (i.e. temporal fluctuations). Instability is due to the environment (either the magnet itself or beyond it) ranging from classical (technical drifts) to quantum\,\cite{zurek1}.
In our experiment the quantum regime has little effect: the high visibility in Fig.\,1b shows spatial decoherence to be small.
We also estimate the decoherence due to entanglement with electrons in the electromagnets to be small \cite{SM}. Finally, as the BEC is in free fall, phase diffusion due to atom-atom interaction is negligible. Our main perturbation is thus classical drifts.
To study this we utilize the half-loop SGI, where visibility is not sensitive to precision as slight changes in momentum and position of the wavepackets after the evolution do not change the visibility but rather induce minor changes to the interference pattern periodicity\,\cite{SM}.
We measure stability by evaluating the shot-to-shot phase difference of the interference pattern via the visibility of a multi-shot sum (Fig.\,1b). Having confirmed that drifts and spatial decoherence are low, we may use spin-coherence in the full-loop as a measure of the third factor, imprecision.

Imprecision of the magnet gives rise to magnetic fields not having the right magnitude and direction throughout the particle's trajectory. This is the focus of the HD effect. Various properties of the wavepacket should be controlled, such as position, central phase, momentum and wavepacket size.  These are related to the phase spread over the wavepacket during the splitting, which is (as argued by Heisenberg and others\,\cite{ESS_1,briegel, Heisenberg}):
$-\delta\phi=-\delta (ET/\hbar)=-(\partial E/\partial z) \sigma_z T/\hbar=FT \sigma_z/\hbar$, where $T$ is the magnetic gradient duration, and $\sigma_z$, $\sigma_{p_z}$, are the initial wavepacket widths in position and momentum. Requiring $FT\gg\sigma_{p_z}$ to ensure wavepacket separation ($F$ is half the differential force), this ``phase dispersion" may be large and has to be undone by the recombination process. In our case, $\delta\phi$ may be as large as several hundred radians. We study this using the full-loop SGI.
Here coherence is determined by the overlap integral (in contrast to the half-loop). As noted by the HD papers a ``microscopic" level of precision is required. This is so as the overlap integral does not change with time, even if the wavepackets expand and overlap after TOF, and a non-negligible value for this integral requires a recombination precision on the scale of the initial wavepacket width in position and momentum.
As in a standard Ramsey procedure, we measure $\phi$ with the help of a second $\pi/2$ pulse followed by a spin population measurement. We measure the visibility by scanning the phase of this $\pi/2$ pulse (Fig.\,1c).

This work utilizes an atom chip\,\cite{keil} with several advantages including strong magnetic gradients created by a source with very low inductance so that the gradients can be switched in micro-seconds.
In addition, the structure and position of the magnet are very precise as it is made of a near-perfect wire\,\cite{SM}.
Furthermore, relative to our previous work where low visibility was obtained\,\cite{machluf}, care was taken to reduce a wide range of hindering effects. For example, a novel method was used to reduce the effect
of current fluctuations by utilizing a 3-wire configuration which produces
a quadrupole and exposes the wavepackets to a weaker magnetic field
while maintaining strong gradients. This reduces phase fluctuations\,\cite{SM}.

Our experiments begin with a superposition of the two spin states $|F,m_F\rangle=|2,2\rangle\equiv |2\rangle$ and $|2,1\rangle\equiv |1\rangle$ of $^{87}$Rb atoms, prepared by a $\pi/2$ RF pulse and split into two momentum components by a magnetic gradient pulse (along the axis of gravity, z).

Our half-loop SGI uses the long TOF ($\sim 10\,$ms) to transform the expanding wavepackets into spatial interference fringes (see Wigner representation\,\cite{SM}).
We present in Fig.\,3 the measured normalized multi-shot visibility as a function of the splitting gradient pulse duration. We are limited to $4\,\mu$m wavepacket separation as our imaging cannot resolve without bias a fringe pattern periodicity below $10\,\mu$m. The latter maximal separation is depicted by the red data point.
The magnetic field curvature of the stopping pulse is responsible not only for stopping the relative wavepacket velocity, but also for focusing each wavepacket to a minimal size of less than $1\,\mu$m along the $z$ direction (Fig.\,2). At this point their separation is 4.5 to 18 times their size\,\cite{SM}.
To examine the effect of instability we present a second data set where noise was injected into the current driving the splitting pulse.
Several theoretical models show good agreement with both data sets. These include an analytical model, and two numerical models: a random vector model where $\epsilon$ is the magnitude of random perturbations (with infinite fluctuation correlation length, shown in Fig. 3, as well as zero correlation length \cite{SM}), and a wavepacket propagation model (dashed and solid red lines), which we consider most detailed in reproducing the experimental conditions\,\cite{SM}. To summarize our half-loop experiment, we find high visibility ($>90\%$) for momentum splitting up to the equivalent of $\sim2\hbar k$ ($\lambda=1\,\mu$m).

\begin{figure}
\centerline{
\includegraphics*[angle=0, width=\textwidth]{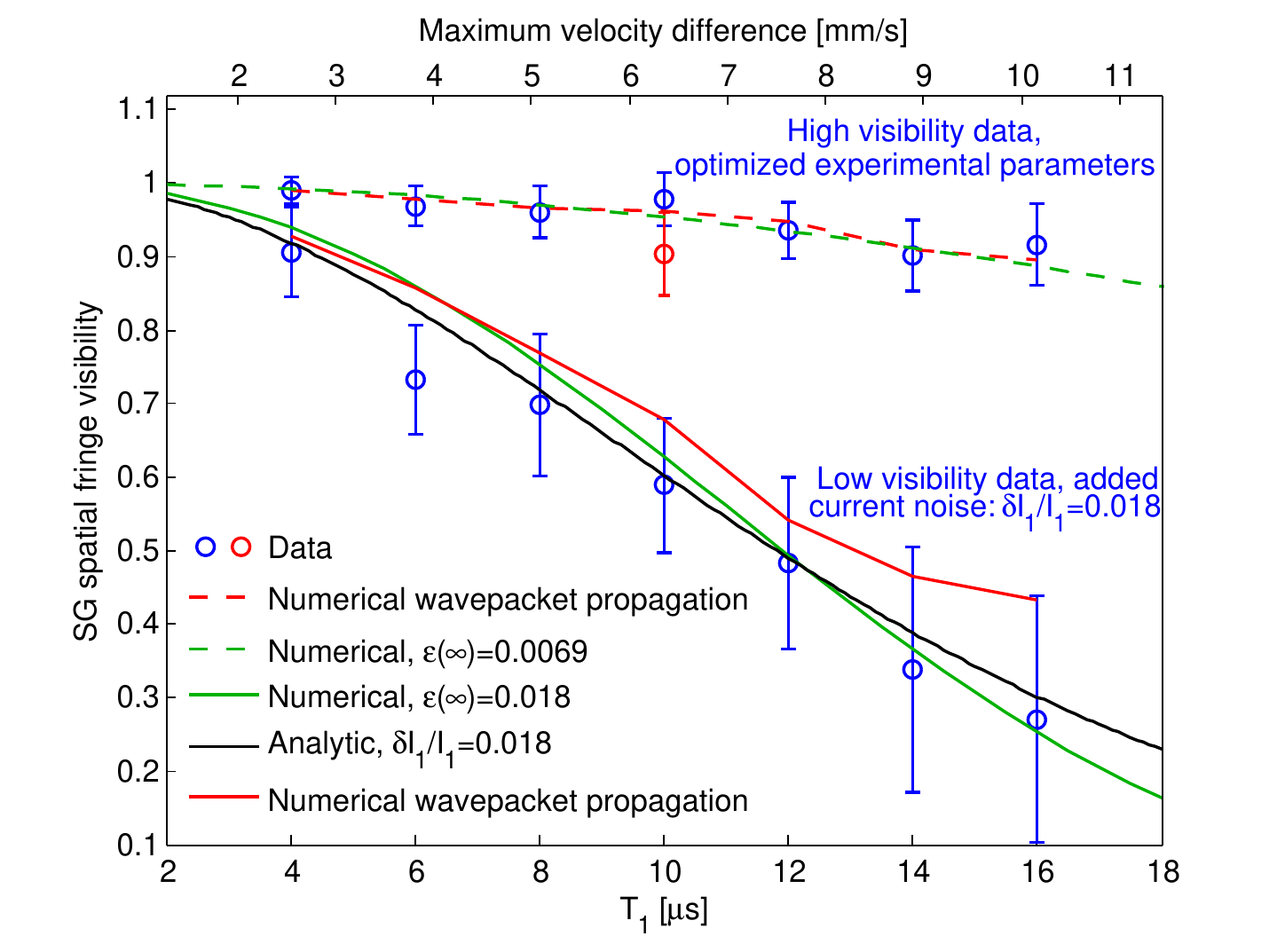}}
\caption{Analysis of stability (half-loop): spatial visibility (multi-shot) vs. splitting pulse duration $T_1$. Data shows the normalized multi-shot visibility of sequences that minimize phase fluctuations (high visibility), and similar sequences where current fluctuations were artificially injected into the splitting pulse (low visibility). The sequences are detailed in \cite{SM}. The raw data for the first data point ($4\,\mu$s) is shown in Fig.\,1b. Error bars include fitting errors for each multi-shot pattern, standard deviation (SEM) of single-shot visibility, and uncertainty due to the finite sample size \cite{SM}, and do not account for long term drifts. Theoretical models are detailed in the text and in\,\cite{SM}.
\label{half-loop}}
\end{figure}

\begin{figure}
\centerline{
\includegraphics*[angle=0, width=\textwidth]{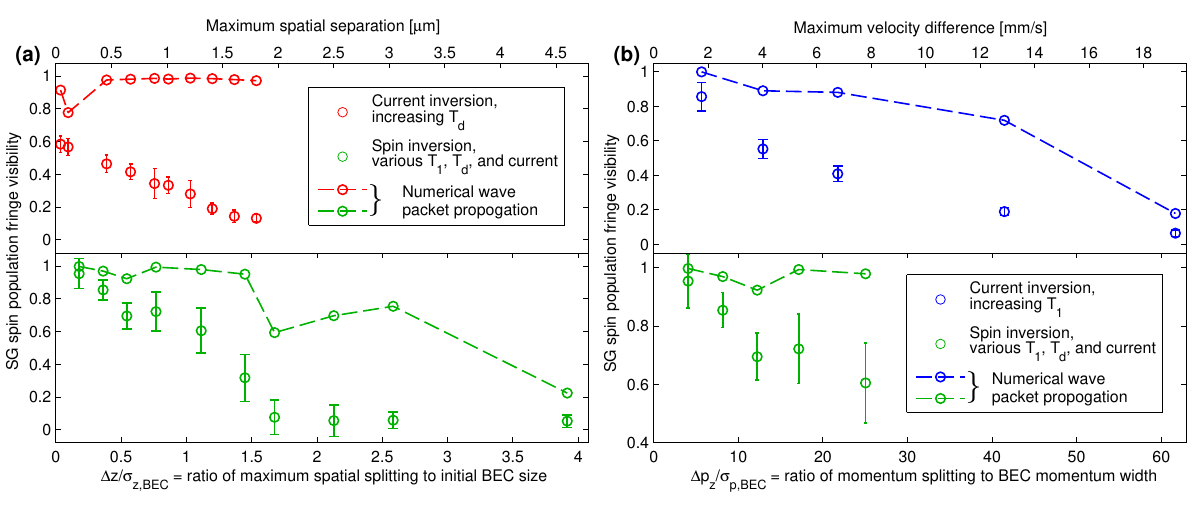}}
\caption{Analysis of precision (full-loop).
(a) Spin population visibility vs. maximal separation $\Delta z$ (in absolute units and in units of the BEC initial width).
Concerning the farthest data point:
although out of the trend of the data we believe it is a valid point\,\cite{SM}. (This is an example of the complex dependence in the 12-dimensional experimental parameter space.) Theory is due to the wavepacket propagation model (dashed lines are a guide for the eye, see text).
(b) Visibility vs. splitting momentum $\Delta p_z$ (in absolute units and in units of the BEC momentum spread). More data points are shown in \cite{SM}. For B-C we use $\sigma_{z,\text{BEC}}\simeq1.2\,\mu$m \cite{SM} (half the Thomas-Fermi width), and $\sigma_{v,\text{BEC}}=\hbar/2m\sigma_z\simeq0.3$\,mm/s (assuming minimal uncertainty wavepacket, and where $m$ is the atomic mass). The raw data for the first green data point (highest visibility) is shown in Fig.\,1c. Some of the points of the green data set (lower panels) presented in (a) have been omitted from (b), since they add no significant new behavior (e.g. some points have the same momentum splitting, and thus represent a vertical line in the momentum graph).
\label{full-loop}}
\end{figure}

\begin{figure}
\centerline{
\includegraphics*[angle=0, width=0.7\textwidth]{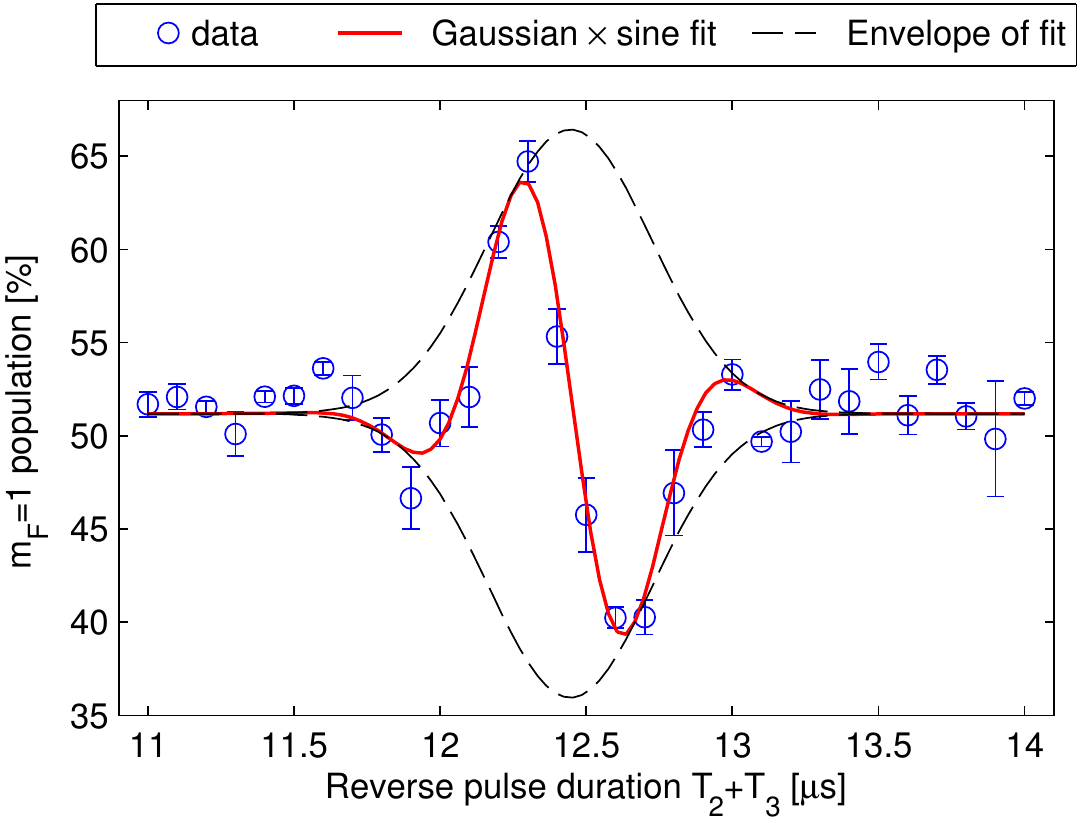}}
\caption{Full-loop optimization procedure: population output as a function of the reverse pulse duration $T_2+T_3$, for $T_1 = T_4 = 6\,\mu$s, $T_{d1}=300\,\mu$s. The population oscillates around the optimal point as expected by a simplified model of a Gaussian times a sine. The peak of the Gaussian envelope corresponds to the time at which the wavepackets' overlap integral at the end of the interferometer is maximized. The width of the envelope is related to the momentum width of the wavepackets, and the sine function corresponds to the added phase between the two interferometer arms, per unit time of reversing pulse.
\label{fig:optimization procedure}}
\end{figure}

Next, we realize the full-loop SGI. Here, the overlap and measurement take place after only a few hundred $\mu$s. To make sure the spin superposition is not dephased due to noise, we also add $\pi$ pulses giving rise to an echo sequence\,\cite{SM}.
To access a larger region of parameter space and to ensure the robustness of our results, we utilize several different full-loop configurations as detailed in\,\cite{SM}. For example, we reverse the sign of the relative acceleration by reversing the sign of the currents in the chip wires, while in other sequences we keep the same currents while reversing the spins with the help of $\pi$ pulses.
We also utilize a variety of magnetic gradient magnitudes, and scan both the splitting gradient pulse duration, $T_1$, and the delay time between the pulses, $T_d$. All results are qualitatively the same.
For weak splitting we observe high visibility ($\sim95\%$), while for a momentum splitting equivalent to $\hbar k$ the visibility is still high ($\sim75\%$) indicating that the magnet precision enabled to reverse the splitting to a high degree.

To test the limits of our precision, in Fig.\,4 we plot the visibility as a function of the normalized values ${\Delta p_z}/\sigma_{p_z}$ and ${\Delta z}/\sigma_z$, where $\Delta p_z$ and $\Delta z$ are the maximal splitting in momentum and position. These parameters are inspired by the parameters appearing in the HD formula\,\cite{briegel,SM}, namely the final imprecisions, where one may assume that a correlation exists between the latter and our values of maximal splitting.
The visibility is normalized to the Ramsey visibility without splitting (i.e., no magnetic gradients), typically $\sim90\%$. As shown previously, the momentum splitting (equalling $2FT$) is the figure of merit in determining the ``phase dispersion", and in our experiment it is as high as ${\Delta p_z}/\sigma_{p_z}=60$ before coherence is lost. In contrast, the visibility is more sensitive to spatial splitting and we achieve ${\Delta z}/\sigma_z=4$, much lower than for the half-loop.

In Fig.\,4 we also present our theoretical prediction, based on our wavepacket propagation model, which accurately simulates the experimental conditions (also used in Fig.\,3).
Notice that when the parameters used in the real experiment were (according to theory) not optimal, the simulation predicts a visibility below one. This may happen when our experimental optimization is imperfect, or if our simulation is inaccurate. When, however, the parameters used in the experiment are exactly those deemed by theory as perfect (e.g. red data set), the visibility is still not improved. This determines the limits of our experiment and theory (see Outlook).
In contrast to the half-loop case, we could not use an analytic model for the full-loop results (HD theory or its extension \cite{SM}) since such a model requires knowledge of the final state of the wavepackets (i.e. their relative separation in space and momentum, and possibly also  the relative phase chirp). These parameters could not be measured directly from the experiment. This is in contrast to the half-loop case, in which no such knowledge is required, as the normalized visibility is not sensitive to these parameters.

In the Fig.~\ref{fig:optimization procedure}, we show an example of a full-loop optimization procedure that we use in order to maximize the interference contrast of the full-loop SGI. In the optimization procedure, we set the durations of the first and last gradient pulses $T_1$ and $T_4$ and also the durations of the delay times $T_{d1}$ and $T_{d2}$ (usually $T_1 = T_4$ and $T_{d1} = T_{d2}$ to begin with). We then measure the output population of the full-loop SGI sequence as a function of the duration $T_2+T_3$ of the second and third gradient pulses, while keeping the total duration $T_2+T_3+T_{d2}$ constant. A typical result is that shown in Fig.~\ref{fig:optimization procedure}, fitted to a Gaussian envelope times a sine function. Ideally for linear magnetic gradients, we would expect the peak overlap to occur when the sequence is symmetric i.e. $T_1 + T_4 = T_2+T_3$. However due to the non-linearity of the magnetic potential created by the chip wires in the z direction, the optimal point is below or above the symmetric time (the specific number depends on the scheme used - spin inversion or current inversion).


Let us note that it is difficult to conclude whether the HD theory over- or under-estimates the loss of coherence as we have no reliable experimental method to determine the HD parameters. (We note that the HD theory is consistent with our numerical calculations, which are based on the estimation of the overlap integral~\cite{SM}.) It is clear, however, that at least qualitatively, we observe the HD effect.

Finally, we briefly compare our experiments to the state-of-the-art\,\cite{OldSG0,OldSG2,OldSG3,OldSG4,OldSG5,OldSG6,OldSG7,OldSG8, Marechal2000,OldSG9,OldSG10}. A detailed comparison is given in \cite{SM}.
While these longitudinal beam experiments did observe spin-population interference fringes, the experiments presented here are very different. Most importantly, as explained in\,\cite{OldSG6} and \cite{Marechal2000}, an analogue of the full-loop configuration was never realized, as only splitting and stopping operations were applied (i.e., no recombination); namely, wavepackets exit the interferometer with the same separation as the maximal separation achieved within. This also means that these experiments could not probe imprecision as an origin of TI or the HD effect.
Furthermore, also within the framework of the half-loop configuration, the differences are significant. Most importantly, the beam experiments could not observe any matter-wave interference fringes due to spatial splitting, as presented here in Fig.\,1. We are able to observe such fringes as we have two well defined wavepackets, almost stationary in the lab frame.

As an outlook (details in\,\cite{SM}), let us note that while we have simulated our experimental conditions with care, and while such simulations accurately describe our previous interferometry results (e.g. \cite{machluf,margalit,zhou}), as well as the half-loop results presented in Fig.\,3 and the basic characteristics of the interferometer\,\cite{SM}, the coherence drop observed in the full-loop experiment is not well described by our theory, as shown in Fig.\,4.  Noting that our experimental precision is 0.1\% (relative charge), we expect the drop to be much weaker. We can only assume that it is perhaps due to more subtle effects such as the fine structure and alignment of the magnets (e.g.\,tilts), as well as minute deviations in the initial position and state of the BEC. Identifying the source of these known effects, which are experimentally minute to the point of being unnoticeable to us, yet are highly dominant in perturbing the evolution, is beyond the scope of this first demonstration. Our extensive efforts to identify the source are described in\,\cite{SM}. As optimizing the visibility in the SGI requires a multi-dimensional scan which is impractical to conduct by hand or even by electronic loops, future use of optimization algorithms may enable further insight into the origin of the coherence loss and the fundamental limits on reversibility and time-symmetry in such systems. Finally, for a quantitative comparison to the HD theory, one would have to formulate this theory with measurable observables, such as those appearing in Figs.\,3,\,4.

To conclude, we have demonstrated for the first time a full-loop SGI, consisting of a freely propagating atom exposed to magnetic gradients, as originally envisioned. Furthermore, we have presented and analyzed for the first time high-visibility spatial fringes originating from SG splitting. We have shown that SG splitting may be realized in a highly coherent manner with macroscopic magnets without requiring cryogenic temperatures or magnetic shielding. Furthermore, we addressed the issues raised theoretically over several decades of whether time reversibility may be achieved in quantum operations performed by classical macroscopic devices giving rise to non-discrete interactions (in contrast to photon based beam splitters for example), and have shown time reversibility to be achievable in a certain range of parameters. In addition, we have qualitatively observed the HD effect, showing a drop in visibility as a function of momentum or spatial splitting.
Finally, achieving this high level of control over magnetic gradients may facilitate technological applications such as large-momentum-transfer beam splitting for metrology with atom interferometry\,\cite{LMT}, ultra-sensitive probing of electron transport down to shot-noise and squeezed currents\,\cite{SubShotNoise}, as well as nuclear magnetic resonance and compact accelerators\,\cite{applications}.

\begin{acknowledgments}
We thank Carsten Henkel, Mark Keil and Shuyu Zhou for helpful discussions. We are grateful to Zina Binstock for the electronics, and the BGU nano-fabrication facility for providing the high-quality chip. This work is funded in part by the Israeli Science Foundation, the ``MatterWave” consortium (FP7-ICT-601180), the DFG through the DIP program (FO 703/2-1), and the program for postdoctoral researchers of the Israeli Council for Higher Education.
\end{acknowledgments}

\pagebreak




\renewcommand{\thesection}{S\arabic{section}}
\renewcommand{\thesubsection}{\thesection.\arabic{subsection}}
\renewcommand{\thetable}{S\arabic{table}}

\renewcommand{\theequation}{S\arabic{equation}}
\renewcommand{\thefigure}{S\arabic{figure}}
\setcounter{figure}{0}



\centerline{Supplementary Materials for}
\bigskip
\bigskip

\centerline{\bf Realization of a complete Stern-Gerlach interferometer}
\bigskip

\centerline{Yair Margalit, Zhifan Zhou, Or Dobkowski, Yonathan}
\centerline{Japha, Daniel Rohrlich, Samuel Moukouri, and Ron Folman$^{\ast}$}


\centerline{$^\ast$Corresponding author. E-mail: folman@bgu.ac.il}
\bigskip
\bigskip
\bigskip
\bigskip
\bigskip
{\bf This PDF file includes:}

\begin{itemize}
\item[] S1. Experimental Procedure
\item[] S2. Data taking and data analysis
\item[] S3. Instability sources and optimization
\item[] S4. Main text figure details
\item[] S5. Theoretical models
\item[] S6. Phase space description of the Stern-Gerlach interferometer (SGI)
\item[] S7. A generalized Humpty-Dumpty theory for an SGI
\item[] S8. Time-irreversibility and the Stern-Gerlach interferometer
\item[] S9. Multi-shot visibility and its standard error
\item[] S10. Dephasing due to electrons
\item[] S11. Comparison to the state-of-the-art (French SG experiments)
\end{itemize}

\pagebreak

\bigskip

\section{Experimental procedure}\label{sec:ExpProcedure}

In the following we describe our two experiments: the half-loop and full-loop Stern-Gerlach interferometers (Fig.\,2 of main text). Both experiments start with the same sequence:
We begin by preparing a BEC of about $10^4$ $^{87}$Rb atoms in the state $\vert F, m_F\rangle =\vert 2,2\rangle$ in a magnetic trap located around $90\,\mu$m below the atom chip surface (accurate numbers are given in the following). The harmonic frequencies of the trap are $\omega_x/2\pi\approx 40$Hz and $\omega_y/2\pi\approx\omega_z/2\pi\approx 126$Hz where the BEC has a calculated edge-to-edge size of 1$\sim 8\,\mu$m along x and $\sim 6\,\mu$m along y and z. The trap is created by a copper structure located behind the chip with the help of additional homogeneous bias magnetic fields in the $x$, $y$ and $z$ directions (see Fig.\,\ref{fig:chip}). The BEC is then released from the trap, and falls a few $\mu$m under gravity (duration of 0.9 ms in the half-loop case, and $T_{d0}$ in the full-loop case, see details in the following). During this time the magnetic fields used to generate the trap are turned off completely. Only a homogeneous magnetic bias field of 36.7 G in the $y$ direction is kept on to create an effective two-level system via the non-linear Zeeman effect such that the energy splitting between our two levels $\vert 2,2\rangle\equiv|2\rangle$ and $\vert 2,1\rangle\equiv|1\rangle$ is $E_{21}$ $\approx h\times$25\,MHz, and where the undesired transition is off-resonance by $E_{21}-E_{10}  \approx h\times $180\,kHz. As the BEC is expanding, interaction becomes negligible, and the experiment may be described by single--atom physics. Next, we apply a radio-frequency (RF) $\pi/2$ pulse (10\,$\mu$s duration) to create an equal superposition of the two spin states,  $|1\rangle$ and $|2\rangle$, and a magnetic gradient pulse (splitting pulse) of duration $T_1=4-40\,\mu$s which creates a different magnetic potential $V_{m_j}(z)$ for the different spin states $m_j$, thus splitting the atomic spatial state into two wave packets with different momentum.

%


\subsection{Half-loop interferometer}
Our experimental sequence for the half-loop interferometer is sketched in Fig.\,2A of the main text. Just after the splitting pulse, another RF $\pi/2$ pulse ($10\,\mu$s duration) is applied creating a wave function made of four wave packets (similar to the beam splitter described in \cite{machluf}):
$|{\bf p}_{\pm}\rangle\equiv \frac{1}{\sqrt{2}}(|{\bf p}_1,{\bf x}_0\rangle\pm |{\bf p}_2,{\bf x}_0\rangle)$, where $|{\bf p},{\bf x}\rangle$ represents a wave packet with momentum ${\bf p}$ and central position ${\bf x}$ (${\bf x}_0$ is the position of the atoms during the splitting pulse) and the plus and minus signs correspond to the final spin states $|1\rangle$ and $|2\rangle$, respectively. In this experiment we choose to work with the momentum superposition of the wave packets having spin $|2\rangle$ [while disregarding the superposition of $|1\rangle$, which after a second gradient and time--of--flight (TOF) is at a different final position]. The time interval between the two RF pulses (in which there are only two wave packets, each having a different spin) is reduced to a minimum ($\sim$~40$\,\mu$s) to suppress the hindering effect of a noisy and uncontrolled magnetic environment so that the experiment does not require magnetic shielding (see Sec.~\ref{sec:instability} for more details). The minimal time between the two RF pulses is determined by a magnetic 'tail' of the gradient pulse, which at shorter times affects the resonance of the two-level system.
After a magnetic gradient pulse of duration $T_2$ designed to stop the relative motion of the two wave packets, the atoms fall under gravity for a relatively long TOF, expand and overlap creating a spatial interference pattern, and after 8-18\,ms (in total since the trap release) we image the atoms by absorption imaging and generate the pictures shown in Fig.\,1A,B of the main text.

Finally, let us note that due to the long TOF expansion, the spatial overlap is ensured even when there is no spatial precision allowing an accurate recombination of small wave packets. As spatial fringes may form even when there is clear momentum separation between two wave packets, this experiment is also not sensitive to the momentum precision of the stopping pulse.


\subsection{Full-loop interferometer}
Our experimental sequence for the full-loop interferometer is sketched in Fig.\,2B of the main text. This configuration is the same as originally envisioned for the SG interferometer (SGI)\,\cite{Wigner}. This configuration is sensitive to the precision with which the final wavepackets overlap in space and momentum (i.e. the visibility is a function of the overlap integral).

In contrast to the half-loop experiment, the second RF $\pi/2$ pulse described above is applied only at the time of measurement (so that the two wave packets have a different spin throughout the propagation), namely it completes the interferometric sequence. This completes a Ramsey sequence. Furthermore, our signal is not a spatial interference pattern but rather an interference pattern formed by measuring the spin population (e.g. starting with $S_x = +1$ and measuring $S_x$). In addition to the splitting and stopping gradient pulses ($T_1$ and $T_2$), we also apply a gradient pulse for accelerating the atoms back towards each other ($T_3$) and then apply a final stopping pulse ($T_4$) so that the two wave packets overlap in momentum and position. These four gradient pulses occur in between the two RF $\pi/2$ pulses.

\begin{figure}
\centerline{
\includegraphics*[width=\textwidth]{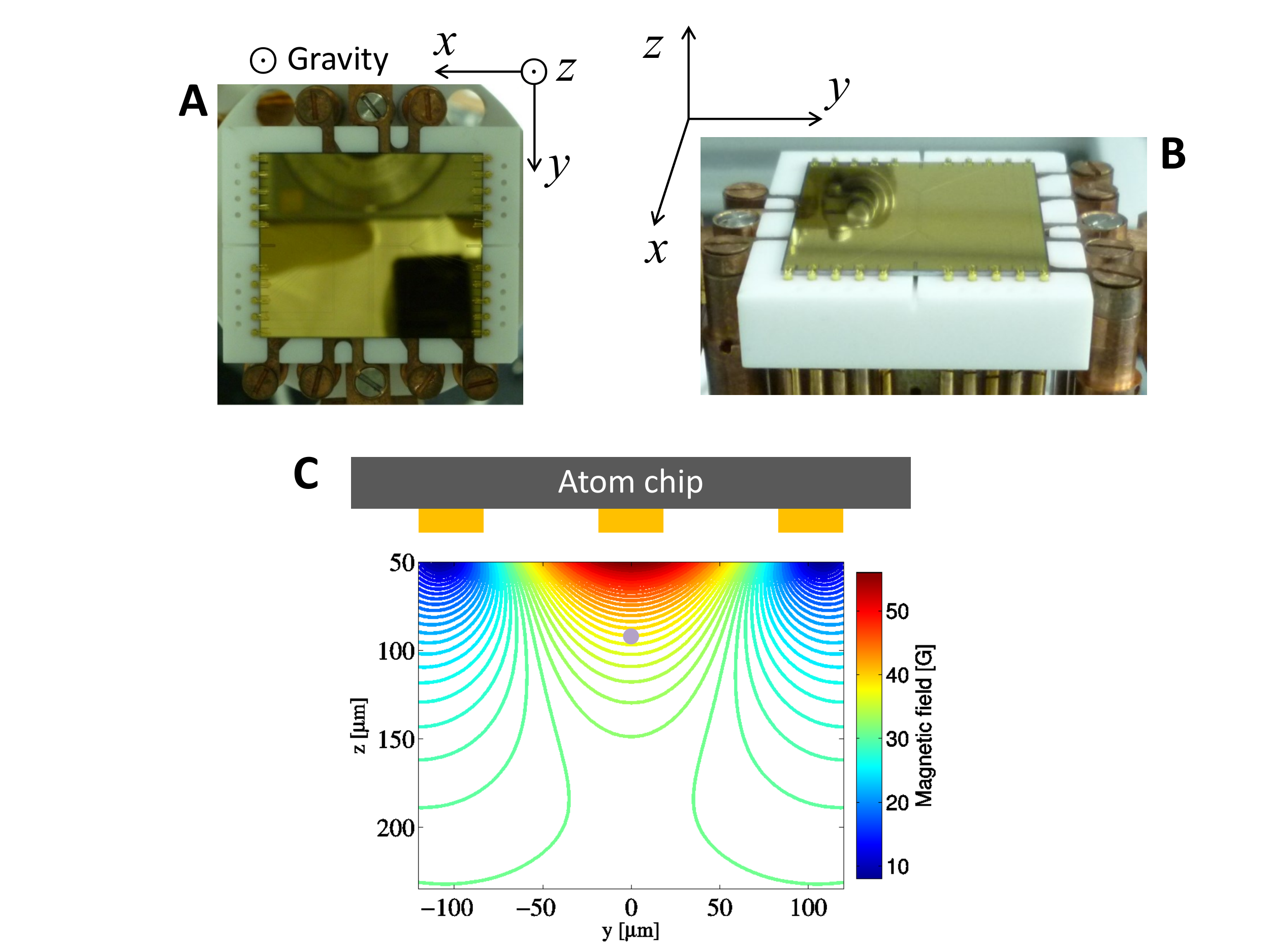}}
\caption{\label{fig:chip} (A,B) Pictures of the atom chip on its mount, with the copper structure visible behind it. Note that its orientation in the experimental setup is face down. (C) Magnetic field strength below the atom chip, generated by the quadrupole field via the chip wires (represented by the orange squares below the chip, see Fig.~\ref{fig:quadrupole}) and an homogeneous bias field $B_y$ generated by external coils. The purple dot shows the location of the trapped BEC, which has, according to simulation, a Thomas-Fermi half-width in the $yz$ plane of about 3\,$\mu$m.}
\end{figure}

In order to create a population interference pattern, the last RF $\pi/2$ pulse is shifted by a phase $\phi$ relative to the first RF $\pi/2$ pulse. This creates population oscillations between the states (as a function of $\phi$), which are later measured by applying a strong magnetic gradient to separate the spin states, and counting the number of atoms in each output state, generating the pictures shown in Fig.\,1C of the main text. The strong gradient is created by running a current in the copper structure behind the chip for a few ms (see Fig.\,\ref{fig:chip}). Note that we also add one or two RF $\pi$ pulses in between the two $\pi/2$ pulses, giving rise to an echo sequence which suppresses the dephasing taking place due to magnetic noise and inhomogeneous magnetic fields in our chamber. This allows us to increase the spin coherence time from $\sim 400$ $\mu$s up to $\sim 4$ ms (depending on the specific sequence used).

Finally, it is worth noting that while all gradient pulses come from the same chip wires, the magnetic pulses may be considered as an analogy of the original thought experiment in which there were different spatial regions with different permanent magnets. This is so as in each pulse the current and duration may be different and have individual jitter, and in addition the atom position and consequently the gradient are different.


\begin{figure}
\begin{center}
\begin{tabular}{c}
\includegraphics[width=0.75\textwidth]{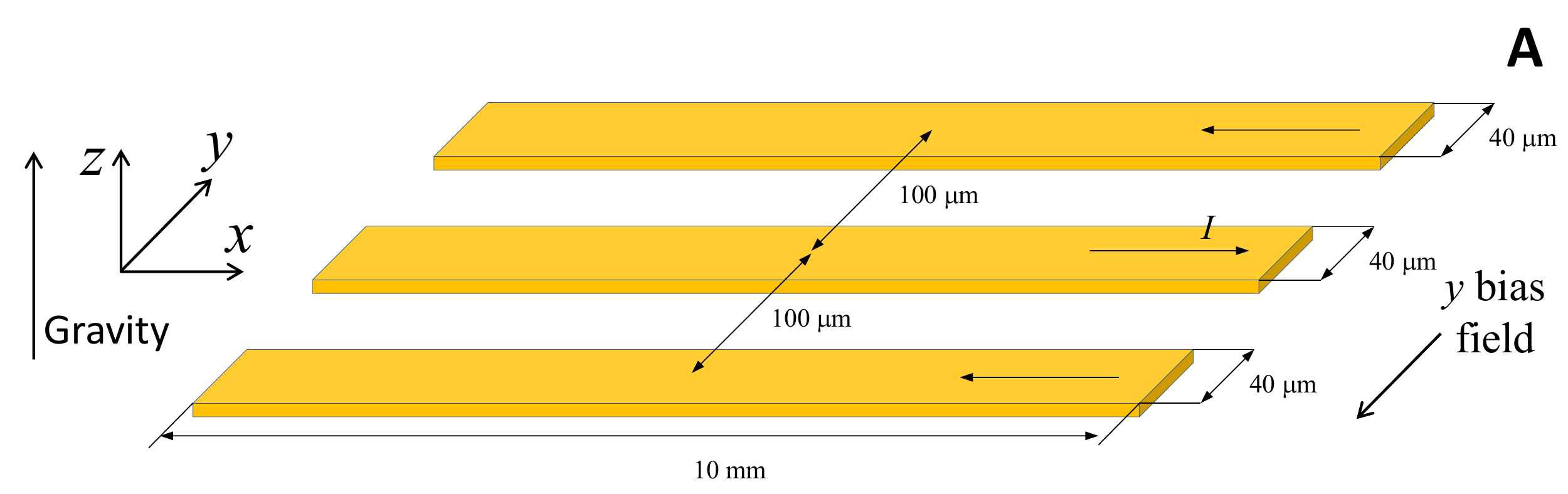}\\
\includegraphics[width=0.7\textwidth]{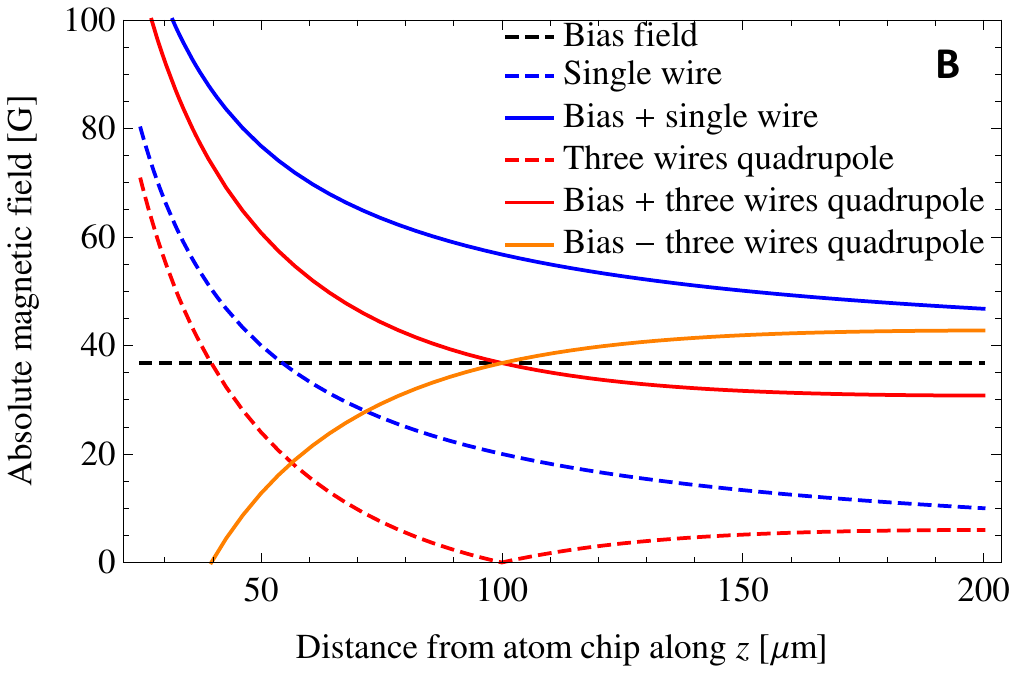}
\end{tabular}\end{center}
\caption{Quadrupole field generation and its benefit. (A) Schematic diagram of the chip wires which are used to generate the quadrupole field. Wires are 10\,mm long, 40\,$\mu$m wide and 2\,$\mu$m thick. The separation of the wires' centers is 100\,$\mu$m, and the direction of the current $I$ alternates from one wire to the next. The wires, being much smaller than the size of the chip (25\,mm $\times$ 30\,mm), are hardly visible in Fig.\,\ref{fig:chip}. (B) While the constant bias magnetic field (dashed black line) is necessary to create an effective two-level system, we do not require any additional bias to be produced by the chip wires during the gradient pulses, but require only the gradient of the field. One can see that the total magnitude of the magnetic field produced by a quadrupole and a bias (red line) is smaller than that produced by a single wire and a bias (blue line, as used in \cite{machluf}), while the gradient (at $\sim$100\,$\mu$m) is the same. Since the phase noise is largely proportional to the magnitude of the magnetic field created during the splitting pulse \cite{machluf}, positioning the atoms near the quadrupole position (98\,$\mu$m below the chip surface) reduces the phase noise. We also present how the opposite gradient is produced by just using a counter-propagating current in the three chip wires.}
\label{fig:quadrupole}
\end{figure}

\subsection{Experimental setup}
In both experiments the setup is the same: the magnetic gradient pulses are generated by three parallel gold wires (along $x$) located on the chip surface (Fig.\,\ref{fig:chip}), which are 10 mm long, 40\,$\mu$m wide and 2\,$\mu$m thick.  The wires' centers are separated by 100\,$\mu$m, and the same current runs through them in alternating directions, creating a 2D quadrupole field (in the $yz$ plane) with its center at $z = 98\,\mu$m below the atom chip. The phase noise is largely proportional to the magnitude of the magnetic field created during the gradient pulse \cite{machluf}, whereas the fluctuations in the very stable current in the external coils giving rise to the homogeneous bias field (along $y$) are relatively small during the short time scale of each experimental cycle. As the main source of magnetic instability is in the gradient pulse originating from the chip currents, positioning the atoms near the middle (zero) of the quadrupole field created solely by the three chip wires 98\,$\mu$m below the chip surface reduces the phase noise (see Fig.\,\ref{fig:quadrupole}). In the same figure we also explain how the reverse current in the three chip wires gives rise to the opposite gradient. This gradient is used in the second and third pulses of the current inversion scheme, in which a negative acceleration between the two wave packets is required in order to close the loop.

The chip wire current is driven using simple 12 V batteries connected in series, and is modulated using a home-made current shutter. To obtain timing resolution of below 1\,$\mu$s, we trigger the shutter using an Agilent 33220A waveform generator, allowing a programming resolution of a few ns. The total resistance of the three chip wires is 13.51\,$\Omega$ (when the chip temperature has stabilized after a few hours of working). Shot-to-shot charge fluctuations are measured to be $\delta I/I= 3.59 \times 10^{-3}$.
The RF signals ($\pi/2$ and $\pi$ pulses) are generated by an SRS SG384 RF signal generator which also shifts the relative phase $\phi$ between the two $\pi/2$ pulses, creating the observed population fringes. The RF signal is amplified by a Minicircuit ZHL-3A amplifier. We modulate the RF power using a Minicircuit ZYSWA-2-50DR RF switch. RF radiation is transmitted through two of the copper wires located behind the chip (with their leads showing in Fig.\,\ref{fig:chip}).

\begin{figure}
\centerline{
\includegraphics[width=0.75\textwidth]{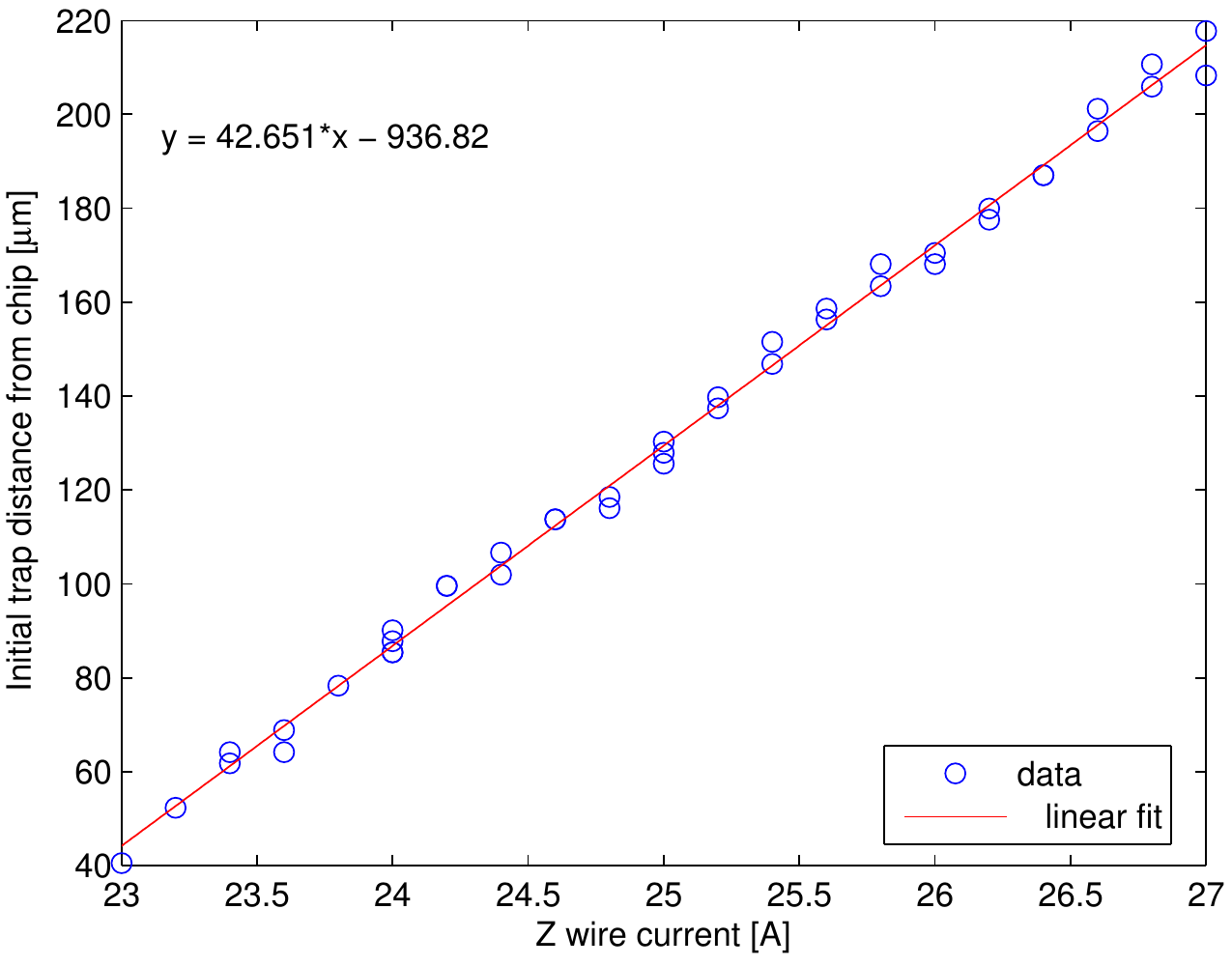}}
\caption{\label{fig:trap position} Initial magnetic trap position (experimental results), as a function of the current in the copper structure (Z shaped wire) under the chip. We determined the y axis values using the following procedure: we loaded atoms into a magnetic guide created by running DC current through the three chip wires used to create the quadrupole. The position of this quadrupole is determined by the chip geometry to be 98\,$\mu$m from the chip's surface, and so the position measured on the CCD of the atoms loaded into this guide was fixed as a reference point. The y axis of the above figure was calibrated to this reference point.}
\end{figure}

\begin{figure}
\begin{center}
\begin{tabular}{c}
\includegraphics[width=0.7\textwidth]{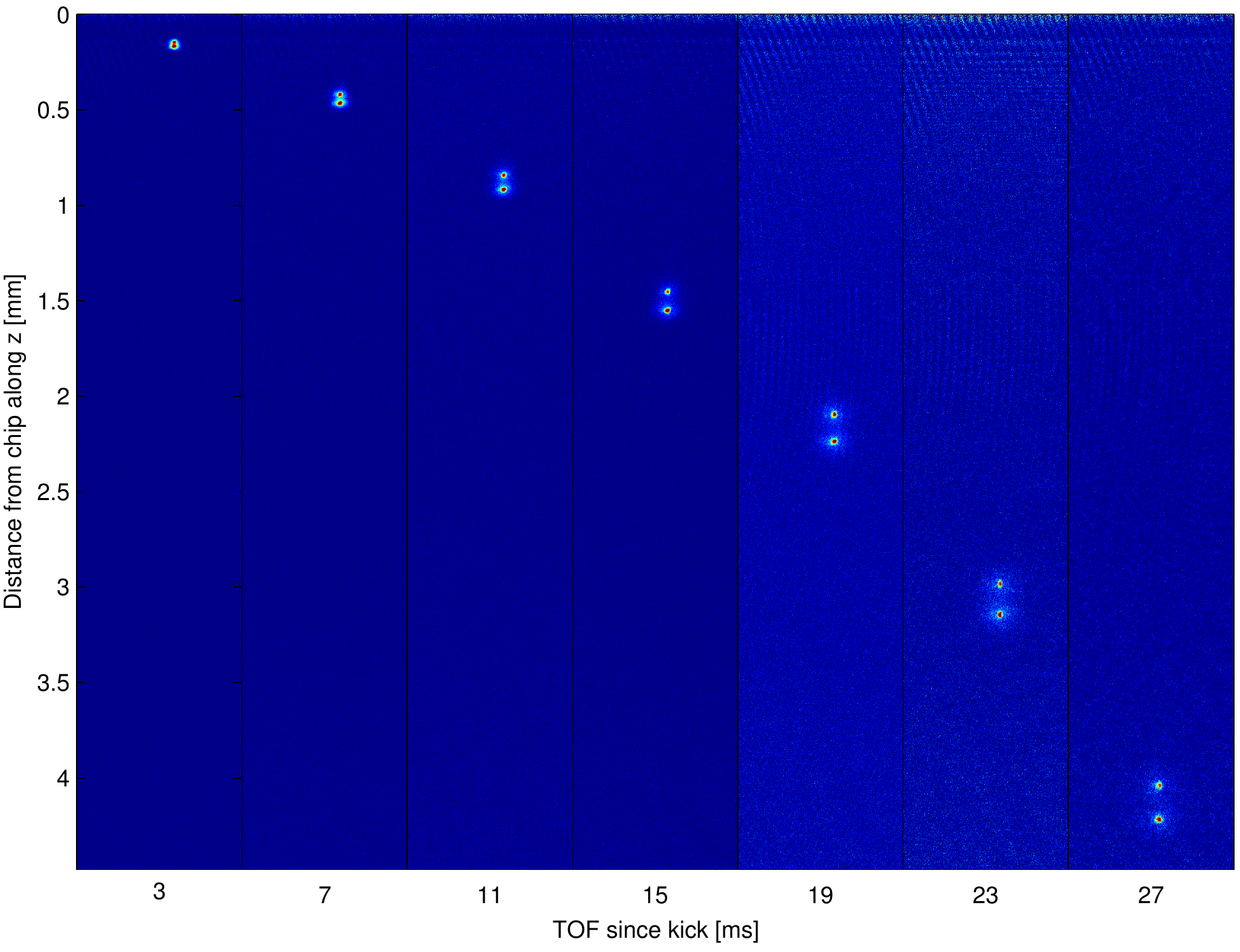}\\
\includegraphics[width=0.5\textwidth]{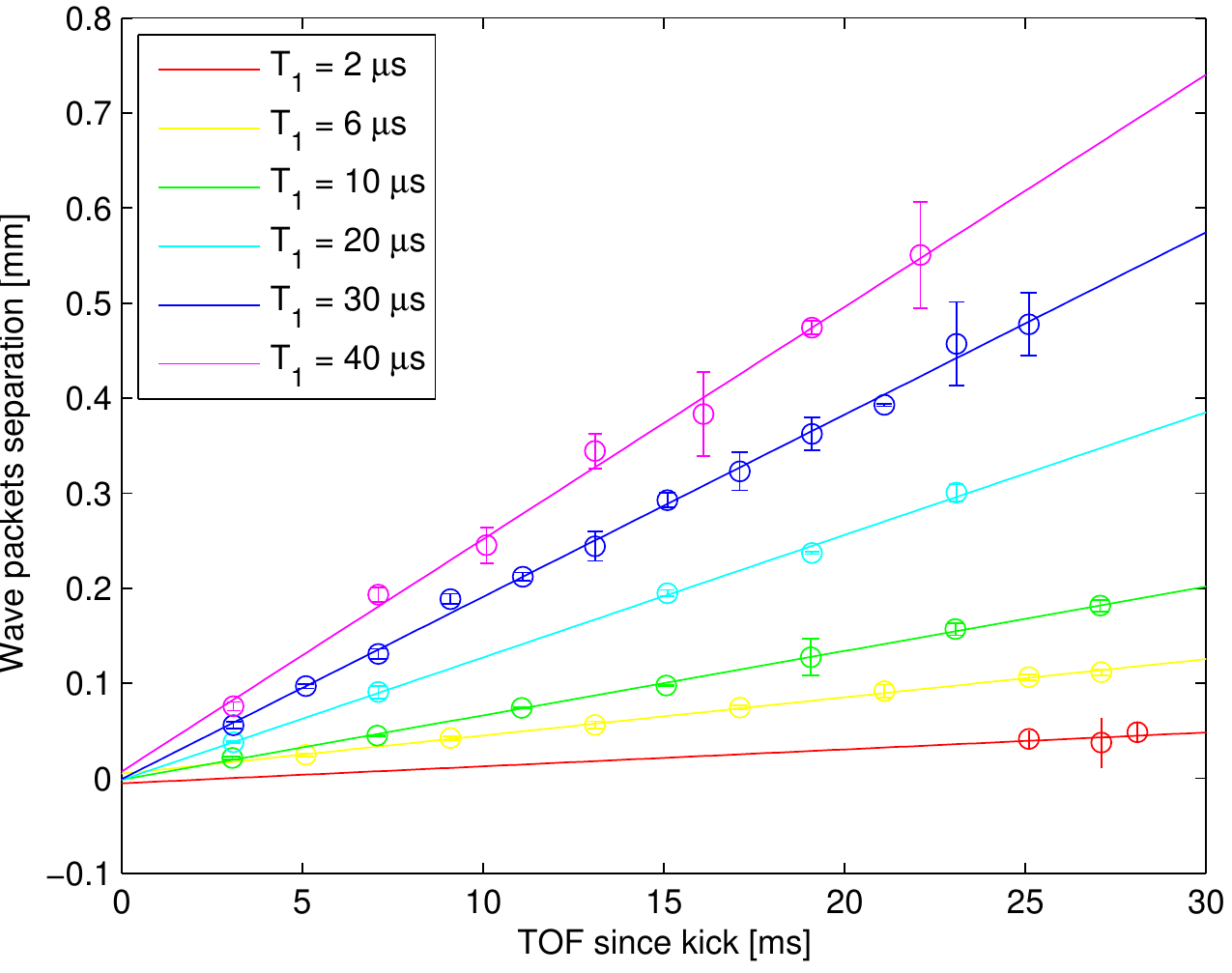}
\end{tabular}\end{center}
\caption{Upper panel: typical data showing a measurement of momentum splitting after an RF $\pi/2$ pulse and a chip magnetic gradient of $T_1=10$ $\mu$s duration. Each subplot shows the wave packets after a different TOF, increasing from left to right. The increasing spatial splitting between the wave packets can be seen. The center positions are extracted using a 1D Gaussian fit. Lower panel: the extracted spatial splitting vs TOF, for different values of $T_1$. The velocity difference is given by the slope of each curve.}
\label{fig:momentum splitting}
\end{figure}

Finally, we show in Fig.\,\ref{fig:trap position} our ability to control the initial trap position by varying the current in the copper structure (Z shaped wire) under the chip. This determines the starting point of our experiments. We also show in Fig.\,\ref{fig:momentum splitting} the measurement of the relative momentum between wave packets following a gradient pulse.

\bigskip
\goodbreak

\section{Data taking and data analysis}
\label{sec:Fig4}
\subsection{Half-loop interferometer}
Here our signal is the multi-shot visibility of an interference pattern made of the summing of many interference patterns one on top of the other with no post-selection or alignment (each interference pattern is a result of one experimental cycle). This visibility is normalized to  the mean of the single-shot visibility. In Fig.\,3 and all half-loop figures in this Supplementary Material the normalized multi-shot visibility is defined as $V_N\equiv V_{\rm av}/\langle V_s\rangle$, where $V_{\rm av}$ is the visibility of the multi-shot pattern obtained by adding shots from many experimental cycles (again, without any post-selection or post-correction), and $\langle V_s\rangle$ is the mean visibility of the single shot patterns, composing the multi-shot pattern. The error bars are estimated by
\begin{equation}
  \Delta V_N = V_N\sqrt{\left(\frac{\Delta V_{\rm a}}{V_{\bf a}}\right)^2
+ \frac{1}{N}\left(\frac{\Delta V_s}{\langle V_s\rangle}\right)^2
+\frac{1}{2N}\left(\frac{1-V_N^2}{V_N}\right)^2},
\label{eq:errorbars}
\end{equation}
where $\Delta V_{\rm a}$ is the fit error of the visibility of the multi-shot image, $\Delta V_s$ is the measured standard derivation of the single-shot visibility ($N$ being the sample size), and the third term under the square root estimates the expected relative standard error of the normalized multi-shot visibility due to the finite sample size [Eq.~(\ref{eq:dVN})].

\begin{figure}
\centerline{
\includegraphics[width=0.7\textwidth]{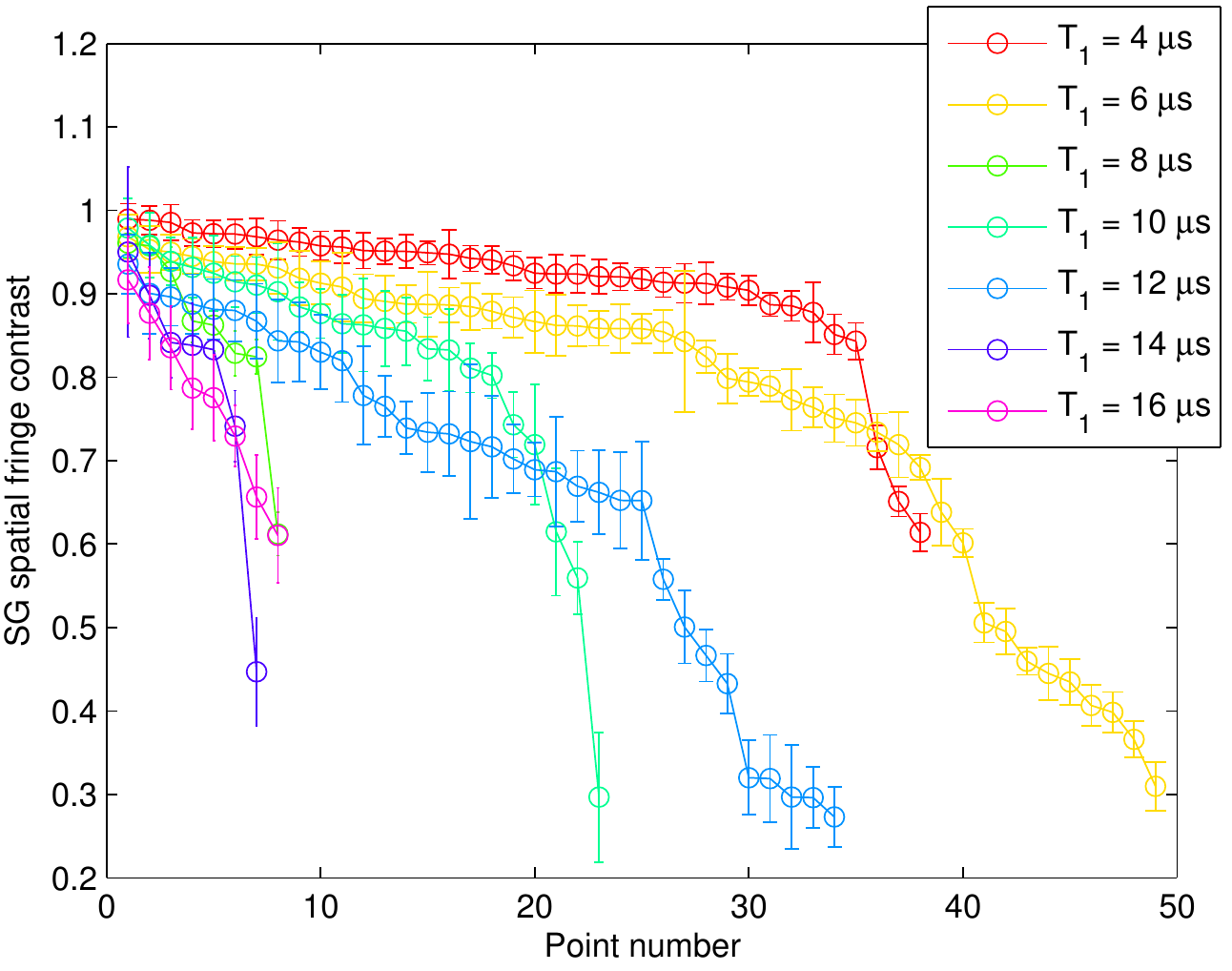}}
\caption{\label{fig:optimization visualization} Complete data set from many months of running showing also lower visibility runs which are a result of the optimization process, or the interferometer drifting away from its optimal point. The data was ordered according to diminishing visibility, and the x axis thus does not represent any time ordering. The mean number of single shots per point is 50.}
\end{figure}

In Table~\ref{table:parameters} we present the parameters used for each data point in Fig.\,3 of the main text. The free propagation time $T_d$ and the second pulse duration $T_2$ were chosen from a large experimental parameter space so as to optimize the normalized multi-shot visibility by minimizing the effect of fluctuations in parameters of the interferometric sequence such as the initial trapping position or the stopping pulse duration, as detailed in section~\ref{sec:instability}. While each data point in Fig.\,3 is a result of continuous data taking in long sessions ranging in duration from an hour to several hours with no post-selection or post-correction, long term drifts of magnetic fields or voltages in the system (e.g. due to warming up of the copper structure under the chip, the coils or the electronics) were addressed by stopping the data taking and re-optimizing the interferometer. These drifts are not taken into account in the error bars presented in Fig.\,3. Nevertheless, significant low-visibility data were taken during optimization sequences. The full data, taken over many months, are shown in Fig.\,\ref{fig:optimization visualization}.

The visibility and phase information of each absorption image is extracted by fitting a 1D cut of the image (along the $z$ direction) to a sine function with a Gaussian envelope, of the form:
\begin{equation}\label{eq:fit function}
n(z) = A\exp\left[-\frac{(z-z_0)^2}{2\sigma^2}\right]\left[1+V\sin\left(\frac{2\pi}{\lambda}(z-z_{\text{ref}})+\phi\right)\right]+c,
\end{equation}
where A is a the amplitude, $z_0$ is the center-of-mass position of the density envelope at the time of imaging, $z_{\text{ref}}$ is a phase reference point (usually taken to be middle of the image), $\sigma$ is the final Gaussian width, $\lambda$ is the fringe periodicity, $V$ is the interference fringes visibility, and $\phi$ is the global phase difference.

\begin{table}
\begin{tabular}{c|ccccccc||c}
$T_1$ [$\mu$s]& 4  & 6  & 8  & 10  & 12  & 14  & 16 & 10 \\   \hline
$T_d$ [$\mu$s]& 116  & 174  & 132  & 90  & 130  & 106  & 114 & 600 \\ \hline
$T_2$ [$\mu$s]& 200  & 150  & 180  & 220  & 200  & 200  & 200 & 70 \\ \hline
$TOF$ & 6760  & 6750  & 8760  & 12760  & 12738  & 13810 & 13800  & 21450 \\ \hline
No. images & 40 & 45 &  42 &  64 &  41 &  43 &  47 & 45 \\ \hline
Scaling factor $\xi$ & 1.18 & 1.37 & 1.22 & 1.11 & 1.22 & 1.15 & 1.18 & 3.36  \\  \hline
exp. $d (\mu$m) & 0.55 & 0.98 & 1.14 & 1.31 & 1.66 & 1.92 & 2.25 & 3.93 \\ \hline
calc. $d (\mu$m) & 0.54 & 0.94 & 1.13 & 1.28 & 1.68 & 1.85 & 2.16 & 3.90 \\ \hline
$\sigma_{\rm min} (\mu$m) & 0.120 & 0.140 & 0.125 & 0.113 & 0.124 & 0.1174 & 0.12 & 0.34 \\ \hline
\end{tabular}
\caption{\label{table:parameters} Parameters of the half-loop data points in Fig.\,3 of the main text: time durations (in $\mu$s) of stages of the interferometer sequence and wave packet widths and separation $d$ achieved during the sequence, measured or calculated analytically from Eqs.~(\ref{eq:xi}) and~(\ref{eq:d}) with the estimated value $\omega=2\pi\times 850$\,Hz for the stopping pulse curvature. The experimental value of $d$ is calculated from the observed periodicity of the interference pattern $\lambda$, and the equation $\lambda = ht/md$. The scaling factor $\xi$ describes the squeezing in phase space (see Sec.\,\ref{sec:phasespace}, Eq.\,\ref{eq:xi}), and $\sigma_{\rm min}$ describes the minimal wave packet width at the focal point (see Eq.\,\ref{eq:sigma_min}). The last column describes the parameters of the large wave packet separation sequence (red data point in Fig.\,3 of the main text).}
\end{table}

\def\arraystretch{0.7}\begin{table}
\begin{center}\resizebox{\columnwidth}{!}{\begin{tabular}{ c | c | c | c | c | c | c | c | c | c | c || c } Point \# & $T_{d0}$ [$\mu$s] & $T_1$ [$\mu$s] & $T_{d1}$ [$\mu$s] & $T_2+T_3$ [$\mu$s] & $T_{d2}$ [$\mu$s] & $T_4$ [$\mu$s] & TOF [$\mu$s] & $T_R$ [$\mu$s] & $z_0$ [$\mu$m] & $I_{c}$ [A] & Contrast\\ \hline{\color{blue} 1} & 920 & 2 & 50 & 4 & 50 & 2 & 502 & 640 & 91.07 & 0.89 & 0.86 $\pm$ 0.08 \\
{\color{blue} 2} & 920 & 6 & 50 & 11.77 & 50.23 & 6 & 486 & 640 & 91.07 & 0.89 & 0.55 $\pm$ 0.05 \\
{\color{blue} 3} & 920 & 10 & 50 & 20.157 & 49.843 & 10 & 470 & 640 & 91.07 & 0.89 & 0.41 $\pm$ 0.05 \\
{\color{blue} 4} & 920 & 20 & 50 & 42 & 48 & 20 & 430 & 640 & 91.07 & 0.89 & 0.19 $\pm$ 0.02 \\
{\color{blue} 5} & 920 & 30 & 50 & 65 & 45 & 30 & 390 & 640 & 91.07 & 0.89 & 0.06 $\pm$ 0.02 \\
{\color{blue} 6} & 920 & 40 & 50 & 85.85 & 44.15 & 40 & 550 & 840 & 91.07 & 0.89 & 0.01 $\pm$ 0.01 \\
 \hline{\color{red} 1} & 920 & 6 & 4 & 11.6 & 4.4 & 6 & 578 & 640 & 96.06 & 0.89 & 0.58 $\pm$ 0.05 \\
{\color{red} 2} & 920 & 6 & 20 & 11.35 & 20.65 & 6 & 546 & 640 & 96.06 & 0.89 & 0.57 $\pm$ 0.05 \\
{\color{red} 3} & 920 & 6 & 100 & 12.1 & 99.9 & 6 & 386 & 640 & 96.06 & 0.89 & 0.46 $\pm$ 0.05 \\
{\color{red} 4} & 920 & 6 & 150 & 12.25 & 149.75 & 6 & 486 & 840 & 96.06 & 0.89 & 0.42 $\pm$ 0.05 \\
{\color{red} 5} & 920 & 6 & 200 & 12.4 & 199.6 & 6 & 586 & 1040 & 96.06 & 0.89 & 0.34 $\pm$ 0.09 \\
{\color{red} 6} & 1440 & 6 & 250 & 12.3 & 249.7 & 6 & 986 & 2060 & 96.06 & 0.89 & 0.33 $\pm$ 0.05 \\
{\color{red} 7} & 1440 & 6 & 300 & 12.45 & 299.55 & 6 & 886 & 2060 & 96.06 & 0.89 & 0.28 $\pm$ 0.08 \\
{\color{red} 8} & 1440 & 6 & 350 & 12.5 & 349.5 & 6 & 786 & 2060 & 96.06 & 0.89 & 0.19 $\pm$ 0.03 \\
{\color{red} 9} & 1440 & 6 & 400 & 12.5 & 399.5 & 6 & 686 & 2060 & 96.06 & 0.89 & 0.14 $\pm$ 0.04 \\
{\color{red} 10} & 1440 & 6 & 450 & 12.5 & 449.5 & 6 & 586 & 2060 & 96.06 & 0.89 & 0.13 $\pm$ 0.02 \\
 \hline{\color{darkGreen} 1} & 960 & 2 & 168 & 3.85 & 164.15 & 2 & 50 & 460 & 96.06 & 0.89 & 0.95 $\pm$ 0.09 \\
{\color{darkGreen} 2} & 958 & 4 & 164 & 7.5 & 164.5 & 4 & 48 & 460 & 96.06 & 0.89 & 0.85 $\pm$ 0.06 \\
{\color{darkGreen} 3} & 956 & 6 & 162 & 11.1 & 160.9 & 6 & 48 & 460 & 96.06 & 0.89 & 0.69 $\pm$ 0.08 \\
{\color{darkGreen} 4} & 956 & 10 & 154 & 18 & 156 & 10 & 46 & 460 & 96.06 & 0.89 & 0.35 $\pm$ 0.05 \\
{\color{darkGreen} 5} & 954 & 12 & 152 & 22 & 154 & 12 & 44 & 460 & 96.06 & 0.89 & 0.15 $\pm$ 0.04 \\
{\color{darkGreen} 6} & 957 & 6 & 161 & 11.1 & 161.9 & 6 & 47 & 460 & 96.06 & 1.81 & 0.47 $\pm$ 0.17 \\
{\color{darkGreen} 7} & 958 & 4 & 164 & 7.4 & 164.6 & 4 & 48 & 460 & 96.06 & 2.73 & 0.6 $\pm$ 0.14 \\
{\color{darkGreen} 8} & 959 & 2 & 167 & 3.8 & 167.2 & 2 & 49 & 460 & 96.06 & 3.73 & 0.72 $\pm$ 0.12 \\
{\color{darkGreen} 9} & 1009 & 2 & 317 & 3.7 & 317.3 & 2 & 99 & 860 & 96.06 & 3.73 & 0.32 $\pm$ 0.14 \\
{\color{darkGreen} 10} & 959 & 2 & 367 & 3.35 & 367.65 & 2 & 49 & 860 & 96.06 & 3.73 & 0.08 $\pm$ 0.1 \\
{\color{darkGreen} 11} & 1059 & 2 & 467 & 3.35 & 467.65 & 2 & 149 & 1260 & 96.06 & 3.73 & 0.06 $\pm$ 0.1 \\
{\color{darkGreen} 12} & 959 & 2 & 567 & 3.35 & 567.65 & 2 & 49 & 1260 & 96.06 & 3.73 & 0.06 $\pm$ 0.05 \\
{\color{darkGreen} 13} & 959 & 2 & 567 & 3.1 & 567.9 & 2 & 49 & 1260 & 96.06 & 5.66 & 0.05 $\pm$ 0.04 \\
 \hline\end{tabular}}\end{center}
\caption{\label{table:full loop parameters} Parameters of the full-loop data points in Fig.\,4 of the main text and in Fig.~\ref{fig:all points}, including timings of the interferometer sequence. Colored point numbers correspond to the colored numbers in Fig.\,\ref{fig:all points} (one color per data set). Some of the points presented here have been omitted from the figures in the main text, since they are either sub-optimal relative to other points (optimization was not good enough), or add no significant new behavior (e.g. all points of the red data set have the same momentum splitting, and thus represent a vertical line in the momentum graph). Parameters are: $T_{d0}$ - time between trap release to first gradient; $T_1$, $T_{d1}$, $T_2$, $T_3$, $T_{d2}$, and $T_4$ are the gradient pulse durations and delay times between the gradients, as shown in Fig.\,2b of the main text; TOF - delay between $T_4$ and the last $\pi/2$ pulse; $T_R$ - total Ramsey interrogation time (i.e. the time between the first and last $\pi/2$ pulses); $z_0$ - initial trap distance from chip (before release); and $I_{c}$ - current in chip wires during gradients. The contrast values are the experimental measured values, including propagated fit error. The value of $z_0$ was calculated using the results of Fig.~\ref{fig:trap position}; uncertainties of a few $\mu$m exit due to position shifts of the copper structure (Z wire) generating the magnetic trap, relative to the chip from which the quadrupole field is generated. Blue and green data sets are measured using current inversion sequence; green data set is measured using spin inversion sequence (see text for details).}
\end{table}

\begin{figure}
\centerline{
\includegraphics[width=0.7\textwidth]{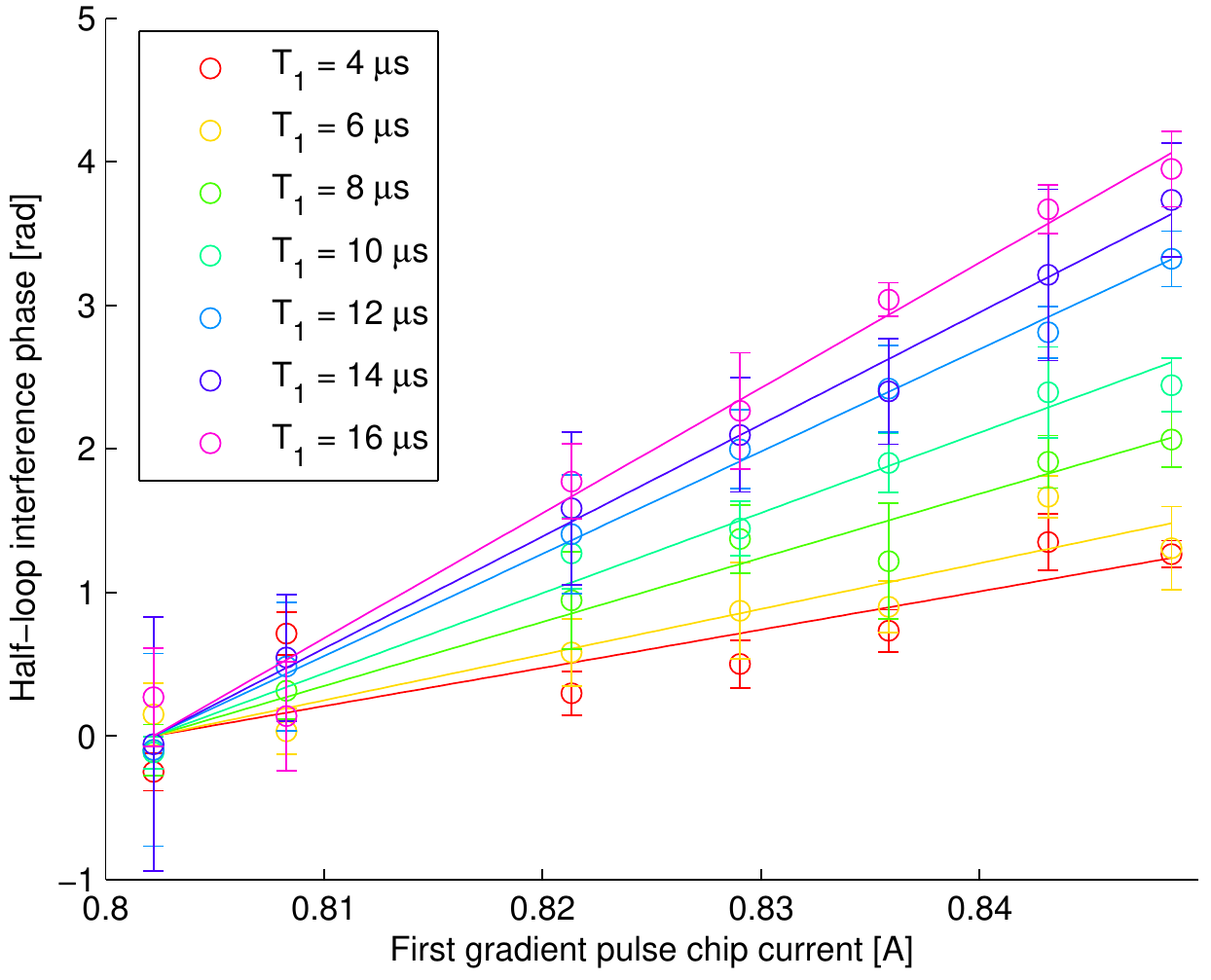}}
\caption{\label{fig:phi vs I1} Half-loop interference phase as a function of the applied chip current during the first gradient pulse $T_1$ (using a voltage stabilizer circuit, see text), used to produce the low visibility data in Fig.\,3 of the main text. The phase data has been shifted vertically so that all data starts at the same initial phase for clarity. The phase is clearly linear with the applied chip current, and we also confirm that the slope divided by $T_1$ of all curves is equal within error bars, as would be expected from theory. The mean slope divided by $T_1$ is $\partial\phi / \partial I_1/T_1 = 6.49$ rad/A/$\mu$s.}
\end{figure}

To study the topic of fluctuations and stability via the multi-shot visibility (Fig.~1B), we examine the randomizing effect in the SGI by varying the magnitude of fluctuations in the magnetic gradient and the coupling time to the atom. The multi-shot signal is sensitive to variations between the experimental cycles in phase, momentum and the spatial separation $d$. The low-visibility data shown in Fig.\,3 were taken by changing the chip voltage driving the splitting pulse using a voltage stabilizer circuit, with corresponding variable values of chip current (the stopping pulse current was kept constant). A different number of images were taken for each value of the current. We then produce a multi-shot fringe pattern resulting from a summation over many measurements with varying splitting pulse currents. In order to properly emulate the spectrum of natural noise, the averaged image was obtained by taking a weighted average such as to create a normal distribution of phases. Since the phase is linear with the applied current (see Fig.~\ref{fig:phi vs I1}), such a distribution corresponds to a normal distribution of currents, where its width was set to $\delta I/I=15.47$ mA / 860 mA$=1.8\%$.

Varying the current giving rise to the first gradient pulse affects both the phase and the periodicity of the fringes (see Eq.~\ref{eq:d}), causing two kinds of chirping effects on the output multi-shot image. The first is a chirp of the interference periodicity, i.e. the image is composed of multiple periodicities (in contrast to a single-shot image or to a high visibility multi-shot image which has only a single periodicity). The second kind is a spatial chirp of the interference visibility, i.e. the visibility is position-dependent. Because of these effects, we cannot extract the visibility of these multi-shot images simply by fitting the pattern to Eq.~\ref{eq:fit function}, and we need to adopt a more general definition of the visibility.

Assuming our interference pattern is composed of an envelope (e.g. a Gaussian) multiplied by some oscillating function, one such possible definition is to take the Fourier transform of the interference pattern. In $k$-space, the result is a sum of three terms: one centered around $k=0$ - representing the envelope, and the two others at $k = \pm k_0=\pm 2\pi/\lambda$ - representing the oscillating terms. The visibility can then be defined as the ratio of amplitudes of the oscillatory components to the amplitude of the zero component, explicitly: $V = [A(+k_0)+A(-k_0)]/A(0)$, where $A(k)$ represent the amplitudes of the Fourier transform at point $k$. The visibility of the output multi-shot images is calculated according to this procedure, using a numerical FFT of each image.

In table\,\ref{table:parameters} we show some wave packet parameters calculated from the experimental parameters or results by using Eqs.~(\ref{eq:xi}) and~(\ref{eq:d}). The wave packet separation at the end of the stopping pulse is calculated from the experimental data by using the measured spatial period of the fringes $\lambda=ht/md$, where $t$ is the TOF. The wave packet separation calculated from Eq.\,(\ref{eq:d}) differs from the values calculated from the fringe periodicity by no more than 4\%. The separation $d$ is larger than the minimal Gaussian wave packet width $\sigma_{\rm min}$ by a factor $4.5-18$.  We note that the spatial separation being much larger than the minimum wave packet width is an experimental fact that is evident from the appearance of multiple fringes in the final interference pattern ($d/\sigma_{\rm min}$ roughly represents the number of interference fringes of a single pattern). However, this wave packet separation is inversely proportional to the spatial period of the fringes, $\lambda$, and as our TOF is experimentally limited by the size of the vacuum chamber and the field of view of the camera and its sensitivity, it follows that the wave packet separation is limited by the practical resolution of the imaging system and cannot exceed the maximal value of about 4\,$\mu$m, which was observed in our experiment.

As noted, we normalize the multi-shot visibility to the mean of the single-shot visibility taken from the same sample. This normalized multi-shot visibility eliminates irrelevant effects such as visibility reduction due to an impure BEC (thermal atoms), lack of perfect overlap between the wave packet envelopes in 3D, as well as imaging limitations such as inaccurate focal point, limited focal depth, spatial resolution, camera speed relative to the speed of the moving fringes, and so on.

Finally, let us also note that as the durations of our interferometer operation and work-cycle are 100\,$\mu$s (without TOF) and 60\,s respectively, and as we take data for several hours in each run, we believe we are sensitive to fluctuations with frequencies lower than about 10\,kHz. As the shortest magnetic gradient pulse is 4\,$\mu$s, we may even be sensitive to frequencies up to 100\,kHz. This captures the dominant part of the 1/f (flicker) noise of electronic systems.

\subsection{Full-loop interferometer}
Here, our signal is the single-shot visibility of spin population oscillations. This visibility is normalized to the Ramsey oscillations' visibility when no magnetic gradients are applied. In Table~\ref{table:full loop parameters} we present the parameters used for each data point in Fig.\,4 of the main text. The atoms were initially trapped as in the half-loop at $90\pm2\,\mu$m, and in later experiments were moved by $5\,\mu$m so that they are closer to the center of the quadrupole (which is at $98\,\mu$m). The drop time (free-fall) before the start of the sequence was $T_{d0}= 0.9-1.4\,ms$ ($4-10\,\mu$m), see table~\ref{table:full loop parameters}.

\begin{figure}
\centerline{
\includegraphics[width=0.8\textwidth]{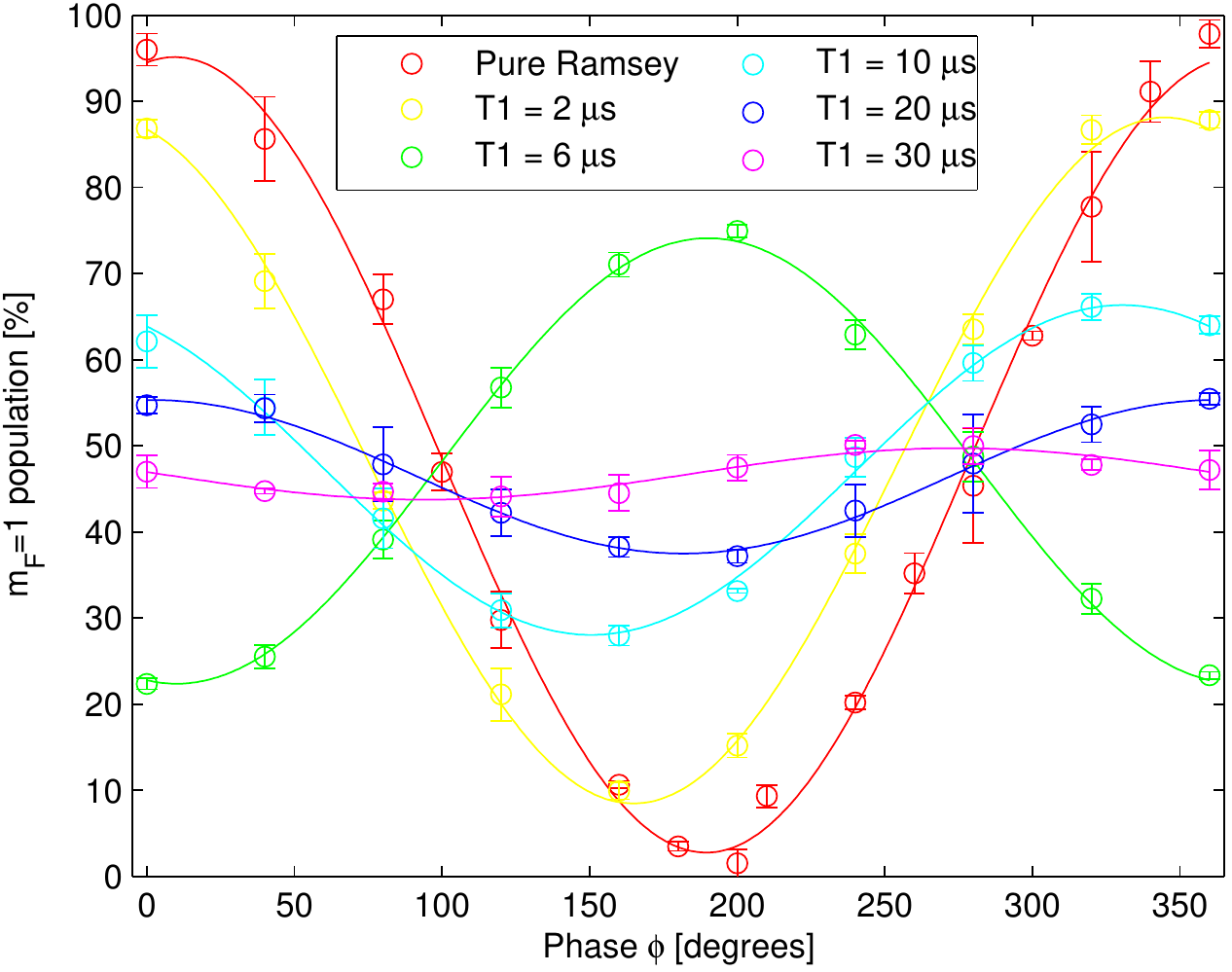}}
\caption{Fractional population in the $m_F=1$ state as a function of the applied phase shift $\phi$ between the Ramsey $\pi/2$ pulses, for different kick times. These data belong to the blue data set as it appears in table~\ref{table:full loop parameters} and Fig.\,4 of the main text. Also shown is the `pure' Ramsey sequence (no gradients), used for the normalization of the contrast values. A fit to a sine function is shown for each curve (see Sec.~\ref{sec:ExpProcedure}).}
\label{fig:example data}
\end{figure}

After achieving a population interference pattern (see Sec.\,\ref{sec:ExpProcedure}), the pattern's contrast is measured by fitting to a sine function of the form $P(\phi) = 0.5 C \sin(\phi+\phi_0)+const$, where $C$ is the contrast, $\phi$ is the applied phase shift, and $\phi_0$ is a constant phase term. As noted, the resulting contrast is normalized to that of a pure Ramsey sequence, i.e., without any magnetic gradients (see Fig.~\ref{fig:example data} for typical data).

\begin{figure}
\centerline{
\includegraphics[width=0.95\textwidth]{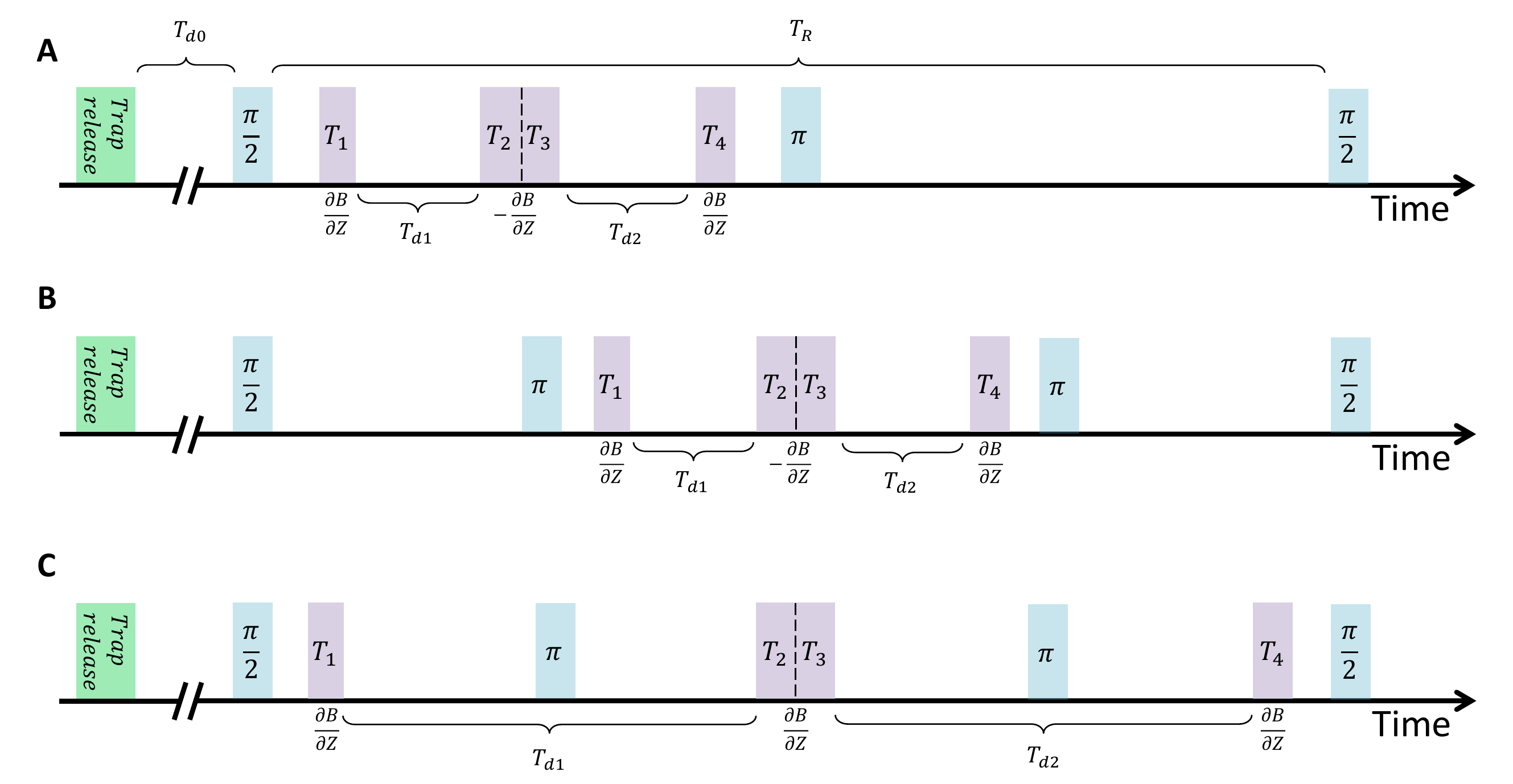}}
\caption{\label{fig:full loop timing} Timing diagram of the full-loop experimental sequences. Symbols are explained in table~\ref{table:full loop parameters}. (A),(B) The current inversion scheme, where the sign of the gradient $\partial B/ \partial z$ is switched during the sequence. The difference between (A) and (B) is the additional $\pi$ pulse before the gradients, used to increase the coherence time. (C) The spin inversion scheme, in which the sign of the gradient $\partial B/ \partial z$ is kept constant, and the relative force between the spin states is inverted by using the $\pi$ pulses in between the gradients, inverting the spin states.}
\end{figure}

The basic experimental procedure used in the full-loop scheme is described in Sec.\,\ref{sec:ExpProcedure}, and its parameters are given in table\,\ref{table:full loop parameters}. Here we describe in more detail the 'current inversion' and 'spin inversion' sequences, used in Fig.\,4 of the main text, beginning with the former sequence. These different sequences are used in order to access a larger region of parameter space and to ensure the robustness of our results (timing diagram is shown in Fig.~\ref{fig:full loop timing}). In the first sequence, after applying a $\pi/2$ pulse and the splitting gradient $T_1$ in one direction (downwards towards gravity), we reverse the sign of the acceleration by reversing the sign of the currents in the chip wires for the stopping and reversing gradients $T_2$ and $T_3$, working in the opposite direction (upwards towards the chip, this is done using two independent current shutters connected to the chip in opposite directions). The opposite gradient causes the relative movement between the wave packets to stop during $T_2$, and to change sign during $T_3$. Finally the wave packets are brought to a relative stop and spatial overlap by the second stopping pulse $T_4$ given in the same direction as the first gradient. The sequence is finished by applying a $\pi$ pulse (to increase coherence time, giving rise to an echo sequence) and a $\pi/2$ pulse, for mixing the different spin states and enabling spin populations interference (the $\pi/2-\pi-\pi/2$ sequence is symmetric in time). The four consecutive gradient pulses used in the current inversion sequence are applied either after the first $\pi/2$ pulse (i.e. only a single $\pi$ pulse is used as described above; used in all points in the blue data set and points \#1-5 in the red data set, see table\,\ref{table:full loop parameters}), or in between two $\pi$ pulses to further increase the coherence time (the $\pi/2-\pi-\pi-\pi/2$ sequence is again time symmetric; used in points \#6-10 in the red data set).

In the second sequence, we keep the same current direction in all gradient pulses, while reversing the spins with the help of two $\pi$ pulses. These pulses are applied, first just before the stopping gradient pulse $T_2$ and second just after the reversing gradient pulse $T_3$. Each $\pi$ pulse causes the spin states to flip sign, thus changing the direction of the applied momentum kicks in the center-of-mass frame (in lab frame, all gradient pulses push the atoms downwards towards gravity). The spin inversion sequence is used in all points of the green data set.

\subsection{Calculation of the BEC wave packet size}

\begin{figure}
\centering\includegraphics[width=0.9\textwidth]{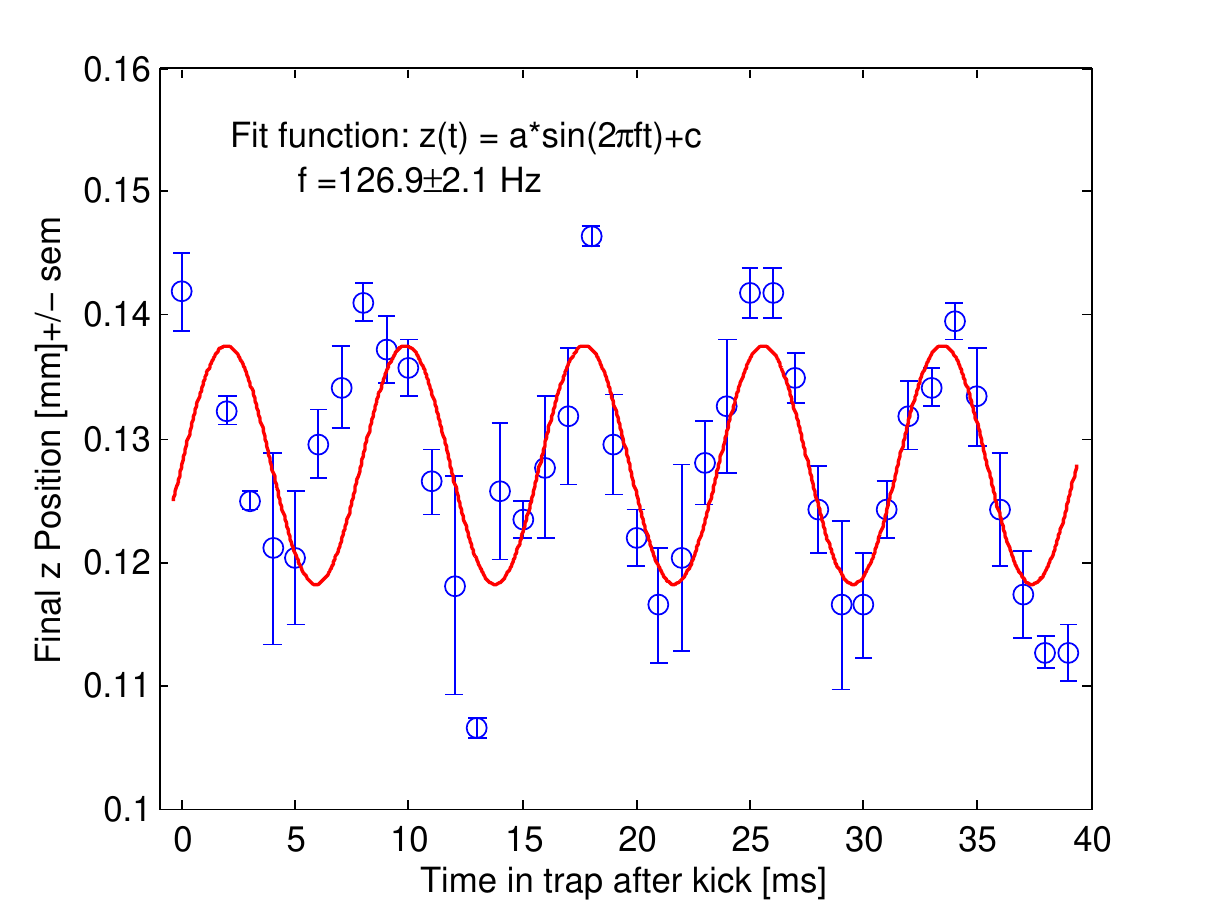}
\caption{\label{fig:trap frequency} Measurement of the trap frequency in the z direction. The x axis is the time the atoms oscillate in the trap, after applying a kick in the z direction, using a gradient from the chip. The y axis shows the final position, measured after trap release and additional TOF. The fit gives a result of $f=126.9\pm2.1$ Hz.}
\end{figure}

We calculate $w_0$, the in-trap Thomas-Fermi condensate half-length, in the z (gravity) direction according to \cite{Ketterle1999}:
\begin{equation}\label{eq:TF wp size}
w_0 = \sqrt{\frac{2\mu}{m}}\frac{1}{\omega_z},
\end{equation}
where the Thomas-Fermi expression for the chemical potential of a harmonically confined condensate is given by \cite{Ketterle1999}
\begin{equation}\label{eq:TF chemical potential}
\mu^{5/2} = \frac{ 15\hbar^2m^{1/2} }{ 2^{5/2} } N_0 \bar \omega^3 a,
\end{equation}
$\bar \omega = (\omega_x \omega_y \omega_z)^{1/3}$ is the geometric mean of the trap frequencies, and $a=98 a_0 = 5.18$ nm is the $^{87}$Rb scattering length.
Using the values: $\omega_z = 127$ Hz (measured, see Fig.~\ref{fig:trap frequency}; we also assume that $\omega_y=\omega_z$); $\omega_x = 38$ Hz (evaluated from magnetic trap simulation), $N_0=$10,000 atoms, we obtain $w_0 = 2.88$ $\mu$m. We then multiply this number by 0.41 to turn it into Gaussian width (see Sec.~S7.2) and obtain $\sigma_z^{TF}\simeq0.41w = 1.2\,\mu$m, which is the number used in Fig.\,4 of the main text. If we assume we have a factor of 2 error in both the number of atoms and in $\omega_x$ (as these parameters have the biggest uncertainties), the resulting error in $\sigma_z^{TF}$ is $\sim 30 \%$.

In order to compare this result to the experiment, we would like to measure the Thomas-Fermi wave packet size directly by imaging. However a 3$\,\mu$m wave packet size is on the edge of our imaging resolution, meaning we cannot expect to see such a small wave packet without bias. Moreover, at short TOF when the cloud is dense and close to the chip, we have some effect which causes negative values of optical density to be measured near the upper and lower edges of the cloud, thus distorting the image. This is possibly due to the cloud diffracting the imaging light, adding more light on the cloud edges (instead of showing positive or zero absorption).

\begin{figure}
\centering\includegraphics[width=0.9\textwidth]{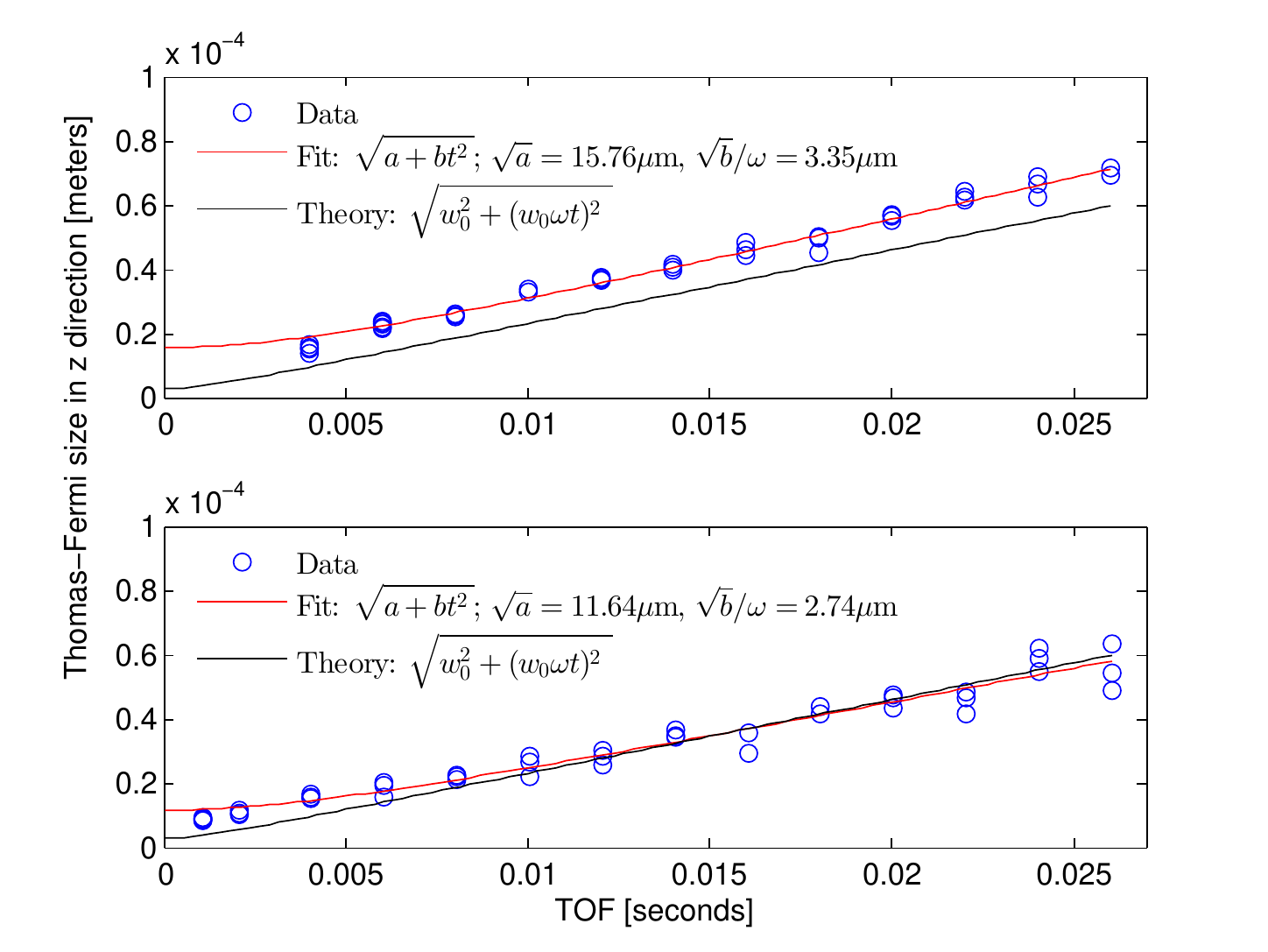}
\caption{\label{fig:WP size} BEC Thomas-Fermi size as a function of TOF, each panel is a different data set. Red line is a fit with two free parameters ($a$ and $b$). Calculating $w_0$ by extrapolating the TOF to 0 (i.e. taking $w_0 = \sqrt{a}$) gives a result far from the theoretical value - much more then the evaluated error of 30\%. However, using the slope ($w_0=\sqrt{b}/\omega$, using the trap frequency $\omega$ which is independently measured), we obtain values close to the theoretical one of  $w_0 = 2.88$ $\mu$m. Black line is the theoretical prediction, no free parameters.}
\end{figure}

These two effects mean that imaging the wave packet at short TOF is not reliable, and we need to use a different technique: we perform a measurement of the Thomas-Fermi wave packet size as a function of TOF (Fig.~\ref{fig:WP size}). The equation of the BEC Thomas-Fermi $w$ size as a function TOF is given by \cite{Ketterle1999}:
\begin{equation}\label{eq:TF size vs TOF}
w(t) = w_0\sqrt{1+\omega^2t^2},
\end{equation}
where $w_0$ is given by Eq.~\ref{eq:TF wp size}. Although we cannot deduce $w_0$ by extrapolating TOF to 0, we can use the slope to determine $w_0$, assuming we know the trap frequency $\omega$. An independent measurement of the trap frequency in z direction is shown in Fig.~\ref{fig:trap frequency}.

Experimental results for the Thomas-Fermi wave packet size as a function of TOF are shown in Fig.~\ref{fig:WP size}. Values are obtained by fitting the imaged atomic density to a bi-modal (Gaussian + Thomas-Fermi) density profile. We fit the resulting wave packet size vs. TOF to an equation of the form $\sqrt{a+bt^2}$, which corresponds to Eq.~\ref{eq:TF size vs TOF} with an extra free parameter for the wave packet size at zero TOF. Using parameter $b$ obtained from the fit, we calculate the initial wave packet size to be $w_0=\sqrt{b}/\omega = 3.35$ and 2.74\,$\mu$m, for the upper and lower panels of Fig.~\ref{fig:WP size}, respectively (each panel represents a different data set). These numbers are close to theoretically calculated value of $w_0 = 2.88$ $\mu$m.

On the other hand, calculating $w_0$ from parameter $a$ gives $w_0=\sqrt{a} = 15.76$ and 11.64\,$\mu$m, for the upper and lower panels, respectively, far from the theoretical value. At a TOF of 4\,ms, theory gives $w(t=4$\,ms$)=10\,\mu$m, while the experimental result gives $15.4\pm0.9\,\mu$m, and $15.74\pm0.7\,\mu$m for the upper and lower panels of Fig.~\ref{fig:WP size}, respectively. The limited imaging resolution alone cannot explain this deviation from theory. Another possible explanation is that we have additional limitations to our imaging system (e.g. cloud is out of focus). Alternatively, it is possible that the experimental results are correct, and the theory misses some other effect. Although the latter case means we spatially split the BEC to less of its spatial extent, it also means we split it in momentum more than the maximal number of $\sim$60 reported in Fig.~4b of the main text (assuming the BEC is a minimal-uncertainty wave packet).


\section{Instability sources and optimization}\label{sec:instability}

\subsection{Half-loop interferometer}
In the half-loop configuration, a major source of phase noise is the shot-to-shot current fluctuations in the chip wires, which cause fluctuations of the magnetic field energy during the time between the two $\pi/2$ pulses, in which the two wave packets occupy two different spin states \cite{machluf}. Using a quadrupole field to create the magnetic gradient reduces this noise (see Fig.\,\ref{fig:quadrupole}).

To suppress the 50\,Hz electrical grid noise which is coupled to the atoms through the bias coils' current supplies, we synchronize the experimental cycle start time to the phase of the electrical grid sine wave (this is done by using a phase-lock loop, which sends a trigger to the experimental control). In our experiment this significantly reduced this type of noise.

Additional technical sources of noise include timing jitter of the magnetic gradient pulses (originating from the experimental control hardware), phase measurement uncertainty (due to the fitting procedure) of about 0.1\,rad, and chip-to-camera relative position fluctuations (along the direction of the fringes). For the latter, assuming a shot-to-shot instability of 1\,$\mu$m and a 31.4\,$\mu$m interference pattern periodicity, these vibrations would create a $2\pi/31.4=0.2$\,rad phase instability.

\begin{figure}
\centerline{
\includegraphics[width=0.55\textwidth]{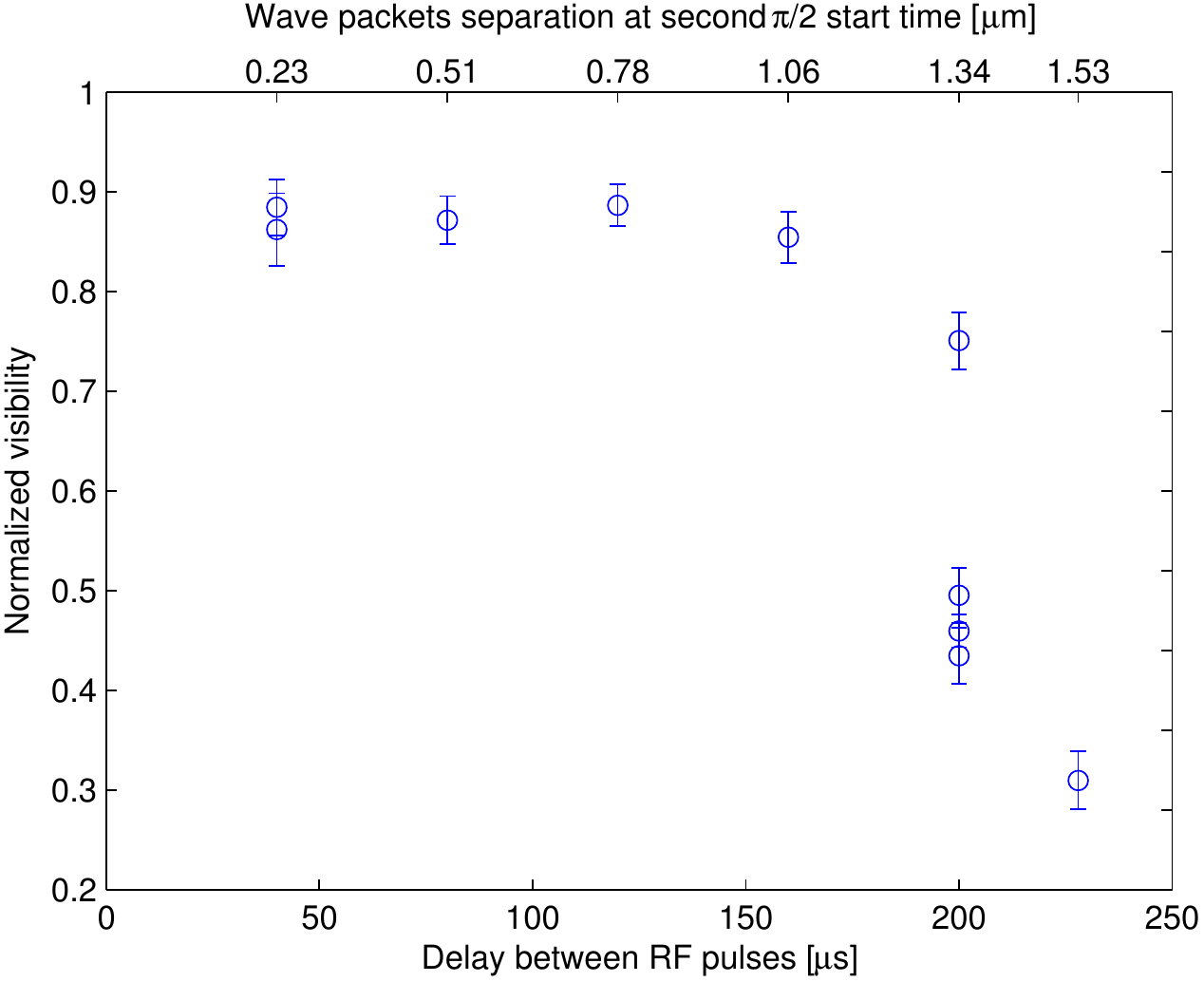}}
\caption{\label{fig:RFdelay}Normalized multi-shot visibility as a function of time interval between the two $\pi/2$ RF pulses (bottom x axis). This is the duration in which there exists a spatial superposition of two different spin states. For times shorter than 150\,$\mu$s, the normalized multi-shot visibility is not sensitive to the interval time. We also show in the top x axis the corresponding calculated separation of the two wave packets. Note that the two x axes do not have exactly the same zero point as the spatial separation starts only with the start of the magnetic gradient pulse and this starts with some delay after the first RF pulse has been completed. For comparison, the Ramsey coherence time (i.e. no spatial separation) is 400\,$\mu$s.  }
\end{figure}

Although the fluctuations coming from the homogeneous bias field are considered to be small, we minimize the time in which the wave packets have a different spin to $40\,\mu$s. In Fig.\,\ref{fig:RFdelay} we show that the visibility is not sensitive to this time interval as long as it is below $150\,\mu$s.
The high normalized multi-shot visibility corresponds to phase fluctuations of less than 0.5~radian for this time period, showing that the fluctuations of the homogeneous bias field are smaller than $\delta B/B\sim 2\cdot 10^{-5}$. However, one should not exclude the possibility that long-time drifts of the value of the bias field do exist and may give rise to changes in the optimal system parameters.

\begin{figure}
\centerline{ \includegraphics[width=0.55\textwidth]{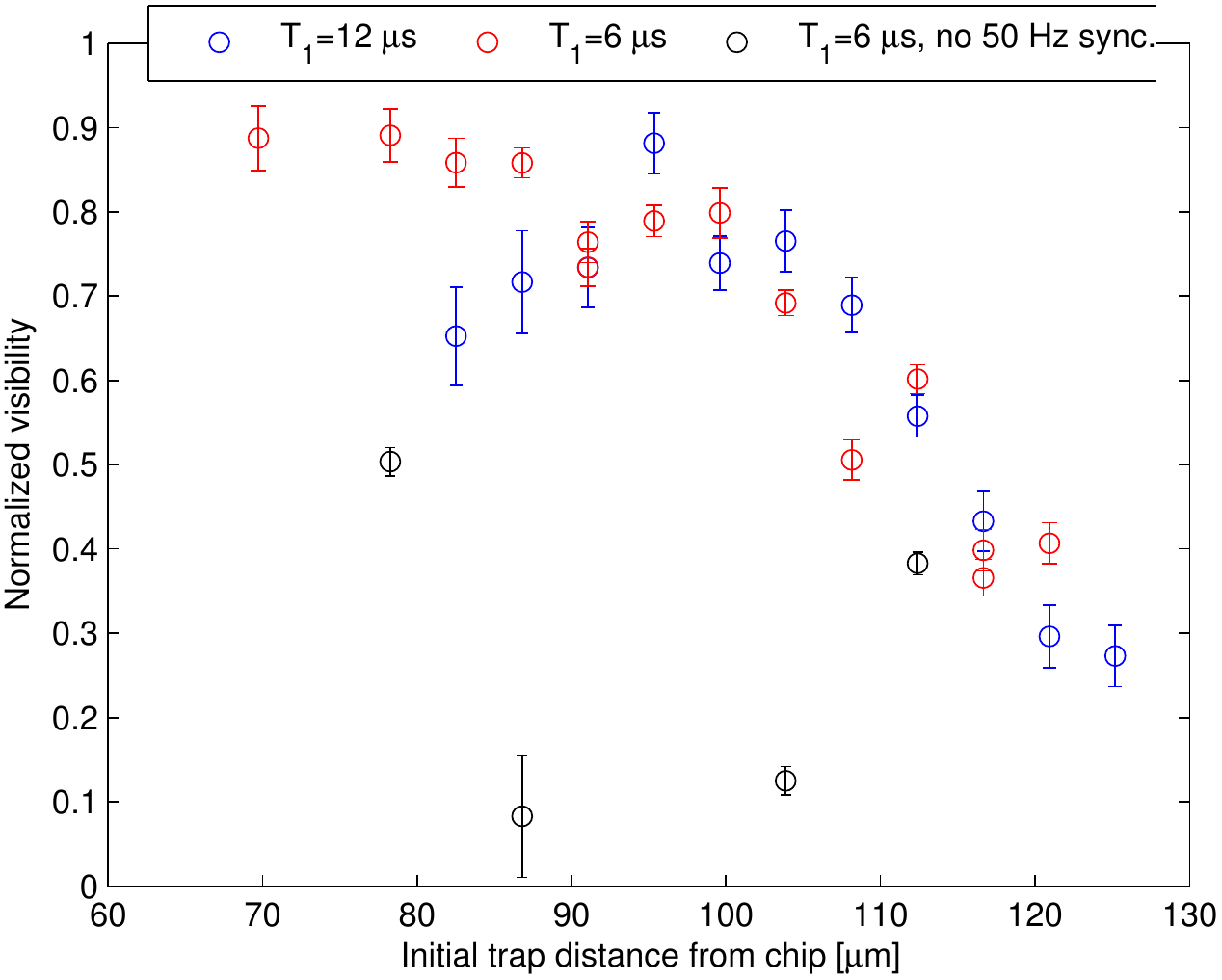}}
\caption{\label{fig:dip}Normalized visibility as a function of the initial trap distance from the atom chip. The other experimental parameters are $(T_1,T_d,T_2)=(6,104,130)\,\mu$s and $(12,158,180)\,\mu$s. Minimal visibility reduction due to magnetic fluctuations during splitting is expected when the splitting position ($4~\mu$m farther than the trapping distance due to free fall between trap release and gradient time) is closest to the quadrupole center (at $z=98~\mu$m). In practice the visibility may also be governed by imperfections in the later stages of the sequence if the parameters are not optimized for each measurement point. The low visibility data in this plot shows the consequence of lack of synchronization with the 50\,Hz electricity grid.}
\end{figure}

To further reduce the phase noise, we optimize the experimental sequence parameters, including the initial distance of the magnetic trap from the chip. The initial trapping distance determines the position of the atoms relative to the magnetic quadrupole created by the chip wires during the gradient pulses.
At a given distance $|z-z_{\rm quad}|$ from the quadrupole center the differential phase fluctuations $\delta\phi$ are proportional to the relative current fluctuations, such that ($\Delta m_F=1$) $\delta\phi(z)=-g_F\mu_B\delta B(z)T_1/\hbar\propto k(T_1)|z-z_{\rm quad}|(\delta I/I)$, where $\hbar k(T_1)$ is the differential momentum kick of the beam-splitter, which is linear with the splitting pulse time $T_1$, and $\mu_B$ is given in units of Joule per magnetic field (in Gauss or Tesla). The latter proportionality to $\delta I/I$ is a result of the fact that $k(T_1)$ is proportional to the magnetic gradient, and the latter times the distance from the quadrupole center equals the magnitude of the magnetic field. These fluctuations never vanish completely, as the wave packets have a finite size (about 6\,$\mu$m Thomas-Fermi edge-to-edge) and their center position also fluctuates from shot to shot (initial trapping position fluctuations are estimated in our system to be $\pm1.5\,\mu$m). Figure\,\ref{fig:dip} shows the dependence of the normalized multi-shot visibility on the initial wave packet position when the other parameters are kept constant. Ideally the visibility would be maximal when the atoms are closest to the quadrupole center at $=98\,\mu$m, namely, when we release the atoms from the trap at $z_{\rm trap}=94\,\mu$m (taking into account a 4\,$\mu$m falling distance before we apply the splitting pulse). However, the initial position of the atoms also affects
the magnitude of the magnetic field gradient (related to the amount of momentum $\hbar k$ imprinted on the cloud) and the field's curvature.
This influences the stability of later stages of the interferometric process, such as relative stopping of the two wave packets by the second gradient pulse, so that the maximal visibility may not occur exactly at the optimal position during the splitting pulse and the highest value of the visibility is not reached. In the half-loop experiment we did not perform a full combined optimization of the initial trapping position and the delay and stopping pulse durations, but rather used a constant trapping position of about $z_{\rm trap}=87.5\,\mu$m (Fig.\,\ref{fig:dip}) and optimized the duration of the delay and stopping pulses, as described below.

\begin{figure}
\centerline{ \includegraphics[width=0.55\textwidth]{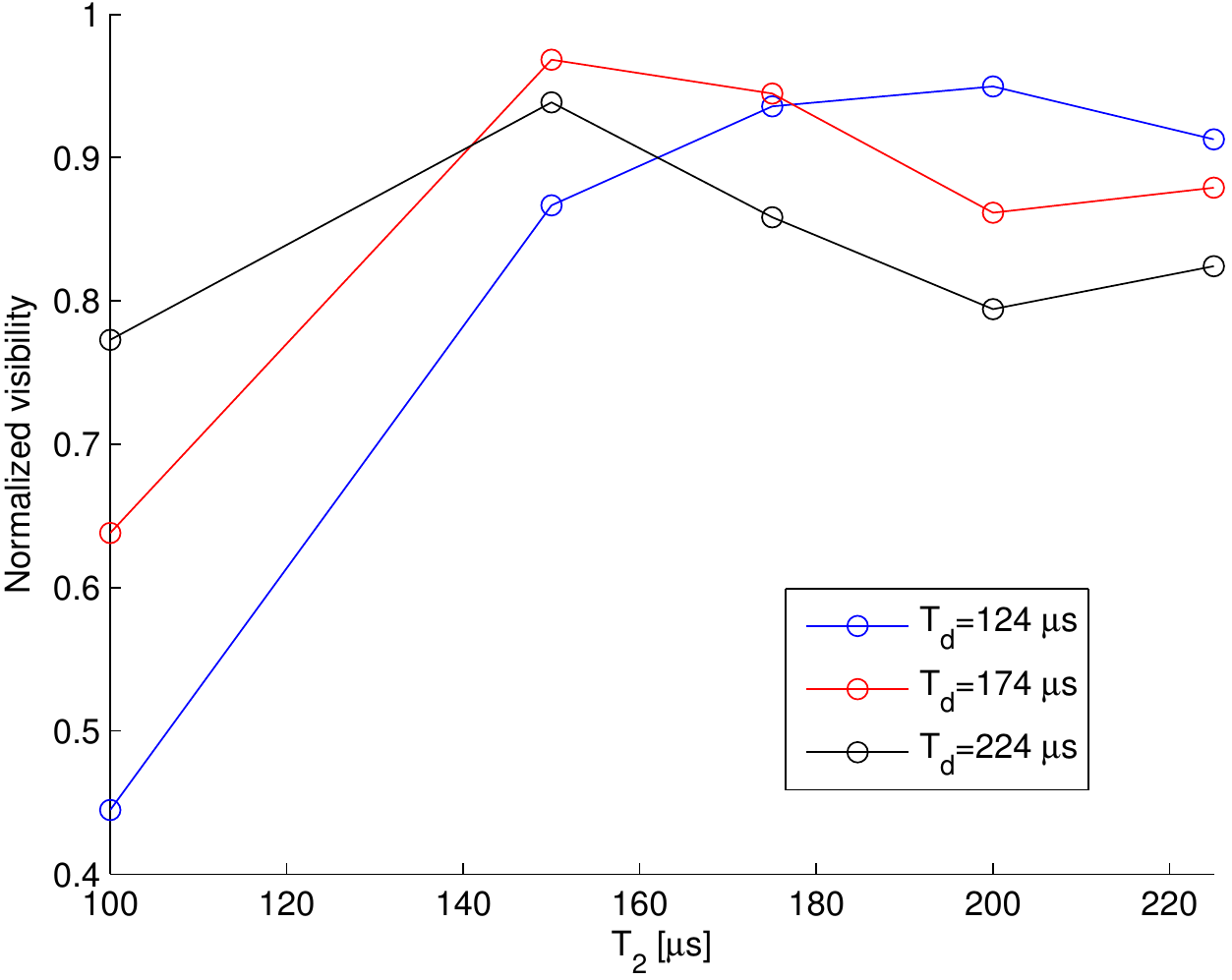}}
\caption{\label{fig:OptimizationT2} Normalized visibility as a function of the stopping gradient pulse duration $T_2$ for a given splitting pulse duration ($T_1=6\mu$s) and a few values of free propagation time between gradient pulses $T_d$. We expect maximal visibility when the stopping pulse is designed to most accurately stop the relative wave packet motion [$T_2={\rm acot}(\omega T_d)/\omega$, see Eq.\,\ref{eq:T2}]. Corresponding nominal optimal values for $T_d=124,174$ and 224\,$\mu$s are $T_2=185, 154$ and 130\,$\mu$s, respectively, for the estimated curvature $\omega=2\pi\times 850$\,Hz.}
\end{figure}

In Fig.~\ref{fig:OptimizationT2} we demonstrate the basis for our main half-loop optimization procedure, which aims to minimize the hindering effects of fluctuations in the later stages of the interferometric sequence: stopping the relative wave packet motion after the free propagation time $T_d$. For a constant initial trapping position (which is relatively close to the quadrupole center) we change the second gradient pulse time $T_2$ for several propagation times $T_d$. For each value of the propagation time $T_d$, maximal normalized multi-shot visibility is observed at a corresponding stopping time $T_2$ for which we believe that a full stopping of the relative wave packet motion is achieved. We note, however, that other factors, such as long-term drifts in the homogeneous magnetic field from the bias coils or voltage supplied to the chip wires, may also affect the stability of the phase, so in order to obtain the data points in Fig.\,3 of the main text we have used the maximal value of the normalized multi-shot visibility taken over many experimental sessions (see Fig.~\ref{fig:optimization visualization}), such that these values represent the minimum effect of noise sources other than the main source: shot-to-shot current fluctuations during the splitting gradient pulse.

We note that if the stopping parameters are optimized, we expect the initial position fluctuations of the atoms to play a very minor role in the final visibility.
As shown in Sec.~\ref{sec:phasespace}, if the stopping pulse is designed to almost completely stop the relative motion of the centers of the two wave packets, then the final shape of the fringe patterns, including their final position, is the same as the shape of the fringe patterns formed just after the splitting pulse, up to a scaling factor. As the phase of the fringes after the splitting pulse, namely the position of their peaks, are determined only by the magnetic field gradient and not by the envelope of the initial wave packet, shifts in the initial position of the wave packets before the splitting kick will not cause any phase shifts.
It follows that the positions (phase) of the observed fringes are expected to be independent of the initial wave packet position, even if the envelopes of the fringe patterns move, as was reported in previous work~\cite{machluf}.

\subsection{Full-loop interferometer}

Table~\ref{table:error budget} lists possible sources of the contrast drop observed in the full-loop scheme (Fig.\,4 in the main text), which is currently unexplained by our theory (see outlook in main text). The table also lists why we ruled out each source, and the expected contrast loss due to each source. Here we explain some of these sources in more detail.

\begin{figure}
\begin{tabular}{cc}
\includegraphics[width=0.5\textwidth]{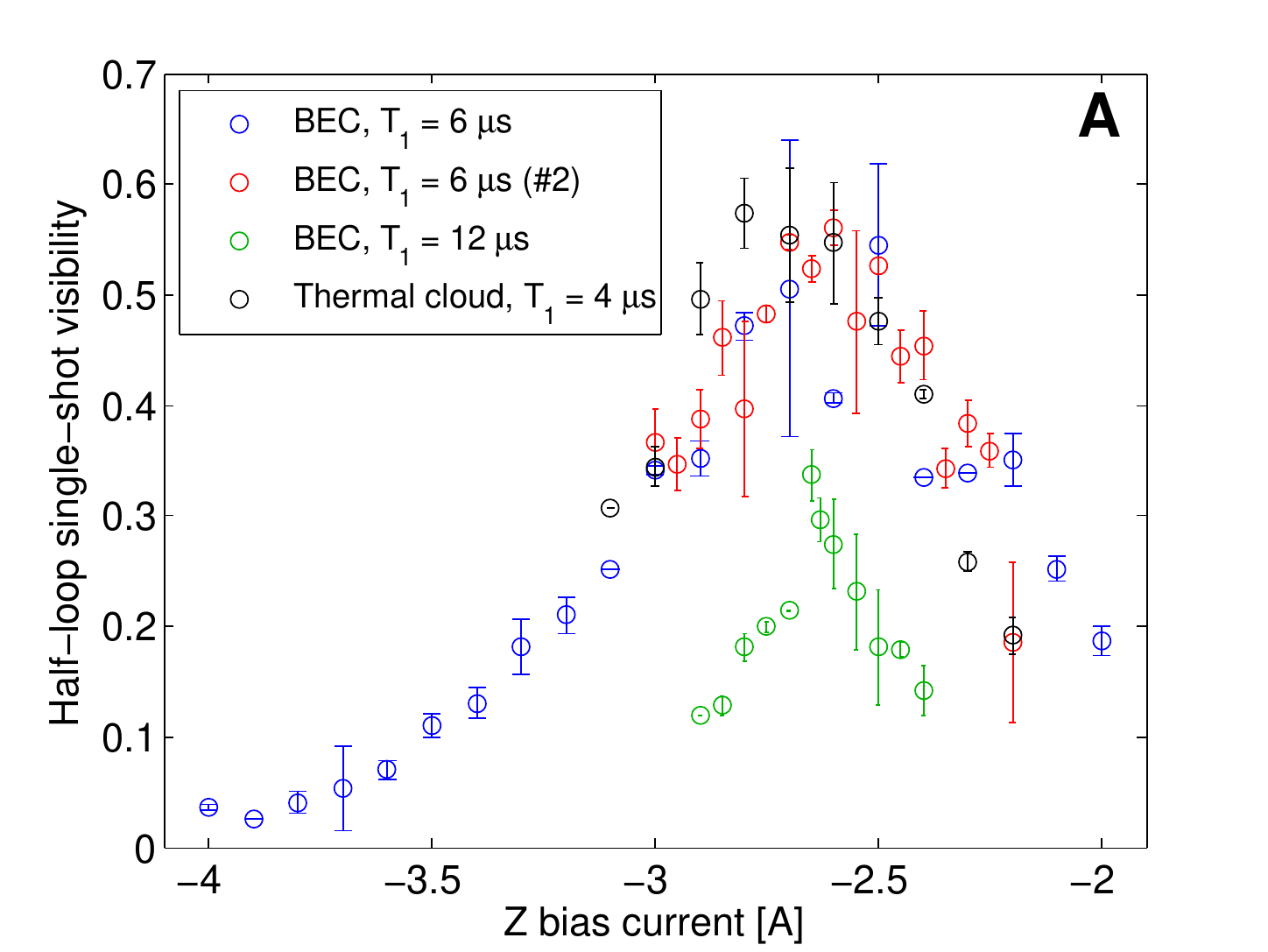}&
  \includegraphics[width=0.5\textwidth]{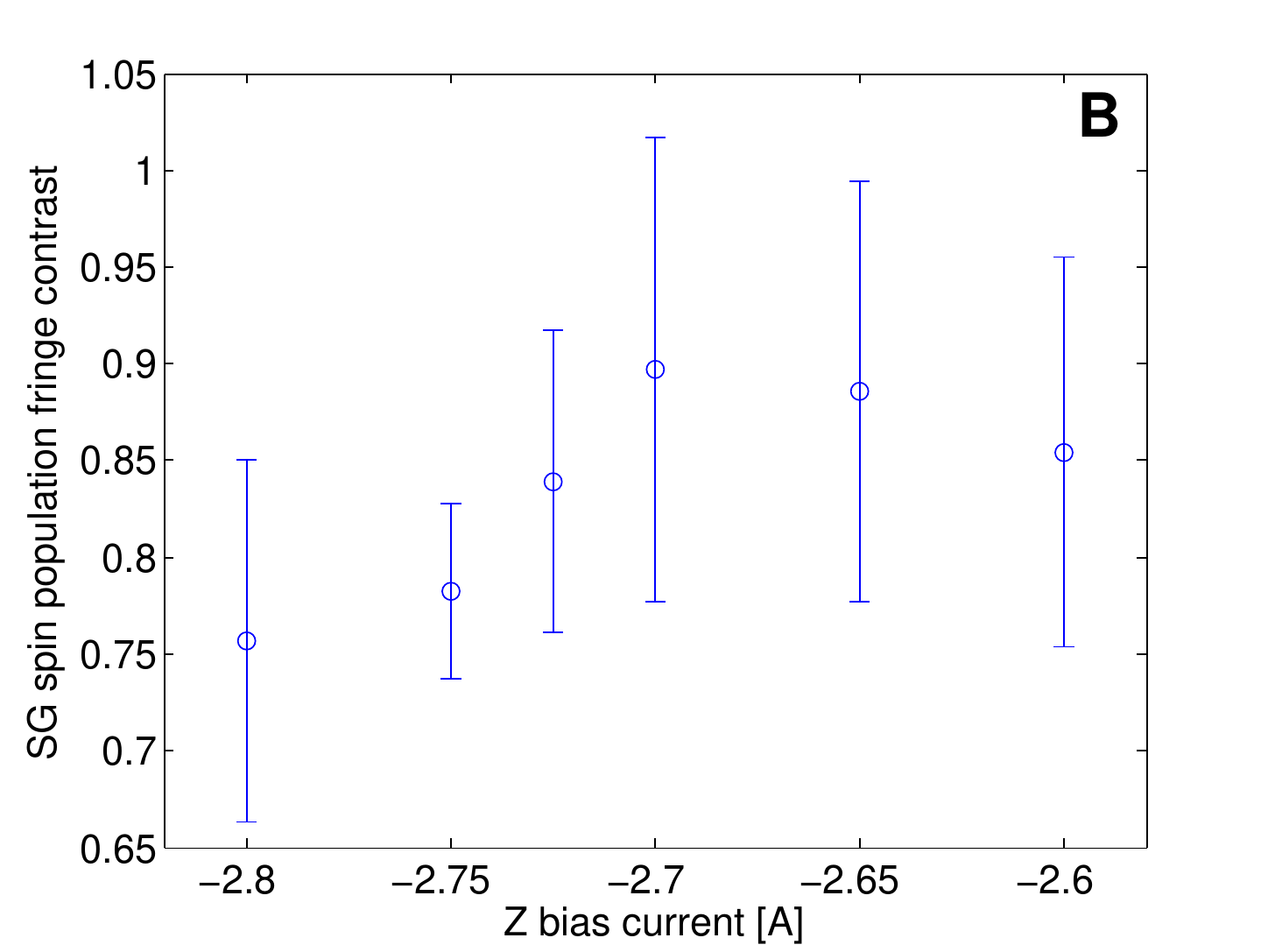} \\
\end{tabular}
\includegraphics[width=0.5\textwidth]{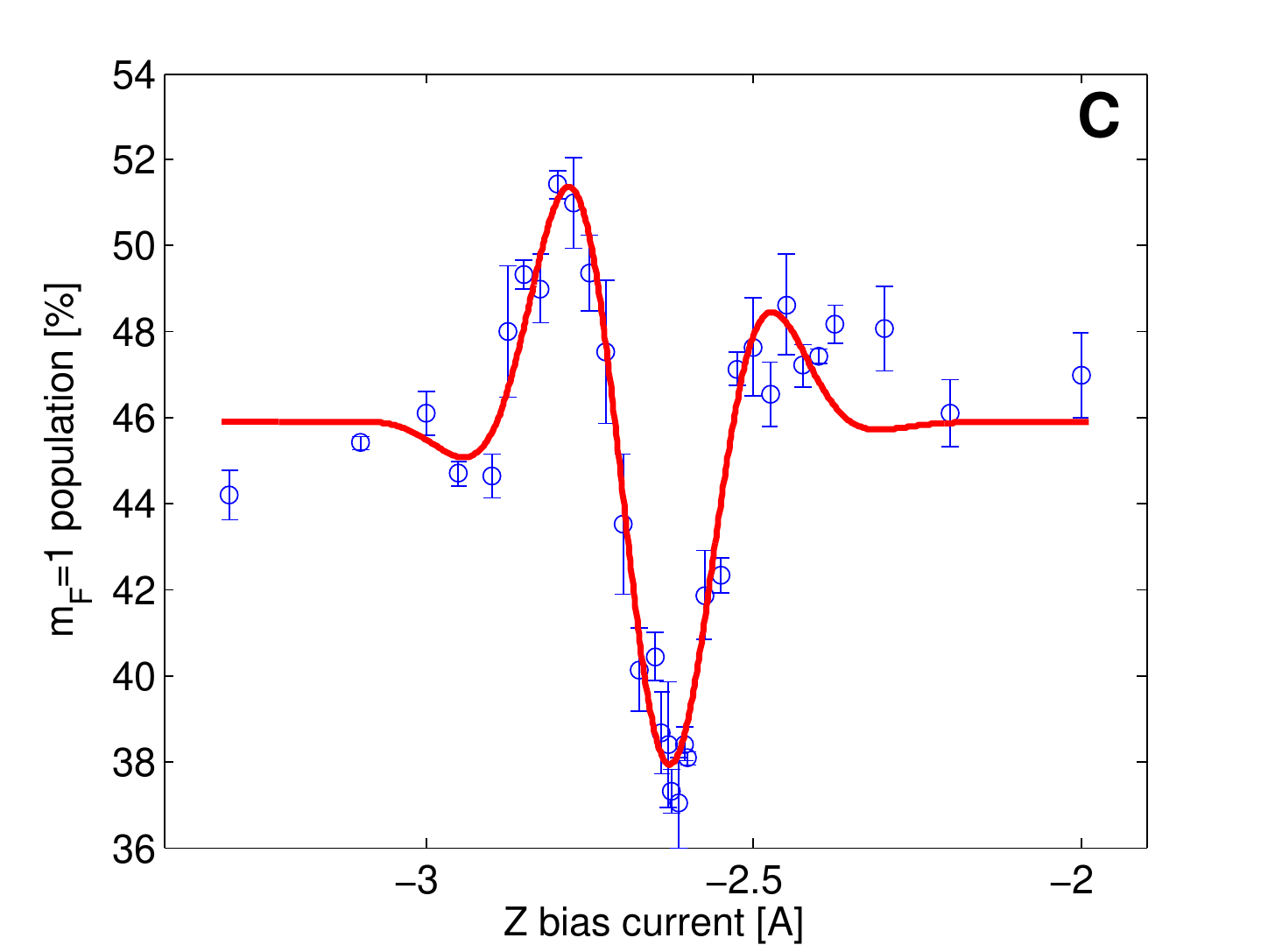}
\caption{\label{fig:Y position optimization} Initial $y$ position optimization using different observables, as a function of the Z bias value. The Z bias causes a $y$ position shift of about 30$\mu$m/A. (A) Single shot visibility in the half-loop sequence of a BEC and a thermal cloud, for different $T_1$ values. (B) Full-loop contrast for $T_1 = 2$ $\mu$s. (C)  Full-loop population oscillations for $T_1 = 20$ $\mu$s. Qualitatively, this result is similar to the result in Fig.\,5 of the main text. All three graphs show a similar result - the optimal point is around -2.65 A, for which the atoms are located directly below the chip center, and thus obtain momentum only in the $z$ direction (see text).}
\end{figure}

The first source on the list is the initial trap position offset in the $y$ direction, relative to the center of the chip. The SGI is sensitive to this source due to the geometry of our system: since the chip wires' dimension in the $y$ direction is 40\,$\mu$m, an offset of a few $\mu$m would mean that during a magnetic gradient kick, the different spins would acquire different momentum not only in the $z$ direction but also in $y$. This would result in loosing any interference signal, either due to momentum orthogonality between the wave packets, or zero overlap in the $y$ direction after some evolution time. On the other hand, we are insensitive to the initial position along $x$ since the wires run along that direction. Figure~\ref{fig:Y position optimization} shows the initial $y$ position optimization for both the half-loop and full-loop schemes. The initial $y$ position is adjusted by changing the $z$ bias magnetic field value before trap release.

Another possible source for contrast loss are constant magnetic gradients that exist in the chamber, either from our bias coils or from stray fields. During a Ramsey sequences the wave packets are found in a superposition of different spin states. A constant magnetic gradient exerts a differential force on the different spin states, thus causing orthogonality between the different spin wave packets and loss of contrast in the Ramsey sequence. In previous work \cite{margalit} we have found that our y bias coils produce a gradient in the z direction. The magnitude of the gradient is evaluated to be $\sim$ 71 G/m, which should induce a Ramsey decoherence rate similar to our measured value of about 400 $\mu$s (within an order of magnitude). However, adding one or two RF $\pi$ pulses (as is done in the full-loop sequence) reduces this effect, enabling to observe much longer coherence times (up to 4\,ms have been observed in our system). This proves that constant gradients should not play a significant role in contrast loss of the full-loop scheme.

Moreover, we can make another claim: if constant gradients are the main source for loss of coherence, we should not observe significant coherence loss when $T_R$ - the total Ramsey time (i.e. the time between the two $\pi/2$ pulses) remains constant, and some other parameter is scanned. However, we see that the contrast decreases rapidly even when $T_R$ is constant - see, for example, points 1-5 in the blue data set and 1-3 in the red data set (table~\ref{table:full loop parameters}). Although all points have the same total Ramsey interrogation time $T_R = 640$ $\mu$s, we see large variation in contrast (0.01-0.86). We can make the same claim using points 1-8 of data set 3.

\setlength\tabcolsep{2pt}

\def\arraystretch{1.1}
\begin{table}\begin{center}
\resizebox{\columnwidth}{!}{\begin{tabular}{p{0.4\textwidth} | p{0.6\textwidth} | p{0.25\textwidth} }
Imperfection causing SGI loss of coherence (in brackets: estimated magnitude of imperfection)& Reason of being insignificant
 & Expected coherence loss \\ \toprule[0.75mm]
Initial offset in y direction ($\sim1.5\mu$m) & Optimized (see Fig.~\ref{fig:Y position optimization} and text for more details); Spin inversion scheme should reduce sensitivity & 2\% for spin inversion \\ \hline
Constant magnetic gradient in x/y/z (71 G/m in z direction) & Spin echo sequences used (see Sec.~\ref{sec:Fig4}) should be insensitive to constant gradients; We see the contrast decreasing rapidly even when $T_R$ is constant; we should see splitting in the imaging x-y plane & negligible \\ \hline
Momentum and position are not optimized in the same value of $T_2$ & Separate optimization did not give any improvement; Since maximum $\Delta z$ is small, effect is negligible& see simulation results \\ \hline
Phase noise (the average population standard deviation of the experimental results is 3.35\%) & We measure the effect of phase noise on the output population (e.g. see Fig.~\ref{fig:example data}, and in main text Fig.\,1C and Fig.\,5), and it is too small  & 3.35\% \\ \hline
Initial wave packet is not close to a minimal uncertainty one, BEC is impure & For most of the data, we observe a BEC with a minimal BEC fraction is 70\%. Thus we don't expect a loss of coherence to below that of the BEC fraction & Reduction to $\sim$70\% of the original contrast. An exact calculation for the thermal part is beyond the scope of this paper \\ \hline
Spin dependent potential curvature, causing differential non-linear relative phase between wave packets& Simulated, should be a small effect & see simulation results \\ \hline
Atom-atom repulsion causing wave packet distortion & Gross-Pitaevskii simulation agrees with Thomas-Fermi and Castin-Dum simulations & negligible \\  \hline
RF pulses $\pi$ or $\pi/2$ are out of resonance, either due to magnetic `tail' of the gradient pulses, or long term drifts & Measured `tail' effect is small as we keep time separation between gradients and RF pulses (see table~\ref{table:full loop parameters}); $\pi$ pulse is calibrated between measurements & Reduction to $\sim$90\% of the original contrast \\ \hline
Fabrication: chip is not symmetric along y & Initial y position optimization should cancel most of the effect; Simulated and does not show strong coherence loss
 & 2\% for spin inversion  \\ \hline
Mismatch in some other dimension? e.g. rotation / tilt / spin... & Below our experimental ability to detect, and beyond the scope of this paper & \\
\end{tabular}}
\end{center}\caption{Possible sources of coherence loss in the full-loop scheme.}\label{table:error budget}
\end{table}

\bigskip
\goodbreak
\section{Main text figures}
In previous sections we have provided numerous details regarding the figures of the main text. In this section we give for completeness some additional details regarding these figures.
\subsection{Figure 1}

Since the thermal interference fringes shown in Fig.~1A suffer from wavelength chirp due to larger thermal cloud size (compared to a BEC), we fit this image by modifying the argument of the sine function in Eq.~\ref{eq:fit function} to
$\phi + \frac{2\pi}{\lambda}(z-z_{\text{ref}}) + k_1(z-z_{\text{ref}})^2$, where $k_1$ is a chirping parameter.

\begin{figure}
\centerline{
\includegraphics[width=0.6\textwidth]{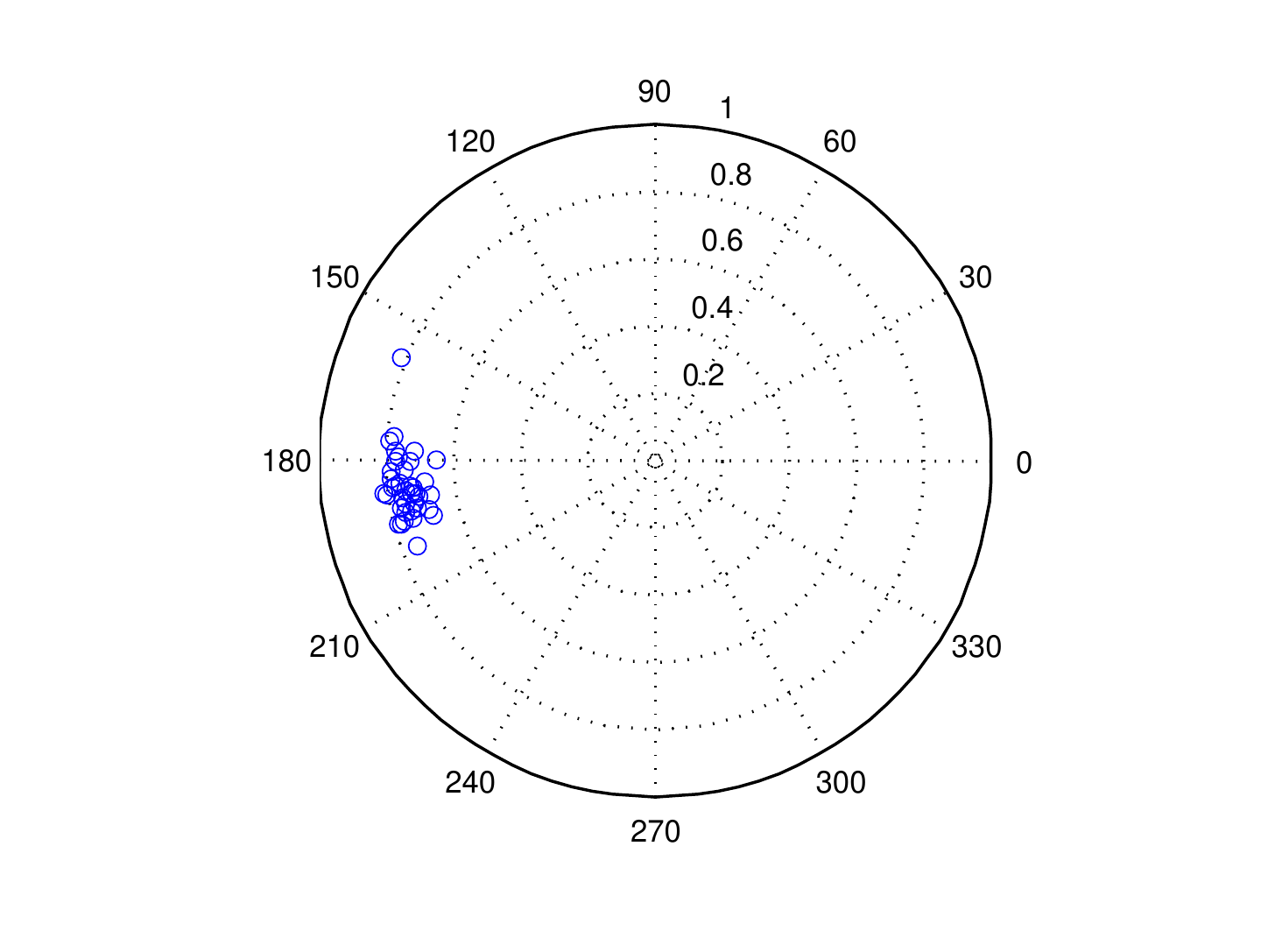}}
\caption{Polar plot of phase vs. visibility (shown as angle vs. radius), of the 40 consecutive half-loop SGI interference images composing the multi-shot image in Fig.\,1B of the main text. The experimental parameters are $(T_1,T_d,T_2)=(4,116,200)~\mu$s.}
\label{fig:polar phase}
\end{figure}

Figure~\ref{fig:polar phase} shows a polar plot of the phase of the 40 consecutive shots composing Fig.\,1B of the main text.
The splitting pulse duration is $T_1=4\,\mu$s and after a delay of $T_d=116\,\mu$s the stopping pulse duration is $T_2=200\,\mu$s. Time-of-flight is $TOF=6810\,\mu$s [the parameters in (A) are almost exactly the same].
As the observed absolute
visibility is $V_{\rm av}=0.76\pm0.01$, and the mean of the single-shot visibility is $\langle V_s\rangle = 0.77\pm0.03$, the normalized
visibility $V_N\equiv V_{\rm av}/\langle V_s\rangle$ is approximately $99\%$ (corresponding to a phase standard deviation of $\sim$0.1\,rad).

In Fig.~1C of the main text we show a high visibility spin population interference pattern. The data shown in 1C is before normalization, and has a visibility (obtained from fit, see Sec.~\ref{sec:Fig4}) of $91.02\pm7.32\%$.  As the contrast of the 'pure' Ramsey sequence (without magnetic gradients) is $95.44 \pm 5.01 \%$, the normalized contrast is $95.37 \pm 9.16 \%$. $T_1=2\,\mu$s, $T_{d1}=168\,\mu$s, $T_2+T_3=3.85\,\mu$s, $T_{d2}=164.15\,\mu$s, $T_4=2\,\mu$s and $TOF=50\,\mu$s (end of $T_4$ to last $\pi/2$ pulse).

\subsection{Figure 2}
In Fig.~2 of the main text we show a schematic position-vs-time diagram for both the half-loop and full-loop longitudinal SGI configurations. Both figures are plotted in the center-of-mass frame, namely, that of an equivalent atom with $m_F = 3/2$ being accelerated by the magnetic gradients and gravity.

As can be seen in Fig.~2A, during the stopping pulse ($T_2$) of the half-loop sequence the two wave packets have the same spin, meaning that the differential force (originating from the different distance of the wave packets from the chip at this stage) is small compared to the force during the first gradient pulse ($T_1$). This means that in order to stop the relative motion of the wave packets, a long stopping pulse is required, giving rise to an harmonic potential, due to the shape of the potential created by the chip wires. This harmonic potential creates a tight focus for the wave packets at time difference $T_d$ from the end of $T_2$. The values of the minimal wave packet width $\sigma_{\rm min}$ at the focal point are listed in table \ref{table:parameters} (calculated using in Eq.\,\ref{eq:sigma_min}). Due to this focusing, we achieve the ratio of 4.5-18 between the wave packet separation and their size, mentioned in the main text.

\subsection{Figure 3}
\begin{figure}
\centerline{
\includegraphics[width=\textwidth]{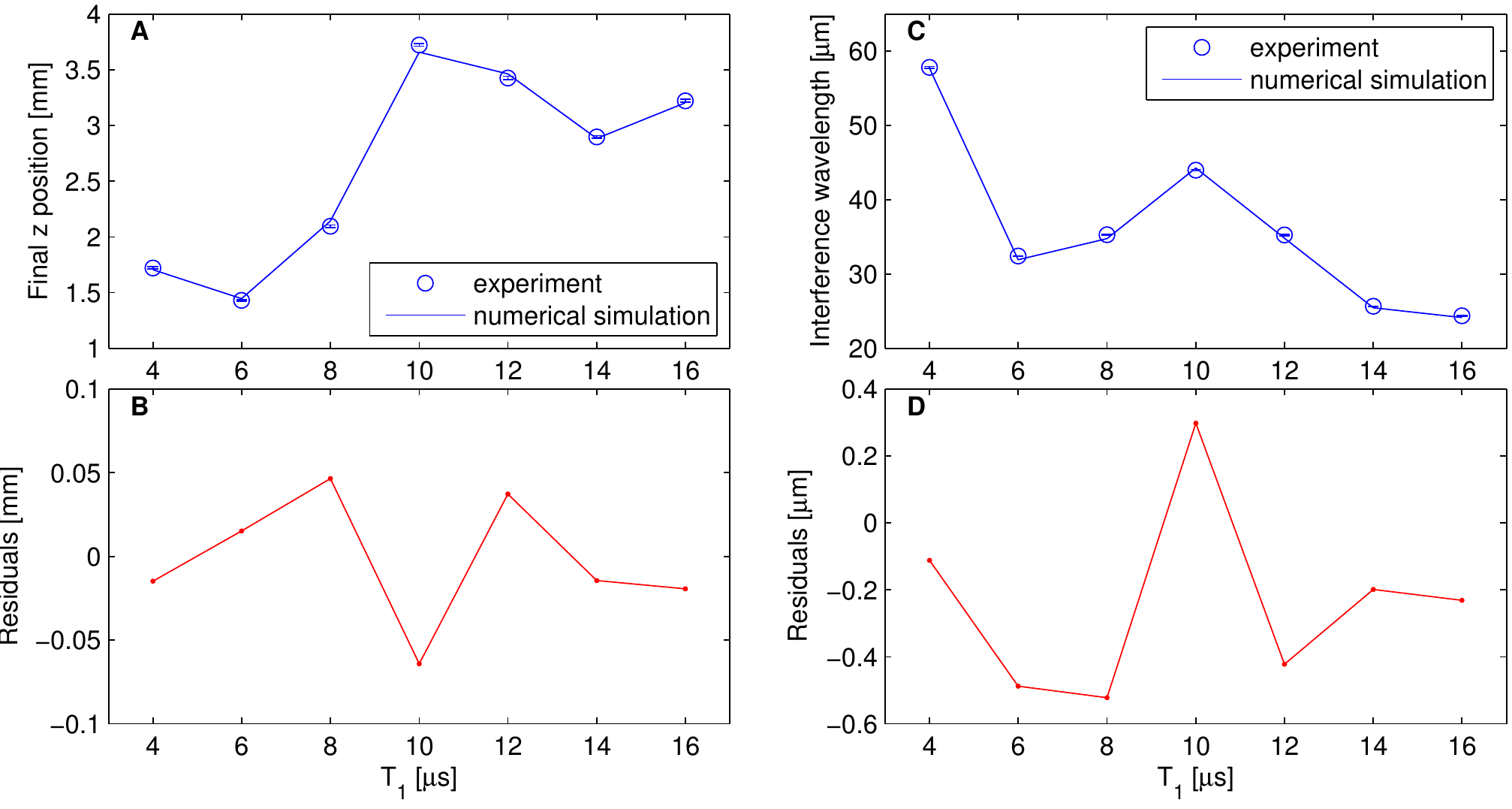}}
\caption{Comparison between the half-loop experimental results and the numerical wave-packet propagation model. (A) Final z position as a function of the first gradient pulse duration $T_1$. (B) Residuals showing the difference between the experiment and the simulation in (A). The mean absolute residual is 1.15\%. (C) Periodicity of the interference pattern as a function of the first gradient pulse duration $T_1$. (D) Residuals showing the difference between the experiment and the simulation in (C). The mean absolute residual is 0.97\%. The lines in (A) and (C) are not monotonic since these are different points in a multi-dimensional space: $T_1$ is not the only parameter that changes between each point, but also $T_d$, $T_2$, and the time-of-flight, due to the complex optimization process.}
\label{fig:half loop comparison}
\end{figure}

Details regarding the theoretical lines in Fig.\,3 are given in the next section. As a preface let us just note here the accuracy of our numerical wave packet dynamics simulation, by showing in Fig.\,\ref{fig:half loop comparison} the 1\% accuracy with which it estimates the basic parameters of the interferometer, specifically the maximal separation and periodicity of the interference pattern.

The random vector model data (green lines) were obtained by solving the time-dependent Schr\"odinger equation Eq.\,\ref{schrodinger} numerically. The random vector was chosen from a Gaussian
distribution by considering the infinite-correlation length limit. The wave packet was chosen to be at a distance of $5 \mu m$ above the
quadrupole center. The chosen parameters were $\epsilon=0.0069$ and $\epsilon=0.018$, respectively for high visibility and low visibility data. The first value was estimated from the wave-packet propagation model described in section \ref{sec:theory}, while the second is the experimental value used for when producing the added noise. A more extensive description of the random vector model is given in section \ref{sec:theory}.

The numerical wave-packet propagation model dashed and solid red lines, for the high and low-visibility runs, respectively) is based on a full simulation of the interferometric process as described in section~\ref{sec:theory}. For the high visibility data the model assumes relative current fluctuations $\delta I_1/I_1=0.3\%$ during the splitting pulse, as measured independently,  and $\delta I_2/I_2=0.14\%$, during the stopping pulse. As the fluctuations in the stopping pulse were not directly measured, we use a number that best fits the experimental data. For the low visibility data we show a line calculated by using the measured fluctuations ($\delta I_1/I_1=1.8\%$) during the splitting pulse. The numerical visibility was normalized to the multi-shot visibility of simulated fringe patterns whose fluctuations are purely due to initial position fluctuations $\Delta z_{\rm initial}$ of 1\,$\mu$m (standard deviation) around $z=87.5\,\mu$m from the chip. The analytical visibility (smooth solid line) was calculated by using Eq.~(\ref{eq:FourierVis}) (see section~\ref{sec:theory} below) with $\sigma_z=1.53\,\mu$m (wave packet width + initial position uncertainty), $\kappa=0.86\,(\mu{\rm m}\cdot \mu{\rm s})^{-1}$ and central distance of $z_0=6.5\,\mu$m from the quadrupole field center.

Here we would like to explain the error bars of the low visibility data in Fig.\,3A, which are much bigger than those of the high visibility data. As noted in Sec.~\ref{sec:Fig4}, the normalized multi-shot visibility is defined as $V_N\equiv V_{\rm av}/\langle V_s\rangle$, and the error bars are estimated using Eq.~\ref{eq:errorbars}. In that equation, the third term under the square root estimates the expected relative standard error of the normalized multi-shot visibility due to the finite sample size [Eq.~(\ref{eq:dVN})], and is given by $\frac{1}{2N}\left(\frac{1-V_N^2}{V_N}\right)^2$. Since for this factor grows larger as $V_N$ approaches 0, it results in a large error estimation, even when the number of single shots $N$ is high ($N=$ 138-267 for the low visibility data).

Since the low visibility data presented in Fig.\,3A were taken a few months after taking the high visibility data, the same experimental parameters gave normalized multi-shot visibility 2-10\% lower than the original data, due to long term drifts. To suppress the effect of this drift on the low visibility data, we normalized each measurement to a corresponding one using the same experimental parameters, in which zero noise was added (i.e. the chip current was held constant throughout the measurement), such that only 'natural' noise affected the results. This normalization means that only the added noise (induced by varying chip current during the first gradient pulse) affected the drop of visibility in the low-visibility data.

\begin{figure}
\centerline{
\includegraphics[width=0.8\textwidth]{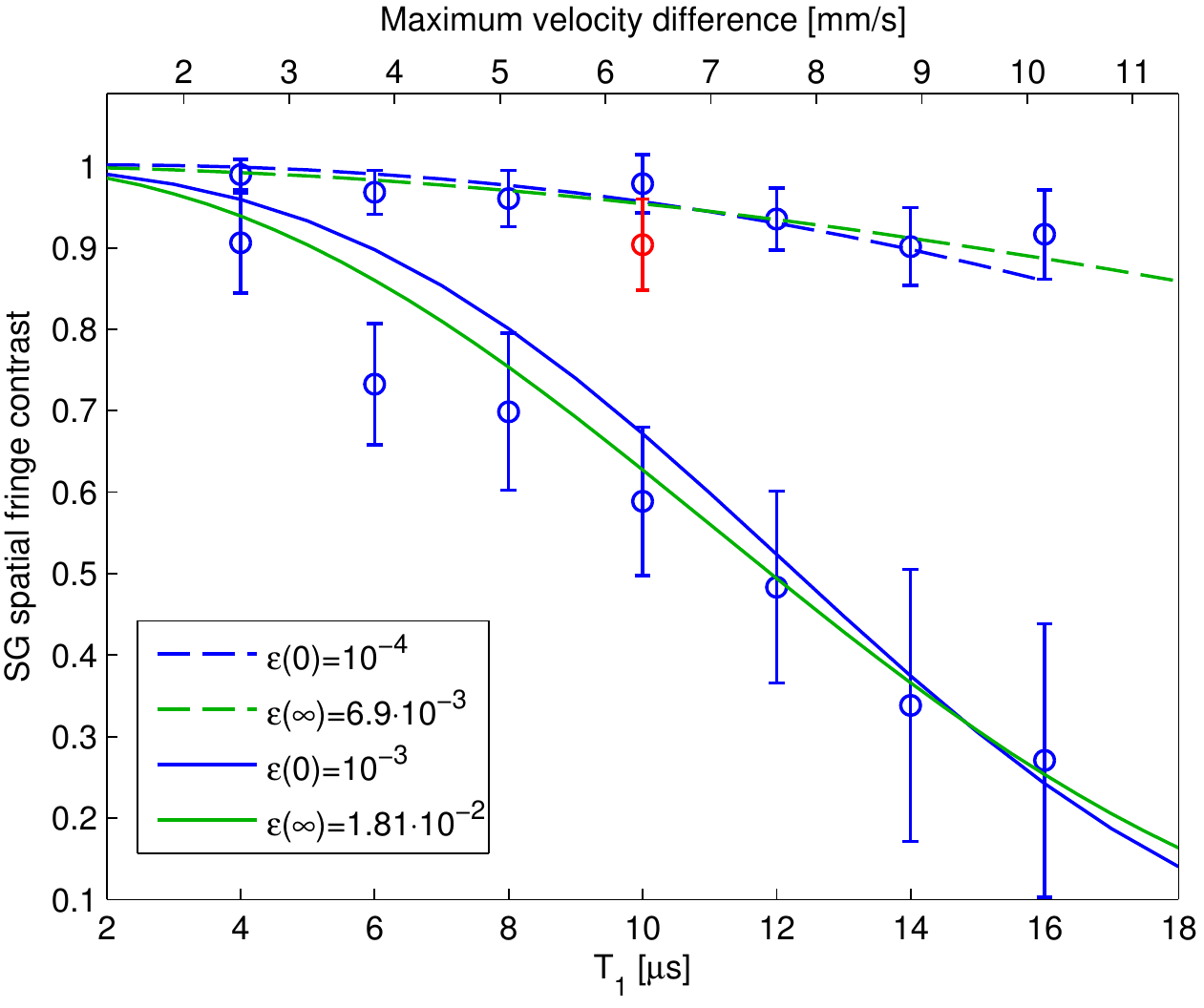}}
\caption{Half-loop results for the random vector model, see section~S5.1.  We numerically solve the model by applying the Crank-Nicolson algorithm \cite{crank-nicolson}, for two extreme cases: The first model assumes fluctuations with zero correlation length. This describes a situation where fluctuations in the magnet are local and the atom is very close to the magnet so that there is no spatial averaging. Such fluctuations may originate in Johnson noise or a local response – e.g. of magnetic domains, impurities and geometrical defects – to global temperature or vibrational fluctuations. The second model assumes the other extreme limit in which the fluctuations are global and the correlation length is infinite. This limit is more relevant to our experimental situation in which current fluctuations are the dominant source of randomness. In the future, as current fluctuations go down to the shot-noise level and below (squeezed currents, e.g. \cite{SubShotNoise})
so that Johnson noise becomes dominant, or when permanent magnets are used, both limits will be important.}
\label{fig:half loop with epsilon model}
\end{figure}

Figure~\ref{fig:half loop with epsilon model} shows a version of Fig.\,3 of the main text which includes theoretical predictions for the random vector model, with both zero-correlation length and infinite-correlation length assumptions for the fluctuation correlation length (see Sec.~\ref{sec:theory}).

\subsection{Figure 4}
\begin{figure}
\centerline{
\includegraphics[width=0.8\textwidth]{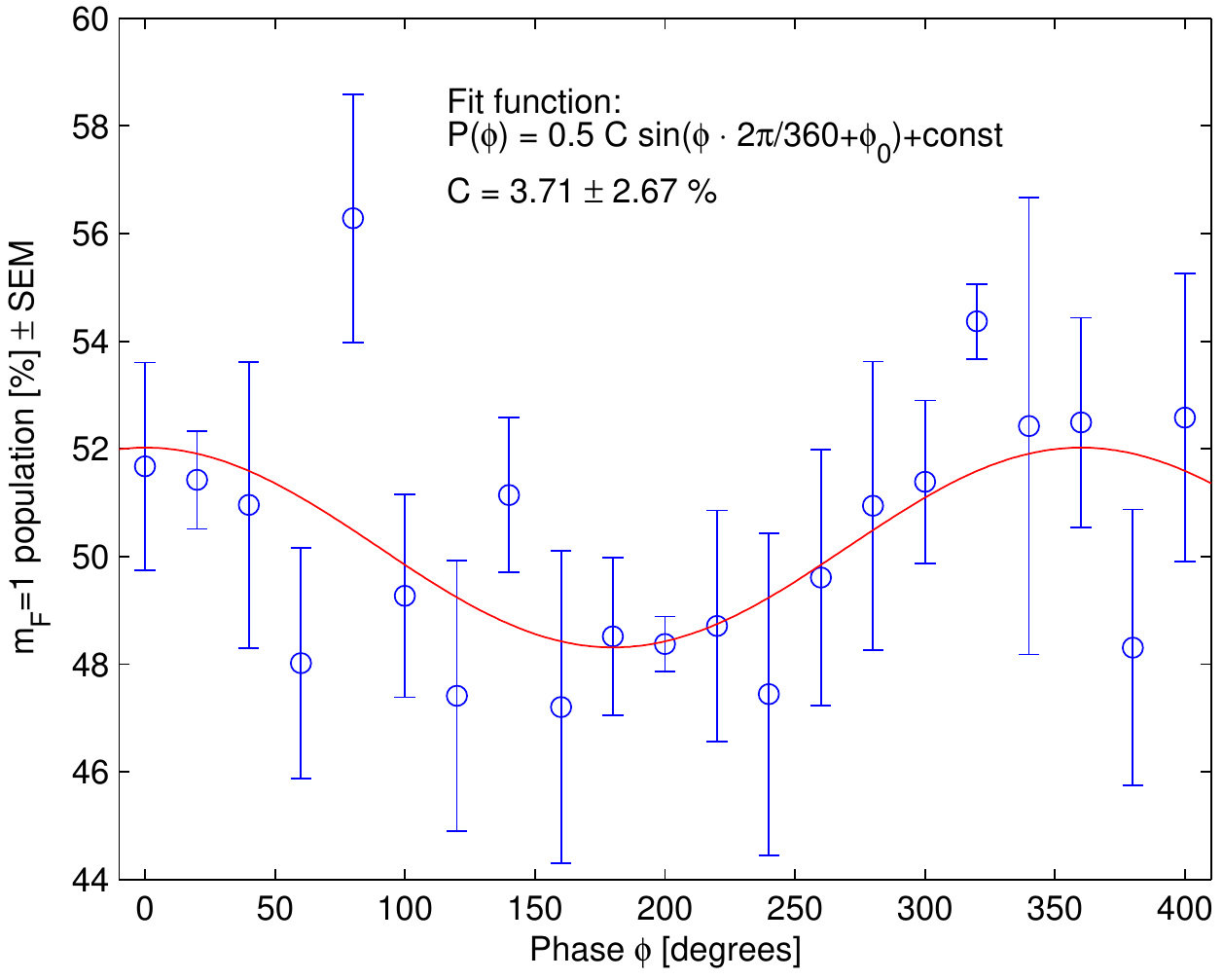}}
\caption{Fractional population in the $m_F=1$ state as a function of the applied phase shift $\phi$ between the Ramsey $\pi/2$ pulses, for data point 13 of the green data set shown in Fig.~\ref{fig:all points} (experimental parameters are listed in table~\ref{table:full loop parameters}). The contrast resulting from the fit is $C = 3.71 \pm 2.67\%$. As the contrast of the 'pure' Ramsey sequence (without magnetic gradients) is $71.20 \pm 6.82 \%$, the normalized contract is $5.21 \pm 3.78 \%$. The number of averaged single shots per point is 4.}
\label{fig:raw data point 13}
\end{figure}

Concerning data point 13 of the green data set ($\Delta_z^{\rm max}/ \sigma_{z,{\rm BEC}} \simeq 4$): although out of the trend of the data we believe it is a valid point. As our interferometer runs on 12 different parameters (4 gradient chip currents, 4 gradient durations, two delay times, and initial distance from chip, initial y position), it is not surprising that individual points appear above the trend due to slightly better optimization. The raw data (population oscillation) of this point are presented in Fig.~\ref{fig:raw data point 13}.

\begin{figure}
\centerline{
\includegraphics[width=\textwidth]{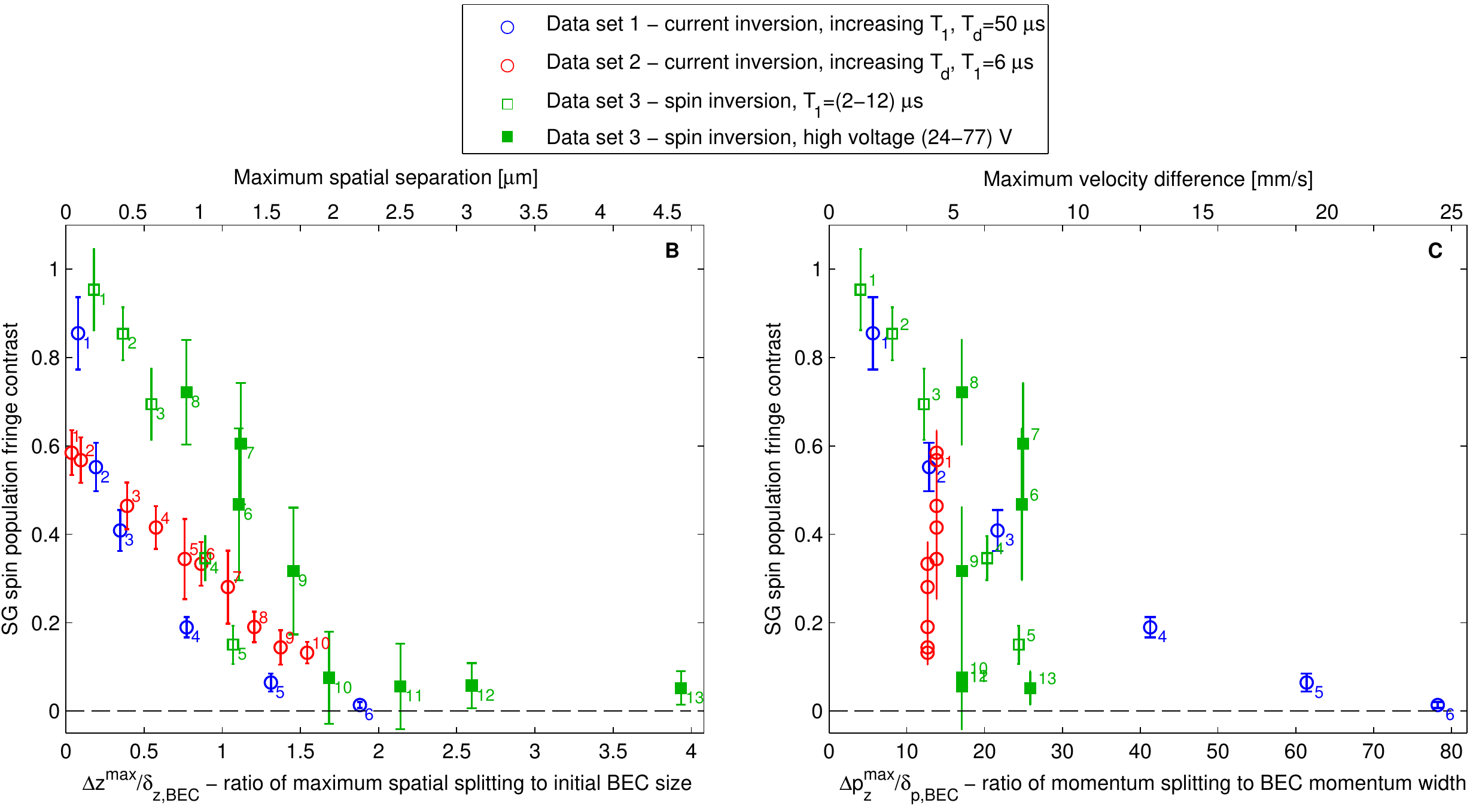}}
\caption{Figure 4a,b of the main text (analysis of stability and precision), including all data points appearing in table~\ref{table:full loop parameters}, some of which have been omitted for clarity from the main text version (as explained in the caption of table~\ref{table:full loop parameters}).}
\label{fig:all points}
\end{figure}

Figure~\ref{fig:all points} shows a full version of Fig.\,4, including all points omitted for clarity from Fig.\,4.

\subsection{Figure 5}

Figure 5 of the main text shows the results of the optimization procedure we use in order to maximize the interference contrast of the full-loop SGI. In the optimization procedure, we set the durations of the first and last gradient pulses $T_1$ and $T_4$ and also the durations of the delay times $T_{d1}$ and $T_{d2}$ (usually $T_1 = T_4$ and $T_{d1} = T_{d2}$ to begin with). We then measure the output population of the full-loop SGI sequence as a function of the second and third gradient pulses $T_2+T_3$ durations, while keeping the total duration $T_2+T_3+T_{d2}$ constant. A typical result is that shown in the inset of Fig.~5, fitted to a Gaussian envelope times a sine function. The Gaussian envelope corresponds to the timing at which the wave packets overlap at the end of the interferometer, where the peak of the envelope is the maximum of the overlap integral, roughly corresponding to zero momentum between the wave packets (although the spatial position also has some contribution). The sine function corresponds to the added phase between the two interferometer arms, per units time of $T_2+T_3$. Ideally for linear magnetic gradients, we would expect the peak overlap to occur when the sequence is symmetric i.e. $T_1 + T_4 = T_2+T_3$. However due to the asymmetry of the magnetic potential created by the chip wires in the z direction, the optimal point is below or above the symmetric time (the specific number depends on the scheme used - spin inversion or current inversion).

The numerical wave-packet propagation model gives similar results concerning the optimal time of $T_2+T_3$, and the population oscillation as a function of $T_2+T_3$. However, there is a discrepancy between the experiment and the simulation regarding the maximum achieved overlap integral and consequently the visibility, as can be seen in Fig.~4 of the main text.

\bigskip
\goodbreak

\section{ Theoretical models }
\label{sec:theory}
In this section we present the analytical and numerical models used for generating the theoretical values appearing in Figs.\,3,\,4.

\subsection{ A random vector model for an SGI} 

We start by utilizing the randomized Hamiltonian model, governed
by a disorder parameter $\epsilon$, to describe TI. We consider two extreme cases. The first
model assumes fluctuations with zero correlation length. This describes a situation where fluctuations
in the magnet are local and the atom is very close to the magnet so that there is no spatial
averaging. Such fluctuations may originate in Johnson noise or a local response – e.g. of
magnetic domains, impurities and geometrical defects – to global temperature or vibrational
fluctuations. The second model assumes the other extreme limit in which the fluctuations are
global and the correlation length is infinite. This limit is more relevant to our experimental
situation in which current fluctuations are the dominant source of randomness. In the future,
as current fluctuations go down to the shot-noise level and below so that Johnson noise becomes
dominant, or when permanent magnets are used, both limits will be important.

The atom-chip SGI presented here is, in a first approximation one-dimensional in the direction of
gravity, perpendicular to the chip plane \cite{machluf,margalit}. This allows for a simple theoretical
study with a tractable model, where we utilize a 1D Schr\"odinger equation for each wave packet
$\psi_j(z,t)$, corresponding to the two spin states. During the pulse duration $T$ the relevant equation is:

\begin{equation}
  i\hbar \frac{\partial \psi_j(z,t)}{\partial t}=-\frac{\hbar^2}{2m}
  \frac{\partial^2 \psi_j(z,t)}{\partial z^2}+V_{m_j}(z)\psi_j(z,t)+
  \epsilon  V_{\epsilon,j}(z)\psi_j(z,t),
  \label{schrodinger}
\end{equation}

\noindent where $V_{m_j}(z)$ is the magnetic potential energy of the wave packet
$j$ due the pulse from the chip wires.
The term $V_{\epsilon,j}$ (a vector with $3000$ elements, where each element corresponds to 5\,nm in real space) describes randomness in
the magnets that results from the effects of the environment, including also the operational limitations
\cite{manfredi}.
This may provide a reasonable model for both the half-loop and full-loop SGI.

However, in the main text we present this model only for the half-loop data as the model does not take into account the specifics of the experiment (in Fig.\,4 we show the theory lines provided by the numerical wave-packet propagation model which is described in the following and which we consider to be more accurate in simulating the specifics of the experiment). The specifics of the experiment which our random vector model does not take into account are the wire widths, the effect of the stopping pulse duration (the second pulse in the half-loop configuration), and the propagation time as well as the time-of-flight. Furthermore, it does not use the Gross-Pitaevskii equation but rather the schr\"odinger equation (this is not a bad approximation as the BEC is expanding in free-fall and the interactions are small). These simplifications were made in order to save run time. Most importantly, we believe the random vector model is less appropriate for the full-loop configuration, as it is not clear how to use a randomization parameter to describe the precision of the magnets, for which we have no good model.
Namely, even if we manage to get a good fit to the data, the interpretation of the model in terms of the actual physical processes taking place will be hard.

We will now concentrate on the model for the half-loop experiment. The average over different random-number seeds simulates the shot-to-shot temporal fluctuations (due to current fluctuations in our apparatus and temperature fluctuations or initial position fluctuations in the permanent magnets).

In our simulation $V_{\epsilon,j}$ is different for different $j=1,2$ as each spin state is in a different region of space and may encounter different randomization. The index $j$ also affects the calibration of the random term such that $V_{\epsilon,j}=V_{\epsilon}V_{m_j}$, where $V_{\epsilon}$ is a random diagonal matrix. We simulated two different limits which are the zero-correlation and the infinite-correlation models. In the zero-correlation model, the elements of $V_{\epsilon}$ are
randomly taken between $-1$ and $1$ with a width equal to $ 1/\sqrt{3}$. In the infinite-correlation model, all the components of $V_{\epsilon}$ are equal to a single random number also chosen between $-1$ and $+1$. In both cases, $V_{\epsilon}$ is reloaded from one shot to another thus mimicking the randomness of our pulses. The magnetic potential in our experiment is $V_{m_j}(z)\approx 10^{-27}$\,J. In our three-layer system (environment--magnet--probe), the signal is a measure of TI in the magnets and is of course also affected by the strength (and duration) of the coupling between the probe atom and the magnet.
Numerous works have directly utilized $V_{\epsilon}$ in relation to TI and the arrow of time (e.g. \cite{waldherr}). Each realization of $V_{\epsilon}$ (single-shot) yields perfect visibility. Like in the experiment, the eventual visibility reduction occurs from averaging over many realizations. Hence, the stochastic time evolution is replaced by an ensemble average. Finally, Eq. (\ref{schrodinger}) is an alternative to the usual density-matrix approach. The effective Schr\"odinger equation used here and the density-matrix approach are known to be equivalent \cite{gaspard2, biele,cucchietti}.

We calculate the interference function between the two wave packets $\psi_j(z,T)$, after mixing the internal spin states with a $\pi/2$ pulse, namely $n_{\epsilon}(z,T)=|\psi_1(z,T)+\psi_2(z,T)|^2$, where $\psi_j$ is the solution of Eq. (\ref{schrodinger}). Only the splitting pulse is accounted for, namely we do not examine the effect of the stopping pulse duration. Furthermore, we do not take into account propagation time or time-of-flight.
This is justified by the optimized recombination sequence mentioned above, which is assumed to represent an almost perfect 180$^\circ$ phase space rotation (Fig.\,\ref{fig:halfloop_Wigner}). This gives rise to a  measured interference pattern having approximately the same shape (in scaled coordinates) as the interference pattern formed right after the splitting, which is what is simulated in our model. The numerical solution of Eq. (\ref{schrodinger}) was obtained by applying
the Crank-Nicolson algorithm \cite{crank-nicolson}. Simulation parameters are chosen so as to match
the experiments. The visibility is calculated by averaging over the different interference patterns
obtained for many realizations of $V_{\epsilon}$.

The source of the magnetic potential $V_{m_j}(z)$ for the two wave packets during the gradient pulse is modeled by three parallel wires extending along $x$, infinitely long and having zero thickness. This gives rise to the analytic form
\begin{equation}
  V_{m_j}(z)=\frac{\alpha_j}{z}(1-\frac{2}{1+(\frac{z_{\rm w}}{z})^2}),
\end{equation}
\noindent where,
\begin{equation}
  \alpha=m_{F_j}g_{F_j}\frac{\mu_B\mu_0I}{2\pi},
\end{equation}
\noindent and where $m_{F_j}=1,2$ for $j=1,2$ are the projections of the hyperfine levels (i.e. Zeeman sub-levels),
$g_{F_j}=\frac{1}{2}$ is the Land\'e factor, the current magnitude is on the order of $I \approx 1A$, and the inter-wire
distance is $z_{\rm w}=100\,\mu$m.
The homogeneous bias field is assumed to be constant and we therefore ignore it.
The center of the quadrupole field (zero magnetic field) in this model is equal to
the inter-wire distance $z_{\rm w}=100\,\mu$m. Since the initial experimental position is about $5\,\mu m$ above the zero of the quadrupole, in this simulation, we choose $z_{\rm quad}=100\,\mu$m and
$z=95\,\mu$m, respectively.

\begin{figure}
\begin{center}
\includegraphics[width=\textwidth, height=5.cm]{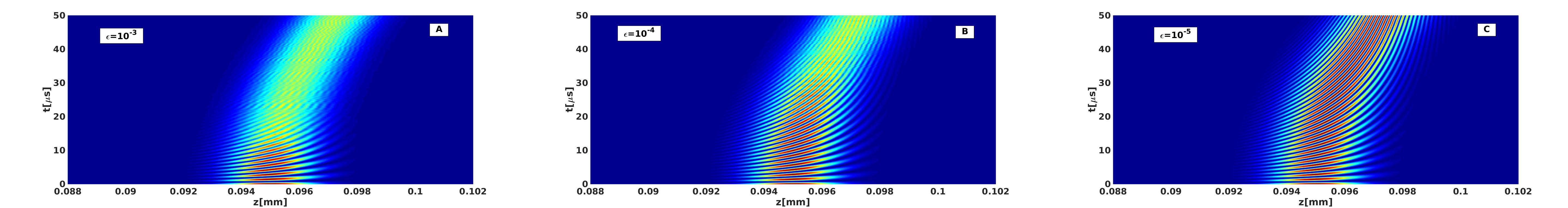}
\end{center}
\caption{Time evolution of the interference pattern $n_{\epsilon}(z,t)$ for
  $\epsilon=10^{-3}$ (A), $\epsilon= 10^{-4}$ (B), and $\epsilon=10^{-5}$
  (C). The center of the cloud starts at $z_0=95\mu$m, $5\mu$m from the simulated quadrupole center. At short times ($<4\mu$s) the total number of atoms oscillates between the two states $|\pm\rangle\equiv 2^{-1/2}(|m_F=2\rangle\pm |m_F=1\rangle)$ and at later times the atomic population of each of these states (specifically $n_{\epsilon}(z)$ of the sate $|+\rangle$ in the figure) creates an interference pattern representing a superposition of two momentum states $e^{ ik_jz}$ ($j=1,2$), where the momentum difference $|k_1-k_2|$ (number of fringes per unit length) grows linearly with time. As $\epsilon$ becomes larger the visibility of these fringe patterns decreases faster in time. As shown in the main text and Sec.~S2 above, the shape of the fringe patterns shown here just after a splitting pulse of a variable duration represents the shape of the corresponding fringe patterns observed after the complete interferometric sequence, given that the dominant source of TI acts during the splitting pulse.  Hence the visibility of the simulated patterns shown here represents the visibility expected in the experiment.
}
\label{fringes_simzc09}
\end{figure}

The visibility is then obtained analogously to the experiment: many single-shot interference patterns (in the simulation, each with its own random vector, and each giving 100\% visibility) are averaged to give a multi-shot pattern $\langle n_{\epsilon}(z,T_1)\rangle$ with reduced visibility due to shifts of the individual patterns.
In Fig.\,\ref{fringes_simzc09}, we show $\langle n_{\epsilon}(z,T_1)\rangle=\langle |\psi_1(z,T)+\psi_2(z,T)|^2\rangle$ for different values of $\epsilon$.
The averaging $\langle \dots\rangle$ is over samples of up to $N_c=1000$
configurations of Gaussian
disorder. The influence of the magnitude
of $\epsilon$ on the interference pattern is evident.
For $\epsilon=10^{-6}$, we did not observe any disappearance of the fringes for
the longest pulse simulated, $T_1=100\,\mu$s. Hence for this $\epsilon$ the system is nearly reversible.
The simulated visibility is obtained from Fig.\,\ref{fringes_simzc09} by a simple fit of the form
\be n_{\epsilon}(z,t)=A\exp[-\frac{(z-z_0)^2}{2\sigma^2}][1+ V_{\epsilon}(t)\cos[k(z-z_0)+\beta(z-z_0)^2+\phi)], \ee
where $k=2\pi/\lambda$ represents the fringe periodicity $\lambda$ and the quadratic term $\beta(z-z_0)^2$ accounts for spatial frequency chirp within the pattern.

\begin{figure}[ht]
\begin{center}
\includegraphics[width=14.cm, height=10.cm]{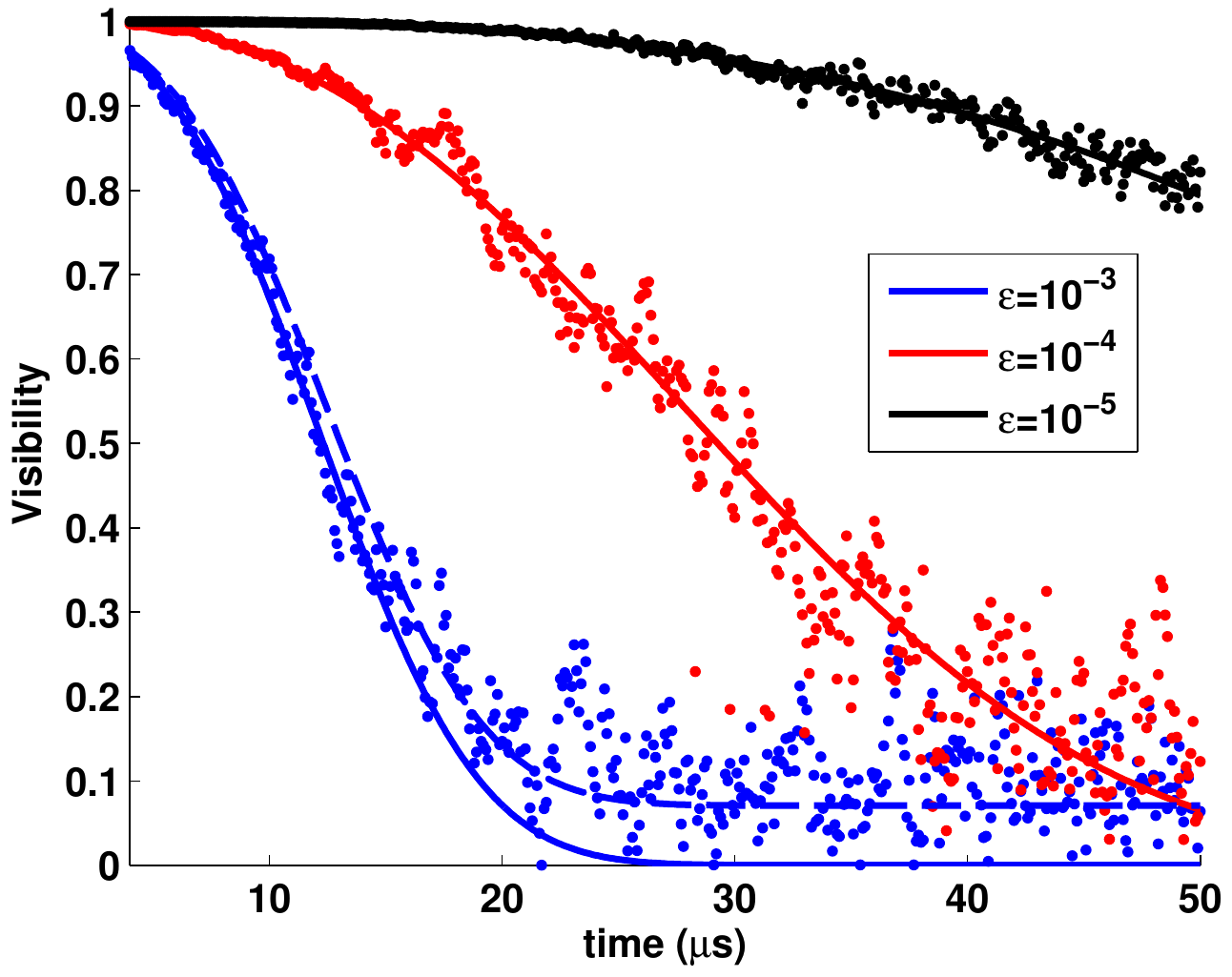}
\end{center}
\caption{The visibility $ V_{\epsilon}(t)$ as a function of $t$ for
different values of $\epsilon$. The solid lines represent a fit to the function $ V_{\epsilon}(t)=\exp[-\sum_{j=1}^3 A_j(\epsilon) t^j]$ (without taking into account the range of low and noisy visibility where $ V_{\epsilon}(t)<0.2$). The
fitting parameters $A_j(\epsilon)$ ($1<j<3$) were used for comparing the random vector model to the experimental results in Fig.\,3 of the main text.
}
\label{fig:vis_theo}
\end{figure}

In Fig.\,\ref{fig:vis_theo}, we show $ V_{\epsilon}(t)$ as
a function of the splitting pulse duration for
$\epsilon=10^{-3}$, $\epsilon=10^{-4}$, and $\epsilon=10^{-5}$.
The finite sample size at each time implies a standard error of the order of $\delta  V_{\epsilon}\sim (1- V_{\epsilon}^2)/\sqrt{2N}$ [see Eq.~(\ref{eq:dVN})]. In order to compare the experimental data (Fig.~3 of the main text) to the theoretical model while eliminating the statistical errors, we fit the visibility $ V_{\epsilon}$ resulting from the simulation to the simple form $ V_{\epsilon}(t)=\exp[-\sum_{j=1}^3 A_j(\epsilon)t^j]$, shown by the solid curves in Fig.\,\ref{fig:vis_theo}.
At low visibilities the numerical points are usually higher than the fit as the statistical noise forces the absolute value of the visibility to saturate at $\langle  V_{\epsilon}\rangle\sim 1/\sqrt{2N}\sim 0.07$. The expected uncertainty of the finite sample visibility is represented by the dashed curve, one standard deviation above the visibility for $\epsilon=10^{-3}$.

\subsection{ Analytical model for half-loop visibility}

We now present a simple analytical model for multi-shot visibility in a half-loop interferometer (namely, spatial interference fringes) with instability. The model was used for the low visibility data in Fig.\,3 of the main text (with no free parameters!). The model assumes fluctuations with an infinitely long correlation length, namely, noise in global parameters of the system.
We assume that the final interference fringes observed in the experiment represent the interference of two similar wave packets $\psi_j$ ($j=1,2$) with fluctuations $\delta\phi$ of the central phase difference, momentum difference uncertainty $\hbar~\delta k$ and center distance fluctuations $\delta z$. If we neglect fluctuations of the delay time between the two pulses and fluctuations in the second (stopping) pulse and assume that the second pulse is optimized to completely stop the relative motion between the two wave packets, then the analysis following the phase space description in section~\ref{sec:phasespace} implies that the fluctuations  in the final fringe correspond exactly to fluctuations in the phase, momentum, and position of the wave packets just after the splitting pulse.  The phase fluctuations $\delta\phi$ are due to the uncertainty in the absolute magnetic field at the wave-packets center during the splitting pulse, and the momentum fluctuations $\hbar\delta k$ are due to fluctuations of the gradient strength during splitting, while position fluctuations $\delta z$ are mainly due to fluctuations in the initial wave packet position. Position fluctuations due to splitting momentum fluctuations are  negligible for a short splitting pulse (as in the experiment) and an optimized stopping pulse so we neglect position fluctuations. Assuming a Gaussian wave packet shape $|\psi_0(z)|\propto \exp[-z^2/2\sigma_z^2]$ and Gaussian distribution of the random variables, we obtain a visibility (see derivation below in section~\ref{sec:Vmulti}, Eq.~(\ref{eq:V_Ntoinf}))
\be V=\frac{\exp\left[-\frac{1}{2}\frac{\langle\delta\phi^2\rangle}{1+\sigma_z^2\langle \delta k^2\rangle}\right]} {\sqrt{1+\sigma_z^2\langle\delta k^2\rangle}}.
\label{eq:FourierVis} \ee
Note that this expression corresponds to a definition of the visibility as the relative amplitude of the Fourier component of the fringe pattern that oscillates in space with a certain periodicity due to the momentum difference between  $\psi_1$ and $\psi_2$. Fluctuations of the momentum between the two wave packets correspond to different periodicities of single fringe patterns such that the oscillation amplitude (local contrast) of the multi-shot fringe pattern varies along the pattern. The expression in Eq.~(\ref{eq:FourierVis}) is based on a Fourier transform of the pattern, taking into account this variation of the oscillation amplitude. In the case of our experiment we believe that the central phase and the momentum fluctuations are correlated since they are both due to current fluctuations in the wire. The ratio between phase fluctuations and momentum fluctuations is $\delta \phi/\delta k=z_0$, where $z_0$ is  the distance of the wave packet center during the splitting from the center of the quadrupole, where the magnetic field of the pulse is zero.
If $\delta k$ and $\delta \phi$ are not correlated then the denominator inside the exponent becomes $1$ and the expression for the visibility can be factored into an exponential term  $\exp(-\langle \delta\phi^2\rangle/2)$ due to phase fluctuations and a term $(1+\sigma_z^2\langle\delta k^2\rangle)^{-1/2}$ due to momentum fluctuations. Note also that if the visibility is defined as the contrast of the fringes at the center of the fringe pattern such that the variation of the contrast along the pattern is neglected, then the visibility is restricted to the exponential term which is generated by the phase fluctuations at the center.

The expression for the correlated  phase and momentum fluctuations in Eq.~(\ref{eq:FourierVis}) implies that when $\sigma_z^2\delta k^2\ll 1$ the visibility is determined by $\delta\phi^2$. However, when $\sigma_z^2\delta k^2>~ 1$
the visibility drops like $\sim 1/\sqrt{1+\delta k^2\sigma_z^2}$.

For generating the analytic curves in Fig.\,3 of the main text we express the visibility in terms of the relative current fluctuations $\eta\equiv \delta I/I$ (or, equivalently, the relative timing fluctuations $\eta\equiv \delta T/T$). The relative final momentum fluctuations $\delta k/k$ are equal to $\eta$, so that $\delta k=\eta \kappa T$, where $\kappa=\partial k/\partial T$ is the final momentum per kick time in for the parameters used in our half-loop SGI (current and distance from the chip).  For a given distance $z_0$ from the center of the quadrupole, since the phase difference due to the splitting pulse is proportional to the momentum kick $\phi(z_0)=z_0\delta k$, we get  $\delta\phi=\kappa z_0\eta T$. It then follows that the visibility is $V(T)=\exp\left[-\frac12\frac{\eta^2\kappa^2z_0^2 T^2}{1+\sigma_z^2\kappa^2\eta^2 T^2}\right]/\sqrt{1+\sigma_z^2 \kappa^2\eta^2 T^2}$.
One can identify two extreme cases: in the first the atoms are close to the center of the quadrupole field responsible for the splitting, and consequently do not suffer from large phase fluctuations as the magnitude of the magnetic field is small. They do however suffer from momentum fluctuations $\delta k$. In the second, the atoms are far from the center of the quadrupole and consequently the phase noise is the main source of randomness.

\subsection{Numerical model for BEC wave-packet propagation}

Here we describe the numerical model that was used for calculating the theoretical values for the visibility (numerical wave packet propagation) in Figs.\,3,\,4 of the main text and some other results given in this Supplementary Material.
These calculations use a wave-packet propagation model in which the center-of-mass coordinates of the two wave packets are calculated by solving the Newton's equations of motion under the influence of gravitational acceleration and the magnetic force (magnetic potential gradient) ${\bf F}_{m_F}({\bf r},t)=-m_Fg_F\mu_B\nabla|{\bf B}({\bf r},t)|$, where $\mu_B$ is the Bohr magneton and $g_F$ is the Land\'e factor. The magnetic field ${\bf B}({\bf r},t)$ is obtained by using the Biot-Savart law from a simulation of the currents in the chip wires and the external bias field. In addition, the size of the wave packets and their quadratic phase are calculated by using a Thomas-Fermi dynamical model for a wave packet evolution in a time-dependent harmonic potential~\cite{castin-dum}. We take the instantaneous harmonic frequency at the location of the center of each wave packet to be given by $\omega_j^2(t)=\frac{1}{m}\partial F_j /\partial x_j$ for the three cartesian coordinates $x_j=(x,y,z)$. This provides an estimation of the expansion or focusing of the BEC wave packets. The center-of-mass phase of each wave packet is given by the action
\be \phi(t)=S(t)/\hbar=\frac{1}{\hbar}\int_0^t dt'[{\bf P}(t)^2/2m-V_m({\bf r}(t),t)]. \ee
This phase together with the linear and quadratic phase along the wave packets determines the relative phase of the spatial fringe patterns in the half-loop experiment. For simulating the multi-shot visibility due to instability of parameters such as the current in the wires (proportional to the magnetic field during the gradient pulse) and instability of the duration of the pulses, we take a Gaussian distribution of such shot-to-shot fluctuations and sum up many fringe patterns calculated from evolution under the influence of the fluctuating parameters and obtain the multi-shot visibility for the given distribution.

The visibility in the full-loop configuration (Fig.\,4) is calculated directly by taking the overlap integral between the two approximate forms of the wave packets after evolution under the influence of the gradient pulses with the given parameters. Each of the wave packet has a scaled Thomas-Fermi shape.

Note that the wave-packet propagation model allows a fast calculation of the dynamics in the three-dimensional space. It is then possible to investigate the expected effects of distortions or tilts of the system beyond the one-dimensional approximation of propagation along the $z$ axis. This allowed us to check hypotheses about the source of the reduced visibility in the full-loop SGI experiment (see section~\ref{sec:instability}).

\begin{figure}
  \includegraphics[width=0.8\textwidth]{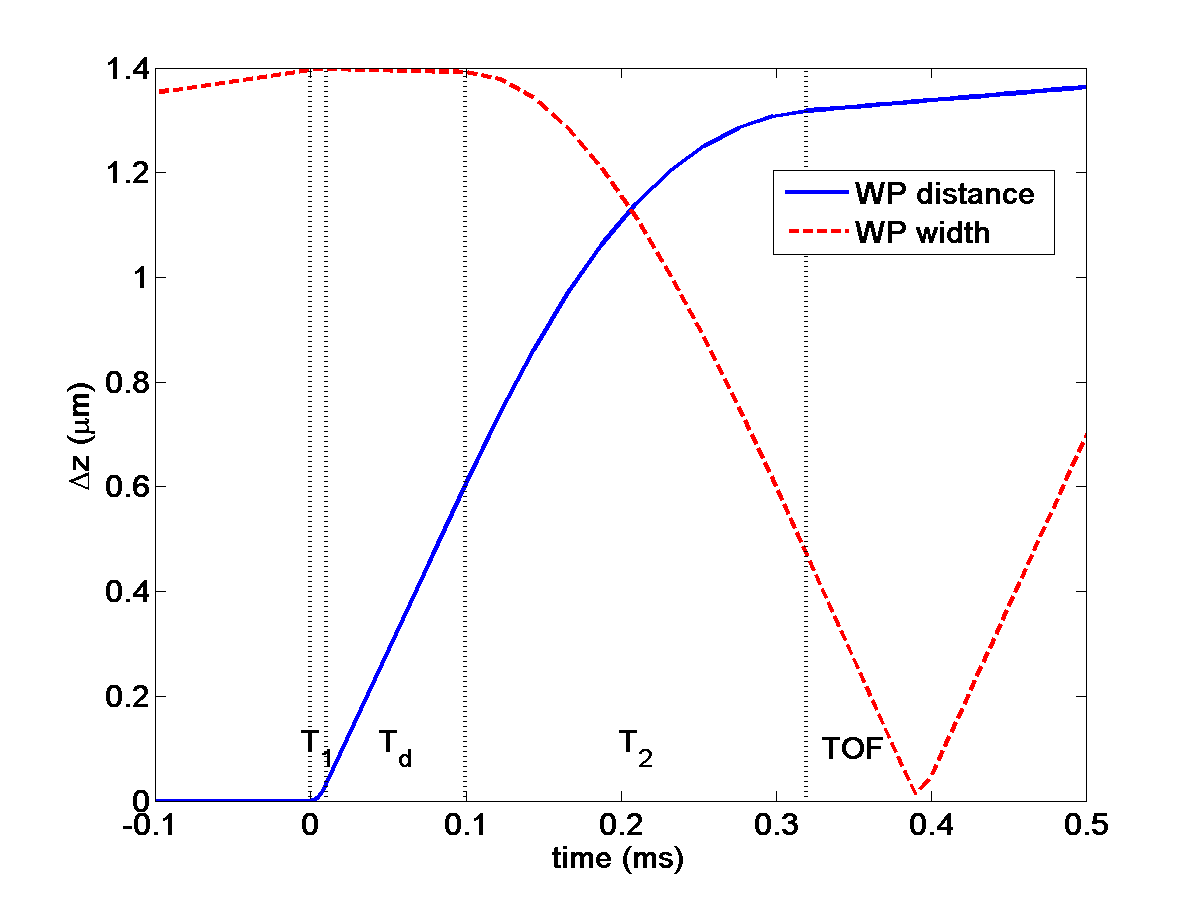}
\caption{wave packets (WP) relative distance (solid line) and size (standard deviation $\sigma_z=\sqrt{ \int dz\, (z- \langle z \rangle)^2|\psi(z,t)|^2 }$, dashed line) along the $z$ direction as a function of time for the data point at $T_1=10\,\mu$s of the high visibility data of Fig.\,3 in the main text (see table~\ref{table:parameters} for parameters). WP distances and widths along the $x$ and $y$ directions are also calculated but not shown.  The size near the focus point ($t\approx 0.39\,$ms) is much smaller than expected because the calculation based on Ref.~\cite{castin-dum} neglects the kinetic term in the Gross-Pitaevskii equation, which is responsible for the minimum WP size.}
\label{fig:WPcalc}
\end{figure}

Fig.~\ref{fig:WPcalc} shows the results of the calculation of the distance between the wave packets and their width (standard deviation) as a function of time for the experimental data point with splitting time $T_1=10\,\mu$s in Fig.\,3 (half-loop SGI). The wave packets (having approximately the same widths) start to expand at the initial time (0.92 ms) after trap release and before the first magnetic gradient pulse. Then the magnetic pulses create a curved potential which causes the focusing of the wave packets, leading to a minimal wave-packet size at a time $T_d$ after the end of the stopping pulse (of duration $T_2$) (see also analytical expressions for the dynamics in section~\ref{sec:phasespace}). The simulation shown in the figure is based on a Thomas-Fermi approximation~\cite{castin-dum}  and neglects the kinetic term whose dynamics are important when the wave packet size is minimal, so the wave packets size near the maximally focused point is imperfectly reproduced by this calculation. See Eq.~~(\ref{eq:sigma_min}) below for an analytic approximation of the minimal size. The wave packet sizes and location along the $x$ and $y$ directions are also calculated but not shown in the figure. The sizes along the $x$ and $y$ directions are much larger than the size along the $z$ direction since the focusing effect occurs only along the $z$ direction.

Note that the above dynamic Thomas-Fermi based approximation does not take into account the repulsive interaction between wave packets that start to separate but still overlap, i.e., it includes only the effect of an atomic distribution having a Thomas-Fermi shape.
In order to verify that this does not reproduce qualitatively different results compared to a full Gross-Pitaevskii calculation taking into account all the repulsive interactions within and between wave packets, we have performed a few calculations with a full solution of the Gross-Pitaevskii equations in either one- or three-dimensions. The results are very similar to those of the approximated method so we are confident that for non-extreme cases as in this work we can use our approximation, which provides a much faster method that can allow calculations of stability and precision in a reasonably short time.

\section{ Phase space description of the Stern-Gerlach interferometer}
\label{sec:phasespace}
\newcommand{\vect}[1]{\left(\begin{array}{c} #1 \end{array}\right)}

\begin{figure}
\centerline{
\includegraphics*[angle=0, width=16.cm,height=8.cm]{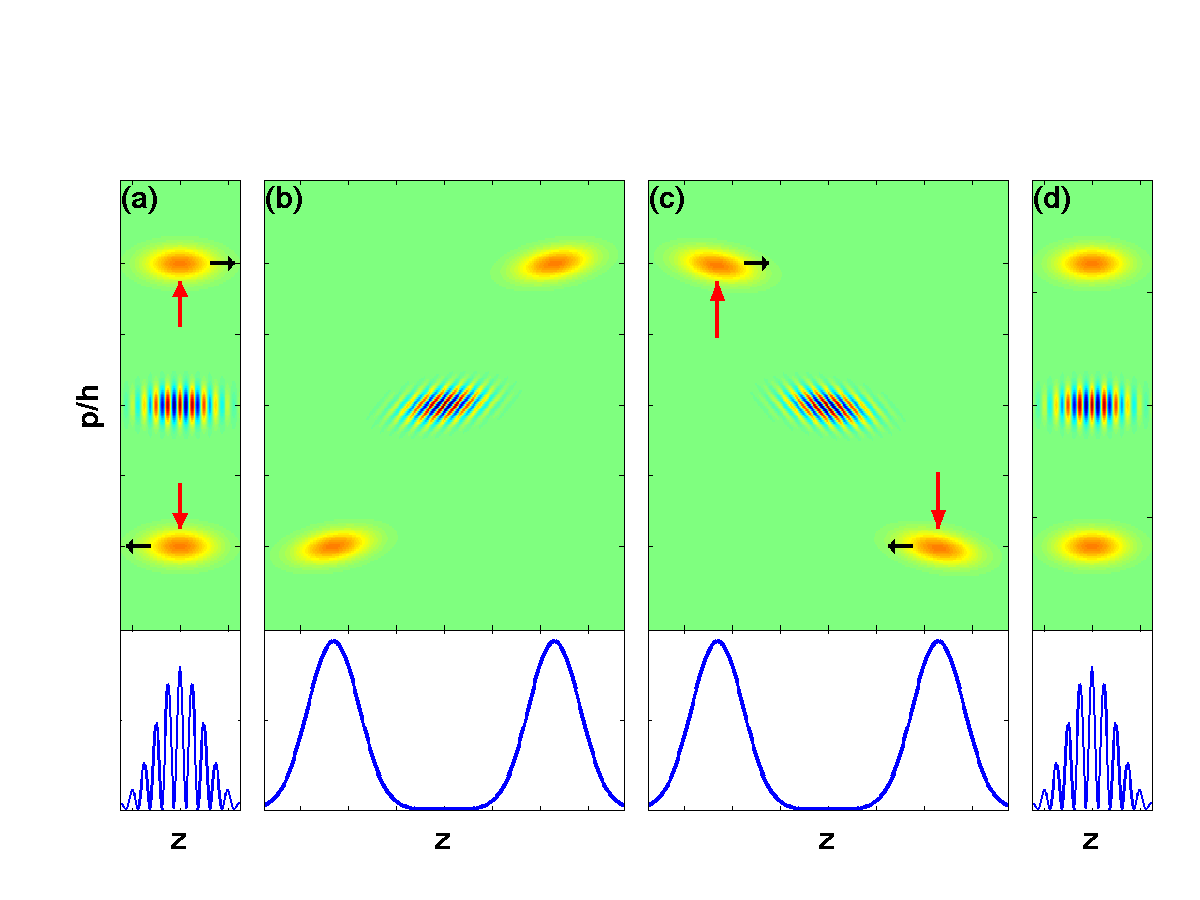}}
\caption{Full-loop SGI scheme in phase space (scalar Wigner function representation obtained by projection into a spin eigenstate of $\hat{S}_x$, red -- positive and blue -- negative). The initial single wave packet (centered at $p=0$ and $z=0$, not shown) is split into two momentum components (a), giving rise to a spatial interference fringe pattern (upon projection onto the position axis, shown at the bottom). Such spatial interference is made possible in reality if the two wave packets are manipulated to have the same spin (effectively undoing the entanglement between path and spin), as done in our half-loop experiment just after splitting. Red arrows correspond to actions driven by the magnetic field giving rise to the observed wave packet position, whereas black arrows correspond to evolution due to free propagation which follows the present wave packet position in the plot and giving rise to the next plot. (b) After some free propagation, the wave packets separate. (c) The momentum of each wave packet is inverted (effectively as done by the inverted current pulses in our full-loop experiment) and the two wave packets propagate back to the original position. (d) After some propagation the wave packets again overlap in space. If the two momentum states are in the same internal atomic state, spatial fringes are again visible.  If the two wave packets have a different internal state and a stopping pulse overlaps them also in momentum space (as in our full-loop SGI experiment) then the final internal state depends on the phase accumulated during the process and the fringes are in spin space. The phase space fringes in (d) are due to a 180$^\circ$ rotation of the fringes in (a).
\label{fig:fullloop_Wigner}}
\end{figure}

\begin{figure}
\centerline{
\includegraphics*[angle=0, width=16.cm,height=8.cm]{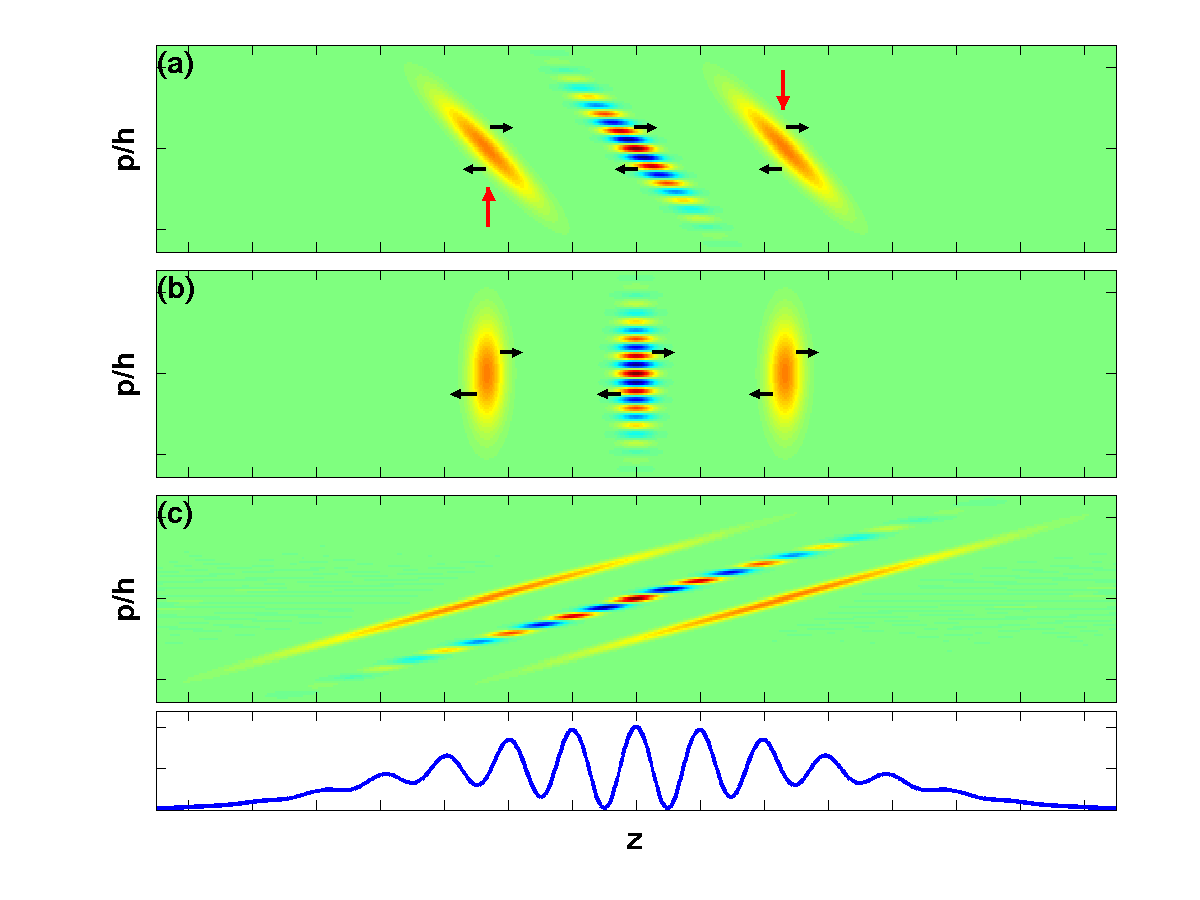}}
\caption{Half-loop SGI phase space dynamics (arrows and color code as in Fig.~\ref{fig:fullloop_Wigner}). We start with the state presented in Fig.~\protect\ref{fig:fullloop_Wigner}(b) (with the two wave packets having the same spin) but, different from Fig.~\protect\ref{fig:fullloop_Wigner}(c), the momentum is not inverted but rather
brought to zero. (a) The two wave packets after they have been brought to a halt by a stopping pulse. The change of direction of the long axis of the wave packets is due to the curvature of the magnetic field. (b) after free evolution for a time equal to the initial propagation time [between Fig.~\protect\ref{fig:fullloop_Wigner}(a) and Fig.~\protect\ref{fig:fullloop_Wigner}(b)], the wave packets are now focused so that they have a minimal
width in real space. (c) further free propagation (TOF) now rotates the phase space image so that spatial fringes become visible (upon projection onto the spatial axis, shown at the bottom). This is our signal.}
\label{fig:halfloop_Wigner}
\end{figure}

In order to gain a qualitative and quantitative understanding of the interferometric sequence in our experiment (specifically the half-loop SGI), we use the position-momentum phase space description in the Wigner representation, which is completely equivalent to the Schr\"odinger picture.
This description allows us to derive some quantitative estimations of the wave-packet width and separation during certain parts of the sequence, as we show below.
Our phase space description of the atomic dynamics is motivated by our previous work~\cite{Shuyu_phasespace}. See also a recent theoretical treatment of the Stern-Gerlach experiment with phase space methods in Ref.~\cite{SGWigner}.

The phase space dynamics in the full-loop SGI and the half-loop SGI is demonstrated in Figs.~\ref{fig:fullloop_Wigner} and~\ref{fig:halfloop_Wigner}, respectively.
For the full-loop configuration the scalar Wigner function representation is allowed by projecting the wave function $\psi_+|+\rangle+\psi_-|-\rangle$, which is an entangled spin-path, into an eigenstate of the spin operator $\hat{S}_x$, giving rise to an equal superposition of $\psi_+$ and $\psi_-$ corresponding to the two spin states.
In the half-loop configuration the wave-function is manipulated to be a superposition of two wave packets with the same spin just after the splitting stage, such that the scalar Wigner function representation is appropriate without further projection. In what follows we provide a mathematical  description of the dynamics in the half-loop configuration and provide some mathematical expressions for the expected wave packet positions and sizes.

We start by defining a phase space rotation represented by the rotation matrix
\be {\cal R}(\omega,t)=\left(\begin{array}{cc} \cos\omega t & \frac{\sin\omega t}{m\omega} \\ -m\omega\sin\omega t & \cos\omega t\end{array}\right). \ee
This rotation describes the time evolution of the phase space variables $\vect{z \\ p}$ in the presence of a harmonic potential with angular frequency $\omega$. In the limit $\omega\to 0$ this rotation describes free space propagation
\be {\cal R}(0,t)\equiv \lim_{\omega\to 0}{\cal R}(\omega,t)=
\left(\begin{array}{cc} 1 & t/m \\ 0 & 1\end{array}\right). \ee

The interferometric scheme starts with two wave packets having the same spatial shape but two different spin states, which we assume here to be opposite spins for simplicity. A splitting pulse of duration $T_1$ splits the two wave packets into two momenta $\pm \hbar k/2$. After this momentum kick the two wave-packet centers move to a distance $\pm \hbar k T_1/4m$ from the initial position. We set the time to be $t=0$ at the middle of the pulse, such that if we reverse the velocity of the wave packets at the end of the pulse and project their position back by free propagation into $t=0$ they would be at the initial point $Z=0$ and having momentum $P=\pm \hbar k/2$. This will be the starting point of our scheme.

After a time $T_d$ of free propagation we apply a stopping pulse of duration $T_2$. In the full-loop SGI this stopping pulse may be given by a homogeneous gradient in a direction opposite to the initial splitting pulse, which will bring back the two wave packet into a relative stop. However, in our scheme we continue the interferometric sequence with the two wave packets in the same spin state and apply a harmonic potential of frequency $\omega$ and duration $T_2$ to stop the relative motion [Fig.\,2A of the main text].
As we see below, this pulse also serves to focus the wave packets into a minimal wave packet size.
After this pulse the center coordinates become
\begin{eqnarray} \vect{ Z(T_d+T_2) \\ P(T_d+T_2)} &=& {\cal R}(\omega,T_2){\cal R}(0,T_d)\vect{0 \\ \pm\hbar k/2} \nonumber \\
&=&  \pm\frac{\hbar k}{2}\vect{\frac{T_d}{m}\cos\omega T_2+\frac{1}{m\omega}\sin\omega T_2 \\  -\omega T_d\sin\omega T_2+\cos\omega T_2}.
\end{eqnarray}
The stopping pulse of length $T_2$ is designed such that $P(T_2)=0$. In this case
\be \sin\omega T_2=\frac{1}{\sqrt{1+\omega^2T_d^2}},\qquad \cos\omega T_2=\frac{\omega T_d}{\sqrt{1+\omega^2T_d^2}}.
\label{eq:T2} \ee

Note that the condition for stopping $\omega T_d\tan \omega T_2=1$ does not depend on the initial momentum kick so that momentum fluctuations are not expected to affect the velocity of the two wave packets after stopping and hence the final position of the interference fringes in our experiment.
As we show below fluctuations in the initial momentum kick are expected to affect the distance between the wave packets after stopping and hence the periodicity of the interference fringes measured in the experiment.

We now look at the state of the atoms after another free-propagation time $T_d$ [Fig.\,2B of the main text].
The combination of the three operations: propagation for a time $T_d$, harmonic stopping pulse for a time $T_2$ satisfying Eq.~(\ref{eq:T2}) and another propagation time $T_d$ transforms a general phase space coordinate as
\be {\cal R}(0,T_d){\cal R}(\omega,T_2){\cal R}(0,T_d)\vect{z \\ p}=
\vect{\frac{\xi}{m\omega}p \\ -\frac{m\omega}{\xi}z}. \label{eq:RRR} \ee
This sequence leads to a complete rotation of phase space by $90^\circ$, while scaling the phase space coordinates by the squeezing factor
\begin{equation} \xi=\sqrt{1+\omega^2T_d^2}.
\label{eq:xi} \end{equation}
At this time the spatial distribution consists of two wave packets at $z=\pm d/2$, where
\begin{equation} \label{eq:d}
d= \frac{\xi}{m\omega} \hbar k,  \end{equation}
and where in the limit of fast stopping relative to the propagation time $\omega T_d\gg 1$ the distance is $d\to \hbar kT_d/m$.
If the initial wave packet at $t=0$ is a minimal uncertainty state where the initial position and momentum uncertainties satisfy $\sigma_{z,0}\sigma_{p,0}=\hbar$, then the distribution corresponding to each of the two wave packets at time $T=2T_d+T_2$ has a minimal spatial width
\begin{equation} \sigma_{\rm min}=\frac{\xi}{m\omega}\sigma_{p,0}=\frac{\hbar\xi}{m\omega\sigma_{z,0}}. \label{eq:sigma_min} \end{equation}

In the limit $\omega T_d\gg 1$ this becomes $\sigma_{\rm min}=\hbar T_d/m\sigma_{z,0}$.
Note that if the initial momentum kick is large enough to fully separate the two wave packets, namely, $\hbar k\gg \sigma_{p,0}$, then after the full stopping at time $t=2T_d+T_2$ the distance between the two wave packets is much larger than their size $d/\sigma_{\rm min}=\hbar k/\sigma_{p,0}\gg 1$.

After stopping the relative motion between the two wave packets we allow them to expand in free space for a long time $t$ until they overlap and form a spatial interference pattern at a large scale [Fig.~\ref{fig:halfloop_Wigner} (c))], which is equivalent to applying ${\cal R}(0,t)$ for $t\gg m \sigma_{\rm min}^2/\hbar$, we obtain a spatial fringe pattern with fringe periodicity
\be \lambda=\frac{2\pi\hbar t}{md} \label{eq:lambda} \ee
and an overall Gaussian envelope width of
\be \sigma_{z,f}=\hbar t/m\sigma_ {\rm min}=\sigma_{z,0} t\omega/\xi.
\label{eq:sigmaf} \ee
The number of observed fringes is then given by $n_{\rm fringes}\approx 2\sigma_{z,f}/\lambda=k\sigma_{z,0}/\pi$, namely, the same number of fringes of the initial microscopic fringe pattern formed just after the momentum kick of the beam splitter.
\bigskip
\goodbreak

\section{ A Generalized Humpty-Dumpty (HD) theory for an SGI}

In this section we derive the expected visibility (spin-coherence)  of a full-loop SGI of the type implemented in this work, using a generalized version of a theory by Englert, Scully, and Schwinger(ESS)\cite{ESS_1}. We first derive a general equation for a general input state. Then we obtain a modified expression for the visibility in a full-loop SGI where the final state is not designed to be exactly similar to the initial state.

\subsection{ Interference visibility for a general initial state}
Here we introduce the theoretical basis for the analysis of interference visibility in the context of our SGI.
We consider an SGI whose input is a cloud of two-level atoms in a given eigenstate of the spin. The initial spatial state is characterized by a single-particle density matrix of the spatial degrees of freedom that can be described as a statistical mixture of wave-function out of an orthogonal set,
\be \rho({\bf r},{\bf r}')=\sum_j P_j \psi_j ({\bf r})\psi_j^*({\bf r}'), \ee
where $\psi_j({\bf r})$ are spatial wave functions of the atom and $P_j$ are the probabilities for an atom to occupy these wave functions.

The interferometric sequence consists of a $\pi/2$ pulse creating a superposition of the two spin states $\frac{1}{\sqrt{2}}(|+\rangle+|-\rangle)$, a magnetic field gradient for splitting, additional gradients for reversing the motion and for stopping and a final $\pi/2$ pulse for mixing the spin population. These operations are represented by the following unitary operator,
\be \hat{U}(t_f,0)=e^{-i\hat{S}_y\pi/2\hbar}(U_+|+\rangle\langle +|+U_-|-\rangle\langle-|)e^{-i\hat{S}_y\pi/2\hbar}, \ee
where the first and the last term of the r.h.s. are the initial and the final $\pi/2$
pulses ($\hat{S}_y=-\frac{i\hbar}{2}(|+\rangle\langle -|-|-\rangle\langle +|)$) and $U_{\pm}$ represent the spatial evolution of the two spin states (with and without the gradient fields). The population difference at the output of the interferometer is then measured. This is represented by the expectation value of the operator $\hat{S}_z=\frac{\hbar}{2}(|+\rangle\langle +|-|-\rangle\langle -|)$,
\begin{eqnarray}
 \langle \hat{S}_z\rangle_f &=& \langle \hat{U}^{\dag}(t_f,0)\hat{S}_z \hat{U}(t_f,0)\rangle_0 \nonumber \\
&=&
\frac{1}{2}\langle [(U_+^{\dag}\langle +|\hat{S}_x |-\rangle U_- + U_-^{\dag}\langle -|\hat{S}_x|+\rangle U_+]\rangle_0=
\frac{\hbar}{4}\langle U_+^{\dag}U_-\rangle_0+{\rm c.c.} ,
\end{eqnarray}
where we have used $e^{i\hat{S}_y\pi/2\hbar}\hat{S}_z e^{-i\hat{S}_y\pi/2\hbar}=\hat{S}_x$ and $\langle \pm|e^{-i\hat{S}_y\pi/2\hbar}|\pm\rangle=\langle \pm|e^{-i\hat{S}_y\pi/2\hbar}|\mp\rangle=1/\sqrt{2}$, assuming that the initial state is a spin $|+\rangle$.

The visibility of the interferometric signal is given by the maximal population difference when the phase between the two arms is set to zero. At this point we expect that the visibility would be unity if the operations on the two spin states during the sequence lead to exactly the same final state at $t_f$, namely $U_+=U_-$. However, due to inaccuracies the operations are not the same and may lead to different results for different initial states. In general, the visibility is given by an integral over the density matrix after the evolution,
\be V=\frac{2}{\hbar}\langle \hat{S}_x\rangle_t^{\rm max}
=\left|{\rm Tr}\{\rho U_+^{\dag} U_-\}\right|=\left|\int d^3r\, \rho_{-+}({\bf r},{\bf r},t)\right|, \ee
where
\be \rho_{-+}({\bf r},{\bf r},t)=\sum_j P_j \psi_{j-}({\bf r},t)\psi_{j+}^*({\bf r},t), \ee
with $\psi_{j\pm}=U_{\pm}\psi_j$. For an initial pure state where only one $P_j$ is unity while the others are zero, the visibility is the overlap integral of the two wave packets when the mutual phase is set to zero (here we assumed that $U_+^{\dag}U_-$ differs from $U_-^{\dag}U_+$ just by a global phase factor).
If instead of recombining the spin state at a specific time $t_0$ we let the two wave packets propagate with a Hamiltonian $\hat{H}_0$ that does not depend on spin and recombine them at time $t$ then the visibility would not change, as for a spin-independent evolution operator $U(t,t_0)=e^{-i\hat{H}_0t/\hbar}$ we have
\be V(t)={\rm Tr}\{U(t,t_0)\rho_{-+}(t_0) U^{\dag}(t_0,t)\}={\rm Tr}\{\rho_{-+}(t_0) U^{\dag}(t,t_0)U(t,t_0)\}={\bf Tr}\{\rho_{-+}(t_0)\}=V(t_0). \ee
This implies that after the sequence of gradient pulses it does not matter when the final $\pi/2$ pulse is performed so that the final state of the atoms at the output of the interferometer does not need to be similar to the initial state. In what follows we derive a more specific expression for the overlap integral as a function of the final wave-packet parameters, regardless of whether the state at the output is a minimal uncertainty state or whether it is similar to the initial state. Before that, we will re-derive the original expression of the HD theory.

\subsection{ The HD theory and its application to the calculation of the visibility}\label{subsec:HD theory}
In Ref.\cite{ESS_1} ESS  assumed a specific situation in which the accurate SGI completes a full reversal of the initial state at the output port or at least brings the two wave packets into a state that is exactly the same as the initial state in a specific frame of reference.
Imperfections are assumed to give rise to relative momentum shifts $\Delta{\bf p}$ and/or position shifts $\Delta{\bf r}$ of the final wave packets at the time $t=t_f$, such that
\be U_+^{\dag}U_-=e^{i(\Delta{\bf r}\cdot{\bf\hat{p}}+{\bf \Delta p}\cdot{\bf\hat{r}})/\hbar}. \label{eq:UpUm} \ee
Then it follows that the population difference at zero phase (the visibility) is given by
\begin{eqnarray} V &=& \frac{2}{\hbar}\langle \hat{S}_z\rangle_f^{\rm max}=\langle e^{i(\Delta{\bf p}\cdot{\bf\hat{r}}+{\bf\hat{p}}\cdot\Delta{\bf r})/\hbar}\rangle_0 \nonumber \\
&=& \int d^3r\, \rho({\bf r},{\bf r}-\Delta{\bf r})e^{-i{\bf \Delta p}\cdot{\bf r}/\hbar}=\int d^3p \rho({\bf p},{\bf p}-\Delta{\bf p})e^{i{\bf p}\cdot{\Delta r}/\hbar} .
\label{eq:HD0} \end{eqnarray}

We first consider the case of a pure momentum shift ($\Delta{\bf r}=0$). In this case the visibility is proportional to the Fourier transform of the density $\rho({\bf r})=\rho({\bf r},{\bf r})$. If the density along the axis of the momentum shift (say, along $\hat{z}$)  is a Gaussian with a standard deviation $\sigma_z$ then the visibility becomes
\be V_G(\Delta p_z)=e^{-\frac12\sigma_z^2\Delta p_z^2/\hbar^2}. \ee
In the case where the initial state is a minimal uncertainty Gaussian state where $\sigma_z\sigma_p=\hbar/2$ we have $V_G=\exp[-\frac{1}{2}(\Delta p/2\sigma_p)^2]$, which is an expression similar to that of Ref.~\cite{ESS_1}, where the momentum separation was defined as half the momentum separation in this derivation, and hence the factor 2 difference in the exponent.

If the initial distribution is a parabolic Thomas-Fermi distribution of a BEC
$\rho({\bf r})\propto {\rm max}\left\{1-\sum_{j=x,y,z} r_j^2/r_{j,\rm max}^2,0\right\}$ then the visibility is
\be V_{TF}(\Delta p_z)=
 \frac{15}{\xi^5} (3\sin\xi-\xi^2\sin \xi  - 3\xi\cos\xi). \ee
where $\xi=\Delta p_z z_{\rm max}/\hbar$.
This function can be fairly well approximated by $V_{TF}\approx e^{-\Delta p_z^2Z_{\rm max}^2/12\hbar}$.  This corresponds to a gaussian approximation for the spatial size $\sigma_z^{TF}\approx 0.41z_{\rm max}$. If the BEC wave packet was considered as a minimal uncertainty wave packet in its Gaussian approximation then we would use the momentum width $\delta p_z\approx 0.41\hbar/2z_{\rm max}$. However, note that a Thomas-Fermi spatial distribution is not a minimal uncertainty state so that the real width of the momentum distribution of this state is larger than $\hbar/2\sigma_z$. In addition, if the BEC is allowed to expand in free space then the atom-atom repulsion would increase the width of the momentum distribution after expansion, while the overlap integral between two BEC wave packets with different momenta does not change. It follows that the relevant momentum distribution for the HD formula remains $\hbar/2\sigma_z$ although the real momentum distribution has a larger size.

In the other case where {\it only} a position shift $\Delta z$ is involved, the visibility according to Eq.~(\ref{eq:HD0}) is a Fourier transform of the momentum distribution $\rho({\bf p})=\rho({\bf p},{\bf p})$.
For a Gaussian momentum distribution we obtain a result similar to the result for the visibility of a Gaussian position distribution
\be V_G(\Delta z)=e^{-\frac12\sigma_p^2\Delta z^2/\hbar^2}. \ee
For a Thomas-Fermi distribution we obtain numerically a visibility that drops like
\be V_{TF}(\Delta z)=e^{-\Delta z^2/2\sigma_{TF}^2}, \ee
where $\sigma_{TF}=0.6249\,z_{\rm max}$, unlike the case of the momentum mismatch.

Note that in Eq.~(\ref{eq:HD0}) $\rho({\bf r},{\bf r}')$ and $\rho({\bf p},{\bf p}')$ are the position and momentum representations, respectively, of the initial density matrix, which is assumed to be the density matrix of the final state in the case of a completely accurate operation. In general, while the visibility or spin-coherence in the SGI does not depend on the time where the final $\pi/2$ pulse is applied, the expectation value of the operator $e^{i(\hat{p}\cdot\Delta{\bf r}+{\bf r}\cdot\Delta{\bf p})/\hbar}$ does depend on time even in the case of propagation in free space represented by an operator $\hat{U}_0=e^{-i\hat{H}t/\hbar}$, as $\hat{U}_0^{\dag}e^{i({\bf p}\cdot\Delta{\bf r}+{\bf r}\cdot\Delta{\bf p})/\hbar}\hat{U}_0=e^{i({\bf p}\cdot(\Delta{\bf r}+\Delta{\bf p}t/m)+{\bf r}\cdot\Delta{\bf p})/\hbar}$. At long enough time even a slight momentum difference would shift the final distance between the two wave packets by a distance that is larger than the their initial width, giving rise to zero coherence according to Eq.~(\ref{eq:HD0}). This result is wrong. As we  show below the distance $\Delta{\bf r}$ should be taken at the time where the two wave packets are at their minimimal width rather than at the time of the measurement.

\subsection{ The generalized HD theory}

As noted above, the ESS theory has used a simplifying assumption that the final form of the wave packets at the output of the SGI is close to the initial state, which is a minimal uncertainty state or a mixture of minimal uncertainty states. It has also assumed that the main effect of inaccuracy is either a position shift or a momentum shift of the wave packets with respect to each other. Here we provide an analytic expression for the overlap integral of more general wave packets, which is consistent with the requirement that the overlap integral does not change in time if the evolution of the two wave packets is spin-independent.

In the more general case we assume a time-dependent form of the wave packets which describes an evolution of the wave packets in free space or in a potential with a quadratic dependence. This form is consistent with a Gaussian wave-packet evolution or with an evolution of a BEC in the approximation used by Castin and Dum~\cite{castin-dum} and applied in our work for the numerical analysis (see section~\ref{sec:theory} of this file). For simplicity we use here a one-dimensional model and assume that the wave packets completely overlap along the other dimensions. We start from the following form
\be \psi_i(z,t)=\frac{1}{\sqrt{\lambda_i}}\psi_0\left(\frac{z-Z_i}{\lambda_i}\right) e^{iP_i(z-Z_i)/\hbar}e^{i\frac{m}{2\hbar}\frac{\dot{\lambda}_i}{\lambda_i}\, (z-Z_i)^2}
\label{eq:CDform} \ee
where $Z_i(t),P_i(t)$ are the central position and momentum coordinates of the two wave packets ($i=1,2$), respectively, and $\lambda_i(t)$ are scaling factors due to expansion or focusing. Here $\psi_0(z)$ is the initial wave-function, which is assumed to have a flat phase. The form of the wave function in Eq.~(\ref{eq:CDform}) is a good approximation for the  solution of the Schr\"odinger equation for each wave packet if the sequence includes time-dependent potential gradients with a linear or quadratic spatial dependence over the volume occupied by the atoms. Note that the same equation is an exact solution for the evolution of a Gaussian wave packet in free-space or under the influence of time dependent quadratic potentials.

Let us first take $\psi_0$ to be a Gaussian of an initial width $\sigma_z$. In this case the overlap integral between the two wave packets becomes
\be \int dz\, \psi_1^*(z,t)\psi_2(z,t)
=\frac{1}{\sqrt{2\pi\sigma_z^2\lambda_1\lambda_2}}e^{-\eta\Delta z^2/4}e^{i\phi}\int dz\, e^{-\eta z^2-i(\Delta P/\hbar-\xi\Delta z)z}, \ee
where $\Delta P=P_1-P_2$, $\Delta z=Z_1-Z_2$,   and
$\phi$ is a phase which is not important here.
The parameters $\eta$ and $\xi$ are
\be \eta=\frac{1}{4\sigma_z^2}\left(\frac{1}{\lambda_1^2}+\frac{1}{\lambda_2^2}\right)+i\frac{m}{2\hbar}\left(\frac{\dot{\lambda}_1}{\lambda_1}-\frac{\dot{\lambda}_2}{\lambda_2}\right), \ee
\be \xi= \frac{m}{2\hbar}\left(\frac{\dot{\lambda}_1}{\lambda_1}+\frac{\dot{\lambda}_2}{\lambda_2}\right) +i\frac{1}{4\sigma_z^2}\left(\frac{1}{\lambda_1^2}-\frac{1}{\lambda_2^2}\right). \ee

By performing the integral and taking the absolute value we obtain
\be V=\sqrt{\frac{2\lambda_1\lambda_2}{\lambda_1^2+\lambda_2^2}}
\left|\exp\left[-\frac{(\Delta P/\hbar- \xi\Delta z)^2}{4\eta}-\eta\frac{\Delta z^2}{4}\right]\right| \ee
 In general, the argument of the exponential is complex. However, if no differential quadratic potentials are applied during the sequence then $\lambda_1=\lambda_2=\lambda$ and the wave packet size $\sigma(t)=\sigma_z\lambda(t)$ is the same for both .  The parameters $\eta(t)=1/2\sigma(t)^2$ and $\xi(t)=m\dot{\sigma}(t)/\hbar\sigma(t)$ are then real and equal for both wave packets. We will now examine this simpler case.

Given a pair of wave-packets with the same size $\sigma(t)=\lambda\sigma_z$ and expansion rate $\dot{\lambda}$ at a given time, we may assume that these two wave-packets evolve in free space. In this case the growth of the wave-packet size is given by
\be \sigma(t)=\sigma_0\sqrt{1+\omega^2 t^2}, \ee
where $t$ is the duration of time since the wave-packets were at their minimal size $\sigma_0$ with a flat phase (if $\dot{\sigma}>0$, expansion) or the duration of time that it will take the wave-packets to reach their minimal size (if $\dot{\sigma}<0$, focusing). In this case $\sigma_0=\sqrt{\hbar/2m\omega}$ and therefore  $\xi=\omega t /2\sigma^2$. We therefore obtain the visibility
\be V=\exp\left[-\frac{\sigma^2}{2}(\Delta P/\hbar- _0\omega t\Delta z/2\sigma^2)^2-\frac{\Delta z^2}{8\sigma^2}\right], \ee
where the argument of the exponent is explicitly time-dependent through $t$ and $\sigma=\sigma(t)$.
If we now use $\Delta z=\Delta z_0+\Delta P t/m$, where $\Delta z_0$ is the wave-packet separation at minimum wave-packet size, then we obtain
\be V=e^{-\sigma_0^2\Delta P^2/2\hbar^2}e^{-\Delta z_0^2/8\sigma_0^2}.
\label{eq:HD_projected} \ee
We have obtained exactly the original HD formula with the values of $\sigma_z$ and $\Delta z$ projected back (or forward) to the point where the two wave-packets are at their minimum size.
The procedure for calculating the overlap integral for wave-packets with the same size and phase curvature is then as follows: if we know the current Gaussian size $\sigma$ of the wave-packet and its phase curvature $\xi=(m\dot{\sigma}/\hbar\sigma$ then we can calculate the wave-packet parameters at the minimum wave-packet size. These are given by
\begin{eqnarray}
\sigma_0 &=& \frac{\sigma}{\sqrt{1+4\xi^2\sigma^4}} \\
t &=& \frac{4m}{\hbar}\frac{\sigma^4\xi}{1+4\sigma^4\xi^2} \\
\Delta z_0 &=& \Delta z-\frac{t}{m}\Delta P
\end{eqnarray}
Once the parameters $\sigma_0$ and $\Delta z_0$ are found, Eq.~(\ref{eq:HD_projected}) can be used for calculating the overlap integral. The overlap integral is exact for a Gaussian wave-packet and approximate for a BEC wave-packet of a Thomas-Fermi shape that can be approximated by a Gaussian.

In order to estimate the expected visibility drop due to imperfections in our interferometric sequence we use the expression in Eq.~(\ref{eq:HD_projected}), which is similar to the original HD expression except that it uses projected values for the wave packet size and the position shift $\Delta z$. We estimate that in most of the experimental situations the final position shift $\Delta z$ is negligible so that the main source of visibility reduction may be the momentum shift $\Delta P$. In our experiment we estimate that relative timing imprecision $\Delta T/T$ and current imperfections $\Delta I/I$ are of the order of $10^{-3}$. If follows that momentum shifts can be minimized, upon a proper optimization procedure, to the order of $\Delta P\sim 10^{-3}P$. A typical value of the momentum in our experiment is $P=2\pi\hbar/1\,\mu{\rm m}$, so that $\Delta P/\hbar\sim 2\pi\cdot 10^{-3}\,\mu{\rm m}^{-1}$.. The initial BEC Gaussian size $\sigma_z$ is of the order of $1\,\mu$m. It follows that the maximum visibility is expected to be
\be[ V_{\rm max}\sim \exp[-(2\pi\cdot 10^{-3})^2/2]\sim 1-2\cdot 10^{-5}. \ee
Even if our optimal $\Delta P$ is larger by an order of magnitude from this estimation we should still expect a negligible reduction of the visibility by less than 1\%. In view of the result, this suggests that our experiment contains some unknown source of imprecision that is yet to be discovered.

\section{Time-irreversibility and the Stern-Gerlach interferometer}

The Stern-Gerlach interferometer (SGI) was used in previous theoretical studies for demonstrating the irreversibility of quantum operations.  In particular, the SGI apparatus was sssumed to have a symmetric structure such that the splitting and stopping operations are reversed by the accelerating and stopping at the second half of the sequence. Ideally the second half of the SGI sequence is expected to bring the two wave packets into their original spatial state where they overlap with each other at the initial position. However, this goal of the reverse operations is not a necessary requirement for spin-coherence. If the two wave-packets overlap at any position and with any wave-packet size and wave-front full spin coherence is expected to be achieved, as long as the overlap integral is 1, namely the two wave-packet have the same spatial wave function which is not necessarily the original one.

Here we show that even if the structure of the SGI is not symmetric as it was envisioned in the past, its coherence still reflects the quality of time reversal in the system, which is determined by the precision and stability of the quantum operations.
Let us consider an initial spatial state represented by a wave packet $\psi_0({\bf r})$ and a spin state that is an equal superposition $\frac{1}{\sqrt{2}}(|+\rangle+|-\rangle)$ of the two spin eigenstates $|\pm\rangle$.
We may represent the ideal operation of the SGI by $|\Psi_{\rm final}\rangle=\hat{U}_{\rm ideal}(t_f,0)|\Psi_0\rangle$, where $|\Psi_0\rangle$ is the initial spin state superposition with the spatial wave function $\psi_0({\bf r})$. The ideal SGI evolution operator can be written as
\be \hat{U}_{\rm ideal}(t,0)=U_{I+}(t,0)|+\rangle\langle+|+U_{I-}(t,0)|-\rangle\langle -|, \ee
where $U_{I\pm}$ are the ideal evolution operators for the two spin states, that take the two wave-packets in two separated paths and are supposed to return them back to the exactly the same spatial state, namely $U_{I+}(t_f,0)=U_{I-}(t_f,0)$ for the final time $t_f$ (but they are different for other times $0<t<t_f$).  Here we also implicitely assume that the SGI ideal operation is insensitive to the initial state $\psi_0$ as it brings the two wave packets into full overlap with each other regardless of the initial state.

Now consider the real (non-ideal) operations $\tilde{U}_+(t_f,0)$ and $\tilde{U}_-(t_f,0)$ which include the effects of imprecision or instability. The visibility of the spin population signal at the end of the process is given by the overlap integral of the two actual wave-functions $\tilde{\psi}_{\pm}({\bf r},t_f)=\tilde{U}_{\pm}(t_f,0)\psi_0({\bf r})$. This overlap integral can be written as
\be \langle \tilde{\psi}_+(t_f)|\tilde{\psi}_-(t_f)\rangle=\langle \psi_0|\tilde{U}_+(t_f,0)^{\dag}\tilde{U}_-(t_f,0)|\psi_0\rangle=
\langle \psi_0|\tilde{U}_+^{\dag}U_{I+}U_{I-}^{\dag}\tilde{U}_-|\psi_0\rangle. \ee
The last step is possible due to the equality of the ideal evolution operators, such that  $U_{I+}U_{I-}^{\dag}=U_{I+}U_{I+}^{\dag}=U_{I+}U_{I+}^{-1}=1$ for the final time $t_f$.

It follows that the overlap integral, which determines the spin coherence of the SGI of an arbitrary symmetry, is determined by the precision of the actual SGI operations $\tilde{U}_{\pm}(t_f,0)$ which is then reversed by the ideal operators $U_{I\pm}(t_f,0)^{-1}=U_{I\pm}(0,t_f)$. If each of the forward and then backward time evolution operations lead to a wave function close enough to the original wave-function $\psi_0$,, {\it i.e.}, if $U_{I+}^{-1}\tilde{U}_+\sim 1$ and $U_{I-}^{-1}\tilde{U}_-\sim 1$ then the sufficient condition for high spin coherence is achieved. Note that this is not a necessary condition as both operations may lead to a wave function that is different than $\psi_0$ but still overlapping between the two spin states. The necessary condition for spin-coherence is in fact that there exist ideal unitary operations $U_{I+}$ and $U_{I-}$ which are close to the actual evolution operators at any time, such that $U_{I+}(t_f,0)=U_{I-}(t_f,0)$ at the final time $t_f$.

\section{ Multi-shot visibility and its standard error}
\label{sec:Vmulti}

\subsection{ Multi-shot visibility for a Gaussian noise}

Multi-shot visibility characterizes an interference pattern obtained by averaging over many interference patterns resulting from possibly different stopping distances $d$, relative velocities (imperfect stopping) and phases of the two wave packets at different experimental shots.
Suppose that a single interference pattern is formed from a superposition of two wave packets $\psi_{j+}({\bf r})$ and $\psi_{j-}({\bf r})$ corresponding to the two spin states during the splitting stage. After the stopping these two wave packets represent two localized atomic states separated by a distance $d$ and after time-of-light (TOF) they represent extended overlapping wave packets with a periodic phase difference.
The averaged interference pattern is then proportional to the sum,
\be \sum_{j=1}^N \left[||\psi_{j+}({\bf r})|^2+|\psi_{j-}({\bf r})|^2+
\psi_{j+}^*({\bf r})\psi_{j-}({\bf r})+\psi_{j-}^*({\bf r})\psi_{j+}({\bf r})|\right], \ee
where the visibility is the magnitude of the last two term (interference term) relative to the first two terms.  If the phase of the individual interference patterns are distributed over some range $\delta\phi$ the sum gives rise to a visibility that it reduced with respect to the visibility of the single shot patterns.
We take the wave packets to be normalized $\int d^3{\bf r}\, |\psi_{j\pm}|^2=1$ and the state in each shot to be an equal superposition of the two wave packets. If the interference pattern in each shot has a large number of fringes (equivalent to a large separation of the wave packets after stopping relative to the minimal size) then we can separate the interference term from the two first terms by performing a Fourier transform of the multi-shot pattern. The Fourier transform at the wave-vector $k=2\pi/\lambda$, where $\lambda$ is the periodicity extracts the interference term while the Fourier transform at $k=0$ extracts the two first terms. The ratio between these two Fourier transforms is the visibility, which is then given by
\be V_N=\frac{1}{ N}\left|\sum_{j=1}^N \int dz\, e^{-ik_tz} \psi_{j+}^*(z, t)\psi_{j,-}(z, t) \right|,
\label{eq:V_N} \ee
where $t$ is the imaging time
and $k_t$ is the mean wave-vector of the single interference patterns.
Here we ignored the transverse coordinates $x$ and $y$ and assumed a meaningful dependence of the wave-function only along the $z$ direction.

Let us now assume that the main source of fluctuations in our system is current variations during the splitting pulse (of duration $T_1$). As we have shown in section~\ref{sec:phasespace} if the stopping pulse is perfect then in the long time-of-flight limit the interfering wave packets have the scaled form as the initial wave packets after splitting, projected backward to the time $t=0$ at the middle of the splitting pulse.  In our case where the atoms move very little during the pulse this time may be taken as the time just after the pulse. At this time the spatial shape of the two wave packets is almost the same while phase difference between them at the $j$'th shot is
\be \Delta\phi_j(z)=\frac{\Delta m_Fg_F}{\hbar}\int dt\, B_j(z,t) \approx k_j(z-z_0)+\delta\phi_j, \label{eq:Deltaphi} \ee
where $\hbar k_j=\Delta F_j T_1$ is the differential momentum applied by the splitting pulse at the experimental shot $j$ ($\Delta F_j=\Delta m_Fg_F\mu_B (\partial B_j/\partial z)_{z=z_0}$ being the differential force applied during the splitting on the atoms of spins differing by $\Delta m_F$). Here $z_0$ is the position of the center of the quadrupole field, where the field from the splitting pulse is zero, and $\delta\phi_j=\Delta m_Fg_F\mu_BB_0T_0/\hbar$ is the  phase from the homogeneous bias field $B_0$, which accumulates during a time $T_0>T_1$ when the atoms occupy different spin states and may differ from shot to shot.

Taking initial shape of the wave packets to be Gaussian, the form of the wave packets just after the projected time $t=0_+$ becomes
\be \psi_{j,\pm}(z,0_+)=\psi_0(z-z_j)e^{i \phi_{j,\pm} (z)}\propto e^{-(z-z_j)^2/4\sigma^2}e^{i \phi_{j, \pm}(z)}, \ee
where $z_j$ is the initial center position of the wave packet in the experimental cycle $j$, which is assumed to have a random distribution of variance $\langle \delta z^2\rangle$, which is uncorrelated with the phase or momentum fluctuations. These phase and momentum shifts are contained in the phase terms $\phi_{j,\pm}$, whose difference $\phi_{j,+}-\phi_{j,-}=\Delta\phi_j$ is given in Eq.~(\ref{eq:Deltaphi}).

We assume that the many experimental shots represent a Gaussian distribution of the momentum and phase fluctuating parameters $k_j$ and $\delta\phi_j$.
For a Gaussian distribution of phase $\phi$ we use the identity
\be \langle e^{i\phi}\rangle=e^{i\langle\phi\rangle}e^{-\langle \delta\phi^2\rangle/2},
\label{eq:eiphi} \ee
where $\delta\phi=\phi-\langle \phi\rangle$. We then obtain in the limit $N\to\infty$
\be \frac{1}{N}\sum_j \psi_{j+}^*(z)\psi_{j-}(z)\to
G_{\bar{\sigma}}(z-\bar{z})e^{i\langle \Delta\phi\rangle}
e^{-\langle \delta k^2\rangle (z-z_0)^2/2}e^{-\langle \delta\phi^2\rangle/2},
\label{eq:V_z} \ee
where $G_{\bar{\sigma}}(z-\bar{z})=\langle |\psi_0(z-z_j)|^2\rangle$ is a Gaussian with extended width $\bar{\sigma}=\sqrt{\sigma^2+\langle \delta z_j^2\rangle}$ centered around the average initial position $\bar{z}=\langle z_j\rangle$, which represents the normalized sum over many Gaussian envelopes each having a width $\sigma$.
This implies that the contribution of current fluctuations leading to momentum fluctuations grows for wave packet parts that are initially located further away from the quadrupole center, giving rise to a chirped visibility pattern.

In order to obtain the overall visibility defined in Eq.~(\ref{eq:V_N}) we Fourier transform Eq.~(\ref{eq:V_z}) and find
\be V_{N\to \infty}=\frac{1}{\sqrt{1+\bar{\sigma}^2\langle \delta k^2\rangle}}e^{-\langle \delta\phi^2\rangle/2}
\exp\left[-\frac{1}{2}\frac{(\bar{z}-z_0)^2\langle \delta k^2\rangle}{1+\bar{\sigma}^2\langle \delta k^2\rangle}\right]
\label{eq:V_Ntoinf} \ee

In Fig.~3 of the main text we have used this equation as the basis for the calculation of the solid theoretical curves. We neglect the fluctuations of the bias field ($\delta\phi=0$) and assume that the momentum fluctuations are caused by current fluctuations in the chip wires.
The momentum uncertainty is then
\be \delta k_{\rm rms}=\frac{\mu_B}{2\hbar}\left.\frac{\partial B}{\partial z}\right|_{z=\bar{z}}T_1\frac{\delta I_{\rm rms}}{I}=\langle k\rangle \epsilon, \ee
where  $\epsilon=\delta I_{\rm rms}/I$ is the relative current fluctuation.
In our case we obtain $\langle k\rangle/T_1=0.86\, (\mu {\rm m} \mu {\rm s})^{-1}$, while $|\bar{z}-z_0|\approx 5\,\mu$m.

\subsection{ Standard error of multi-shot visibility}

Let us now consider the uncertainty $\delta V_N$ which is the standard error of the multi-shot visibility $V_N$ due to the finite number $N$ in a sample. For simplicity we consider only global phase fluctuations, such that the multi-shot visibility is
\be V_N=\frac{1}{N}\left|\sum_{j=1}^N e^{i\phi_j}\right|\equiv \frac{1}{N}\left|S_N\right|. \ee

We have $V_N^2=N^{-2}[({\rm Re}S_N)^2+({\rm Im}S_N)^2]$, which has the explicit form
\be \frac{1}{N^2}[{\rm Re}S_N]^2=\frac{1}{4N^2}
\sum_{j,k}[e^{i(\phi_j+\phi_k)}+e^{i(\phi_j-\phi_k)}+{\rm c.c.}] \ee
\be \frac{1}{N^2}[{\rm Im}S_N]^2=\frac{1}{4N^2}
\sum_{j,k}[-e^{i(\phi_j+\phi_k)}+e^{i(\phi_j-\phi_k)}+{\rm c.c.}]. \ee
By summing over the real part and imaginary part and separating the sum $\sum_j e^{i(\phi_j-\phi_k)}$ into the case $j=k$ and $j\neq k$ we obtain
\be V_N^2=\frac{1}{N}\left[1+\frac{1}{2N}
\sum_{j\neq k}(e^{i(\phi_j-\phi_k)}+{\rm c.c.})\right]. \ee
By taking an average over ensembles with Gaussian distribution of the phases and using Eq.~(\ref{eq:eiphi}), the average over each of the $N(N-1)$ terms in the sum over $j\neq k$ becomes  $e^{-\langle \delta\phi^2\rangle^2}=\langle V_N\rangle^2$ and we have
\be \langle V_N^2\rangle-\langle V_N\rangle^2=\frac{1}{N}(1-\langle V_N\rangle^2). \label{eq:dVN} \ee
When the distribution of phases is narrow and $\langle V_N\rangle\sim 1$ the standard error of the multi-shot visibility is small, but when the visibility is small the standard error goes to the limit $\delta V_N\sim 1/\sqrt{N}$.

\section{ Dephasing due to electrons}

Let's begin by calculating how a single atom interacts with a single electron.  The electron produces a current in the atom-chip wire, and the atom is in a superposition of spin states:  its magnetic moment is either pointing towards the electron or away from it. Let ${|e \rangle}$ be the initial state of the electron, and $\left[ \ap +\am \right]/\sqrt{2}$ the initial state of the atom, before they interact and entangle.  When the interaction between the atom and electron turns on, it entangles them and their state is

\begin{equation}
  \vert \Psi \rangle = \left[ \em\ap +\ep\am \right]/\sqrt{2},
\end{equation}

\noindent where the combinations $\em\ap$ and $\ep\am$ remind us that if the atom gains momentum then the electron looses momentum (along the same axis) and vice versa. But while the states $\ap$ and $\am$ are orthogonal atom spin states, $\ep$ and $\em$ are not necessarily orthogonal. The absolute value of their inner product $|\langle e-|e+ \rangle|$ could range from 0 to 1 depending on the strength of the
entangling interaction.

To get the interference pattern, we have to calculate the absolute value squared of $\vert \Psi\rangle$.  The bra-ket notation is not especially convenient, but we know that the result is that the interference pattern of the atom will be multiplied by $\vert {\langle} {e-} | e + \rangle \vert$, which can equal 1 (the electron carries away no information about the atom superposition and does not affect its visibility) or can be less than 1 (in which case the visibility cannot be 1, regardless of relative clock times, etc.).  Let's assume that the atom has picked up (lost) momentum $p_z$, in which case the electron has lost (picked up) momentum $p_z$.  Taking 20 T/m for the gradient, 10$^{-23}$ J/T for the Bohr magneton, and $g_F = -1/2$, we get $10^{-22}$ J/m for the force; and if the force lasts about 20 $\mu$s, then $p_z \approx 2 \times 10^{-27}$ kg m/s.

The next step is to consider $N$ atoms instead of one atom.  These atoms are initially in a product state $\left[ \ap +\am \right]^{\otimes N} 2^{-N/2}$.  If we expand this expression in a binomial expansion, the important terms are not those with the largest magnetic moment ($N$ times a single magnetic moment) but those with the greatest degeneracy ($\sqrt{N}$ times a single magnetic moment).  Thus the total momentum exchange $P_z$ is not $p_z$ and not $Np_z$ but $P_z \approx \sqrt{N} p_z$. If $N\approx 10^4$, then $P_z \approx 100 p_z \approx 2 \times 10^{-25}$ kg m/s.  Now, if all this momentum were transferred to a single electron, it would be a very serious effect, as follows.  The width of the wire in the $z$ direction is 2 $\mu$m, thus $\Delta z\le 2 \mu$m.  Therefore $\Delta P_z \ge \hbar/(2\times 2~ \mu$m) $\approx 3 \times 10^{-29}$ kg $\cdot$ m/s, which is four orders of magnitude smaller than $P_z$, so the effect of the momentum exchange should be clearly visible.  But the $N$ atoms do not all couple to one electron.  There are about $10^{14}$ electrons around, each with $\Delta z\le 2~ \mu$m, and the average momentum gain or loss of each electron is about $3 \times 10^{-39}$ kg $\cdot$ m/s, undetectable according to the uncertainty principle.  This division by $10^{14}$ is probably not justifiable, because most of the electrons are deep in the Fermi sea and don't absorb or lose any momentum.  But taking the estimate that only about 1\% of the electrons are near the Fermi level and only these interact, we still get $10^{12}$ electrons and an average momentum gain or loss per electron of about $3 \times 10^{-37}$ kg $\cdot$ m/s, still undetectable according to the uncertainty principle.  Hence there is no dephasing of the atoms via the electrons, because no measurement procedure on the electrons can reveal the spin state of the atoms.

\section{ Comparison to the state-of-the-art (French SG experiments)}

For completeness, we compare our experiments to previous SG type interferometry experiments although these were significantly different\,\cite{OldSG0,OldSG2,OldSG3,OldSG4,OldSG5,OldSG6,OldSG7,OldSG8, Marechal2000,OldSG9,OldSG10}. Baudon, Robert, and colleagues, have realized a series of elaborate SGI experiments over a period of 15 years and, more recently, even applied the SG effect to the study of twin-atom angular momentum coherence in the photodissociation of $H_2$ \cite{Carlos}. While their longitudinal beam interferometer did observe interference fringes, this interferometer is very different from the interferometers presented here. Most importantly, as explained in detail in one of their papers\,\cite{OldSG6}, their experiment is not an analogue of the full-loop configuration as only splitting and stopping operations were realized (i.e., no recombination); namely, wavepackets exit the interferometer with the same separation as the maximal separation achieved within.
Fig.\,2 of Ref.\,\cite{OldSG6} shows the scheme of the beam experiment. As can be seen, only a splitting and a stopping pulse are applied. This creates what the authors call, a ``beaded atom" \cite{OldSG2}. We have not found anywhere in the many papers published by this group (only some of which we referenced) evidence of four operations being applied as required for a full-loop configuration, whether the experiment was with longitudinal or transverse gradients. This also means that these experiments could not probe imprecision as an origin of TI or the HD effect.

Furthermore, also within the framework of the half-loop configuration, there are many differences from our experiment. Mainly, the beam experiments could not image high visibility spatial interference fringes, as presented in Fig.\,1 of our main text. In fact, we believe that no spatial interference fringes were observed at all, and that the spatial modulation presented in \cite{OldSG8,OldSG9} is an ensemble of many trajectories, each undergoing Ramsey interferometry, and not a result of any coherent spatial splitting. Let us explain why. First we divide the explanation into two scenarios: a longitudinal effect and a transverse effect. The French group studied both effects.

Longitudinal effect: It is clear from their papers that no spatial fringes have been observed in the longitudinal direction when the splitting was longitudinal (i.e. along the propagation of the beam). In contrast, such longitudinal spatial fringes are precisely what we observe in our experiment.  Indeed, this would have been an impossible task in the beam experiments taking into account the beam velocity spread and the fact that the signal is a sum of an enormous amount of wave packet pairs, situated at different positions along the beam (and indeed there is no claim in the papers that longitudinal spatial fringes were observed). Similarly, the beam experiments could not verify the existence of two independent and separated wavepackets, while in our case we image them explicitly (see experimental sections of this SM).
Let us also mention that the achieved separation in the French work for seeing any interference pattern (they used a spin population signal) was only a few Angstrom\,\cite{OldSG6} while in our experiment it is a few $\mu$m (they report inducing much larger separations, but with no matter-wave interference pattern\,\cite{OldSG4}). More importantly, their separation is the same as the wavepacket width (or coherence length), while ours went up to 18 times the wavepacket width.

Transverse effect: Reading the abstract of Ref.\,\cite{OldSG9}, one may be slightly confused to think that a spatial interference pattern has been observed: ``When a static radial magnetic gradient is used, the beam profile is modulated by interference. The transverse pattern, which can be translated at will by adding a homogeneous field, is observed for the first time using a multi-channel electron multiplier followed by a phosphor screen and a CCD camera."
Indeed, Figs. 12 and 13 in \cite{OldSG9} show beautiful spatial modulations of the signal in the plane perpendicular to the beam propagation axis. However, we believe these are not spatial interference fringes (i.e. originating from a spatial splitting), and we find in the published studies no proof or clear statement that they are.

What we believe is the dominant effect in producing the observed transverse modulation is that the atomic beam gives rise to many parallel (with a slight diverging angle) trajectories, each undergoing Ramsey interferometry. The quadrupole field (in the plane transverse to the beam propagation, see Fig.\,9 of\,\cite{OldSG9}) acts as a spatially varying ``phase plate" which gives a different internal phase to each of the Ramsey interferometers, thereby giving a different spin population at the output of each interferometer. This modulation of the spin output then appears as a spatial modulation in the plane perpendicular to the beam propagation axis. Indeed, Fig.\,10 shows that a full-cycle (2$\pi$) spin-population oscillation requires a change of 14\,mA, corresponding to 6.3\,mG (the caption states 0.45\,mG/mA). On the other hand, the caption of Fig.\,12 states that the 10\,mm-diameter of the phosphor screen corresponds to 2.4\,mm at the interferometer mid-point (10\,mm x 314\,mm/1202\,mm from Fig.\,1).  A gradient of 12.75\,mG/mm (fourth panel of Fig\,12) produces a pattern with a modulation periodicity of 2.45\,mm (peak-to-peak distance measured along the y-axis of the MCP image).  In addition, 2.45\,mm at the MCP corresponds to 0.59\,mm at the interferometer mid-point (= 2.45\,mm x 2.4\,mm/10\,mm), which in turn corresponds to a change in field of 7.5\,mG (= 0.59\,mm x 12.75\,mG/mm), which is very close to the 6.3\,mG obtained above for the longitudinal full-cycle spin-population oscillation.  Similarly, eqns.(7) and (9) of Ref.\,\cite{OldSG9} both yield phase shifts $\varphi$ of $\pi$ radians using the experimental values for the homogeneous field $B_H$ [eqn.(7)] and the gradient G [eqn.(9)].
We have also calculated the transverse magnetic force acting on the two spin wavepackets (by using the above gradient stated by the authors), and find that indeed a very small separation is induced (on the order of an Angstrom) so that the two wavepackets still transversely overlap at the analyzer such that a Ramsey sequence could be completed, the latter being the essence of the longitudinal SGI when no significant splitting is applied. We believe this explains the main features observed.

For the previous experiment, we also considered the possibility of a spatial interference scenario, i.e., in which a spatial splitting is the source of the observed fringes. Even if we consider the above separation of an Angstrom as a two-point source (e.g., like a double slit experiment), there are many two-point sources identical to this one which are slightly shifted in their position along the transverse dimension up to the width of the beam of a few millimeters. As this is also the periodicity of the observed fringes, any fringe should be washed out and we cannot see how any visibility may survive.

Consequently, to the best of our understanding, the only previous SG spatial interference pattern was achieved in our own work \cite{machluf}, in which low visibility fringes were observed.
To conclude, it is clear that the new experiments reported here go well beyond the previous state-of-the-art.



\begin{thebibliography}{99}

\bibitem{stern-gerlach} W. Gerlach and O. Stern, Z. Phys. {\bf 9}, 349 (1922).
\bibitem{SLB} M.O. Scully, W.E. Lamb Jr., and A. Barut, Foundations of Physics {\bf 17}, 575 (1987).
\bibitem{Wigner} E. P. Wigner, Am. J. Phys. {\bf 31}, 6 (1963). The device is mentioned earlier by D. Bohm in [5].

\bibitem{briegel} H.J. Briegel, B.-G. Englert, M.O. Scully, and H. Walther, in {\it Atom Interferometry} (ed. Berman, P. R.), 240 (Academic Press, 1997).
\bibitem{Bohm} D. Bohm, {\it Quantum Theory}, 604-605 (Prentice-Hall, Englewood Cliffs, 1951).
\bibitem{ESS_1} B.-G. Englert, J. Schwinger, and M. O. Scully, Found. of Phys. {\bf 18}, 1045 (1988).
\bibitem{ESS_2} J. Schwinger, M. O. Scully, and B.-G. Englert, Z. Phys. D {\bf 10}, 135 (1988).
\bibitem{ESS_3} M.O. Scully, B.-G. Englert, and J. Schwinger, Phys. Rev. A {\bf 40}, 1775 (1989).
\bibitem{englert_1} B.-G. Englert, Z. Naturforsch A {\bf 52}, 13 (1997).
\bibitem{englert_2} B.-G. Englert, Eur. Phys. J. D {\bf 67}, 238 (2013).
\bibitem{SM} See Supplementary Material.
\bibitem{caldeira} T. R. de Oliveira and A. O. Caldeira, Phys. Rev. A {\bf 73}, 042502 (2006).
\bibitem{Devereux} M. Devereux, Can. J. of Phys., {\bf 93}, 1382 (2015).

\bibitem{OldSG0} J. Robert et. al., Europhys. Lett. {\bf 16}, 29 (1991).
\bibitem{OldSG2} Ch. Miniatura et al., Journal de Physique II {\bf 1(4)}, 425 (1991).
\bibitem{OldSG3} Ch. Miniatura et al., App. Phys. B {\bf 54}, 347 (1992).
\bibitem{OldSG4} J. Robert et al., Journal de Physique II {\bf 11(2)}, 601 (1992).
\bibitem{OldSG5} Ch. Miniatura et al., Phys. Rev. Lett. {\bf 69}, 261 (1992).
\bibitem{OldSG6} S. Nic Chormaict et. al., J. Phys. B: At. Mol. Opt. Phys. {\bf 26}, 1271 (1993).
\bibitem{OldSG7} J. Baudon, R. Mathevet, and J. Robert, J. Phys. B: At. Mol. Opt. Phys. {\bf 32}, R173 (1999).
\bibitem{OldSG8} M. Boustimi et al., Phys. Rev. A {\bf 61}, 033602 (2000).
\bibitem{Marechal2000}E. Mar\'echal, R. Long, T. Miossec, J. L. Bossennec, R. Barb\'e, J. C. Keller, and O. Gorceix, Phys. Rev. A  {\bf 62}, 53603 (2000).
\bibitem{OldSG9} B. Viaris de Lesegno et al., Eur. Phys. J. D {\bf 23}, 25 (2003).
\bibitem{OldSG10} K Rubin et al., Laser Phys. Lett. {\bf 1}, 184 (2004).

\bibitem{Monroe1996} C. Monroe, D. M. Meekhof, B. E. King, and D. J. Wineland, Science {\bf 80}, 1131 (1996).

\bibitem{Mandel2003} O. Mandel, M. Greiner, A. Widera, T. Rom, T. W. H\"ansch, and I. Bloch, Phys. Rev. Lett. {\bf 91}, 010407 (2003).

\bibitem{Steffen2012} A. Steffen, A. Alberti, W. Alt, N. Belmechri, S. Hild, M. Karski, A. Widera, and D. Meschede, Proc. Natl. Acad. Sci. {\bf 109}, 9770 (2012).

\bibitem{Mizrahi2013} J. Mizrahi, C. Senko, B. Neyenhuis, K. G. Johnson, W. C. Campbell, C. W. S. Conover, and C. Monroe, Phys. Rev. Lett. {\bf 110}, 203001 (2013)

\bibitem{Kienzler2016}D. Kienzler, C. Fl\"uhmann, V. Negnevitsky, H. Y. Lo, M. Marinelli, D. Nadlinger, and J. P. Home, Phys. Rev. Lett. {\bf 116}, 140402 (2016).

\bibitem{Johanning2009} M. Johanning, A. Braun, N. Timoney, V. Elman, W. Neuhauser, and C. Wunderlich, Phys. Rev. Lett. {\bf 102}, 073004 (2009).

\bibitem{Werth2002} G. Werth, H. H\"affner, and W. Quint, Adv. At. Mol. Opt. Phys. {\bf 48}, 191 (2002).
\bibitem{zurek1} W.H. Zurek, Physics Today {\bf 44}, 36 (1991); updated version: arXiv:quant-ph/0306072 (2003).
\bibitem{Heisenberg}  W. Heisenberg, {\it Die physikalischen Prinzipien der Quantentheorie}, 33-34 (Hirzl, Leipzig, 1930).
\bibitem{keil} M. Keil et al., J. Modern Optics {\bf 63}, 1840 (2016).
\bibitem{machluf} S. Machluf, Y. Japha, and R. Folman, Nature Comm. {\bf 4}, 2424 (2013).



\bibitem{margalit} Y. Margalit et al., Science {\bf 349}, 1205 (2015).
\bibitem{zhou} S. Zhou et al., Phys. Rev. A {\bf 93}, 063615 (2016).

\bibitem{LMT} Paul Hamilton et al., Phys. Rev. Lett. {\bf 114}, 100405 (2015).
\bibitem{SubShotNoise}  D. V. Strekalov, N. Yu and K. Mansour, NASA Tech Briefs (2011); http://www.techbriefs.com/component/content/article/11341.
\bibitem{applications} E. Danieli, J. Perlo, B. Bl\"umich, and F. Casanova, Phys. Rev. Lett. {\bf 110}, 180801 (2013).


  \bibitem{Shuyu_phasespace} S. Zhou et al., Phys. Rev. A {\bf 90}, 033620 (2014).
\bibitem{SGWigner} P. Gomis and A. P\'erez, Phys. Rev. A {\bf 94}, 012103 (2016).

\bibitem{manfredi} G. Manfredi and P.-A. Hervieux, Phys. Rev. Lett. {\bf 100}, 050405 (2008).

\bibitem{waldherr} G. Waldherr and G. Mahler, Eur. Phys. Lett. {\bf 89}, 40012 (2010).
\bibitem{gaspard2} P. Gaspard and M. Nagaoka, J. Chem. Phys. {\bf 111}, 5676 (1999).
\bibitem{biele} R. Biele and R. D'Agosta, J. Phys. Cond. Matt. {\bf 24}, 273201 (2012).
\bibitem{cucchietti} F.M. Cucchietti, D.A.R. Dalvit, J.P. Paz, and W.H. Zurek, Phys. Rev. Lett. {\bf 91}, 210403 (2003).
  \bibitem{crank-nicolson} J. Crank and P. Nicolson, Proc. Cam. Phil. Soc. {\bf 43}, 50 (1947).
\bibitem{castin-dum} Y. Castin and R. Dum, Phys. Rev. Lett. {\bf 77}, 5315 (1996).

\bibitem{Carlos} Carlos R. de Carvalho et al., Euro. Phys. Lett. {\bf 110}, 50001 (2015).

\bibitem{Ketterle1999} W. Ketterle, D.S. Durfee, and D.M. Stamper-Kurn,
arXiv:cond-mat/9904034 (1999).
  \end{thebibliography}
\end{document}